\documentclass[12pt,titlepage]{utarticle}

\usepackage{amsmath}
\usepackage{amsfonts}
\usepackage{amssymb}
\usepackage{amsthm}
\usepackage{mathtools}
\usepackage{graphicx}
\usepackage{color}
\usepackage{ucs}
\usepackage[utf8x]{inputenc}
\usepackage{hyperref}

\definecolor{aqua}{rgb}{0, 1.0, 1.0}
\definecolor{fuschia}{rgb}{1.0, 0, 1.0}
\definecolor{gray}{rgb}{0.502, 0.502, 0.502}
\definecolor{lime}{rgb}{0, 1.0, 0}
\definecolor{maroon}{rgb}{0.502, 0, 0}
\definecolor{navy}{rgb}{0, 0, 0.502}
\definecolor{olive}{rgb}{0.502, 0.502, 0}
\definecolor{purple}{rgb}{0.502, 0, 0.502}
\definecolor{silver}{rgb}{0.753, 0.753, 0.753}
\definecolor{teal}{rgb}{0, 0.502, 0.502}

\makeatletter
\newdimen\itex@wd%
\newdimen\itex@dp%
\newdimen\itex@thd%
\def\itexspace#1#2#3{\itex@wd=#3em%
\itex@wd=0.1\itex@wd%
\itex@dp=#2ex%
\itex@dp=0.1\itex@dp%
\itex@thd=#1ex%
\itex@thd=0.1\itex@thd%
\advance\itex@thd\the\itex@dp%
\makebox[\the\itex@wd]{\rule[-\the\itex@dp]{0cm}{\the\itex@thd}}}
\makeatother

\makeatletter
\newif\if@sup
\newtoks\@sups
\def\append@sup#1{\edef\act{\noexpand\@sups={\the\@sups #1}}\act}%
\def\reset@sup{\@supfalse\@sups={}}%
\def\mk@scripts#1#2{\if #2/ \if@sup ^{\the\@sups}\fi \else%
  \ifx #1_ \if@sup ^{\the\@sups}\reset@sup \fi {}_{#2}%
  \else \append@sup#2 \@suptrue \fi%
  \expandafter\mk@scripts\fi}
\def\tensor#1#2{\reset@sup#1\mk@scripts#2_/}
\def\multiscripts#1#2#3{\reset@sup{}\mk@scripts#1_/#2%
  \reset@sup\mk@scripts#3_/}
\makeatother

\makeatletter
\newbox\slashbox \setbox\slashbox=\hbox{$/$}
\def\itex@pslash#1{\setbox\@tempboxa=\hbox{$#1$}
  \@tempdima=0.5\wd\slashbox \advance\@tempdima 0.5\wd\@tempboxa
  \copy\slashbox \kern-\@tempdima \box\@tempboxa}
\def\slash{\protect\itex@pslash}
\makeatother

\def\clap#1{\hbox to 0pt{\hss#1\hss}}

\let\oldroot\root
\def\root#1#2{\oldroot #1 \of{#2}}

\DeclareSymbolFont{symbolsC}{U}{txsyc}{m}{n}
\SetSymbolFont{symbolsC}{bold}{U}{txsyc}{bx}{n}
\DeclareFontSubstitution{U}{txsyc}{m}{n}

\DeclareSymbolFont{stmry}{U}{stmry}{m}{n}
\SetSymbolFont{stmry}{bold}{U}{stmry}{b}{n}

\makeatletter
\def\re@DeclareMathSymbol#1#2#3#4{%
    \let#1=\undefined
    \DeclareMathSymbol{#1}{#2}{#3}{#4}}
\re@DeclareMathSymbol{\neArrow}{\mathrel}{symbolsC}{116}
\re@DeclareMathSymbol{\neArr}{\mathrel}{symbolsC}{116}
\re@DeclareMathSymbol{\seArrow}{\mathrel}{symbolsC}{117}
\re@DeclareMathSymbol{\seArr}{\mathrel}{symbolsC}{117}
\re@DeclareMathSymbol{\nwArrow}{\mathrel}{symbolsC}{118}
\re@DeclareMathSymbol{\nwArr}{\mathrel}{symbolsC}{118}
\re@DeclareMathSymbol{\swArrow}{\mathrel}{symbolsC}{119}
\re@DeclareMathSymbol{\swArr}{\mathrel}{symbolsC}{119}
\re@DeclareMathSymbol{\nequiv}{\mathrel}{symbolsC}{46}
\re@DeclareMathSymbol{\Perp}{\mathrel}{symbolsC}{121}
\re@DeclareMathSymbol{\Vbar}{\mathrel}{symbolsC}{121}
\re@DeclareMathSymbol{\sslash}{\mathrel}{stmry}{12}
\re@DeclareMathSymbol{\invamp}{\mathrel}{symbolsC}{77}
\re@DeclareMathSymbol{\parr}{\mathrel}{symbolsC}{77}
\makeatother

\makeatletter
\DeclareRobustCommand\widecheck[1]{{\mathpalette\@widecheck{#1}}}
\def\@widecheck#1#2{%
    \setbox\z@\hbox{\m@th$#1#2$}%
    \setbox\tw@\hbox{\m@th$#1%
       \widehat{%
          \vrule\@width\z@\@height\ht\z@
          \vrule\@height\z@\@width\wd\z@}$}%
    \dp\tw@-\ht\z@
    \@tempdima\ht\z@ \advance\@tempdima2\ht\tw@ \divide\@tempdima\thr@@
    \setbox\tw@\hbox{%
       \raise\@tempdima\hbox{\scalebox{1}[-1]{\lower\@tempdima\box
\tw@}}}%
    {\ooalign{\box\tw@ \cr \box\z@}}}
\makeatother

\makeatletter
\def\udots{\mathinner{\mkern2mu\raise\p@\hbox{.}
\mkern2mu\raise4\p@\hbox{.}\mkern1mu
\raise7\p@\vbox{\kern7\p@\hbox{.}}\mkern1mu}}
\makeatother

\newcommand{\gt}{>}

\theoremstyle{plain}

\theoremstyle{definition}

\theoremstyle{remark}

\begin{document}

%-------------------------------------------------------------------

\preprint{
UTTG-11-10 \\
TCC-020-10 \\
}
\title{Tinkertoys for Gaiotto Duality}

\author{Oscar Chacaltana and Jacques Distler
     \oneaddress{
      Theory Group and\\
      Texas Cosmology Center\\
      Department of Physics,\\
      University of Texas at Austin,\\
      Austin, TX 78712, USA \\
      {~}\\
      \email{oscarch@physics.utexas.edu}\\
      \email{distler@golem.ph.utexas.edu}\\
      }
}

%\date{\today}
\date{August 30, 2010}

\Abstract{
We describe a procedure for classifying $\mathcal{N}=2$ superconformal theories of the type introduced by Davide Gaiotto. Any curve, $C$, on which the 6D $A_{N-1}$ SCFT is compactified, can be decomposed into 3-punctured spheres, connected by cylinders. We classify the spheres, and the cylinders that connect them. The classification is carried out explicitly, up through $N=5$, and for several families of SCFTs for arbitrary $N$. These lead to a wealth of new S-dualities between Lagrangian and non-Lagrangian $\mathcal{N}=2$ SCFTs.
}

\maketitle

\thispagestyle{empty}
\tableofcontents
\vfill
\newpage
\setcounter{page}{1}

\section{Introduction}

Four-dimensional theories with $\mathcal{N}=2$ supersymmetry have long been considered an interesting laboratory to study a variety of field theory phenomena, as their IR description can, in principle, be computed by Seiberg-Witten theory. Not long ago, Gaiotto \cite{Gaiotto:2009we} showed that the marginal deformations of certain four-dimensional $N=2$ superconformal quiver gauge theories could naturally be identified with $\mathcal{M}_{g,n}$, the moduli space of a curve, $C$, of genus $g$, with $n$ punctures. Various degeneration limits of $C$ correspond to different weak-coupling limits, of the gauge theory, related by S-duality.  Gaiotto's work generalizes the classic examples of $\mathcal{N}=2$ S-duality of Argyres and Seiberg \cite{Argyres:2007cn} to a much larger class of 4D $\mathcal{N}=2$ superconformal field theories.

These 4D theories can be identified as the low-energy limit of $N$ M5-branes, wrapped on $C$, with the punctures corresponding to points where non-compact M5-branes intersect $C$. The Seiberg-Witten curve of the 4D theory is realized as an $N$-sheeted cover of $C$ (with a canonical choice of Seiberg-Witten differential), with the punctures being the branch points.

Equivalently, one can consider the (2,0) $A_{N-1}$ SCFT, in 6D (the low-energy limit of the theory on the $N$ M5-branes), compactified on $C$. With a certain twist, to preserve $\mathcal{N}=2$ supersymmetry \cite{Gaiotto:2009hg}, the Kaluza-Klein reduction to 4D is the $\mathcal{N}=2$ gauge theory of interest. The punctures are the locations of certain defect operators in the 6D theory. In the Kaluza-Klein reduction, the gauge invariant fields ($\sim \operatorname{Tr}(\Phi^k)$) which parametrize the Coulomb branch of the 4D gauge theory correspond to (meromorphic) sections of the bundle of $k$-differentials on $C$. In the presence of the defect operators, the latter can have poles, of prescribed orders, at the punctures. These, in turn, determine the Seiberg-Witten curve as a branched cover of $C$.

In previous works, the $T_N$ series of 4D $\mathcal{N}=2$ interacting SCFTs played a particularly central role. This was true when investigating the gravity dual \cite{Gaiotto:2009gz} and when studying these theories as webs of 5-branes in IIB string theory \cite{Benini:2009gi}. In both approaches, the $T_N$ theory played a central role; other fixtures could be obtained by deforming along the Higgs branch. While this is formally true, we find it a not-so-useful viewpoint, as the global symmetry groups, that arise, are not strict subgroups of the ${SU(N)}^3$ symmetry of the $T_N$ theory. Rather, various enhanced global symmetries appear. There is nothing surprising, or contradictory, about the existence of special points, of enhanced global symmetry, on the Higgs branch. But this does somewhat militate against the ``special" role of the $T_N$ theories. We take a more ``democratic" approach, in our analysis. This will be particularly important when extending our $\mathcal{N}=2$ analysis to $\mathcal{N}=1$ theories (for some preliminary work, in that direction, see \cite{Benini:2009mz}).

In this paper, we set out to classify the basic building blocks, i.e., the 3-punctured spheres and the cylinders connecting them, that comprise the various degeneration limits of the curve, $C$. Not every imaginable combination of three punctures appears in a 3-punctured sphere, and not every pair of punctures can be connected by a cylinder. For each combination which \emph{is} allowed, we give its 4D interpretation. Some of the 3-punctured spheres correspond to theories of free hypermultiplets, while others correspond to intrinsically interacting SCFTs, which may or may not include additional free hypermultiplets. 

In \S\ref{setup}, we present our rules for constructing punctured surfaces, $C$. We introduce the notions of ``regular" and ``irregular" punctures. The former are in 1:1 correspondence with $SU(N)$ Young diagrams (and, in turn, to the defect operators of the 6D description); the latter admit higher-order poles of the $k$-differentials on $C$, but of a prescribed sort. We explain which combinations of punctures form admissible 3-punctured spheres (``fixtures") and which cylinders (gauge groups) are allowed to connect pairs of punctures.  In \S\ref{symmetries}, we present formul\ae\ for the central charges, $(a,c)$, in terms of the genus of $C$ and the pole structures of the $k$-differentials at the punctures; later these formul\ae\ will provide a useful check on our identifications of the 4D interpretations of the fixtures.

In \S\ref{identifying}, we discuss the systematics of identifying the nature of each allowed fixture in the compactification of a given $A_{N-1}$ theory. We then go on, in \S\ref{taxonomy} to present an exhaustive catalogue of the fixtures and cylinders that appear in the $A_1$, $A_2$, $A_3$, and $A_4$ cases, and their 4D interpretations. We perform many S-duality checks on our results. One key feature is the global symmetry group associated to a given fixture. As a further check on our identifications, in \S\ref{mirrors}, we dimensionally-reduce to 3 dimensions, and exploit some features of the 3D mirror theories, to determine the global symmetry group for a fixture corresponding to an interacting SCFT.

Having presented an exhaustive taxonomy, for $N\leq 5$, we proceed, in \S\ref{series} to construct some infinite families (analogous to the $T_N$ series) of fixtures corresponding to interacting SCFTs. These provide an interesting new set of S-dualities, at arbitrary $N$ (vastly generalizing the rank-1 case discussed in \cite{Argyres:2007tq}). Finally, in \S\ref{irregulartheories}, we show that our recipe allows for a small number of additional theories, where $C$ has (at most) one irregular puncture.
 
\section{Setup}\label{setup}

We study the $A_{N-1}$ (2,0) 6d theory compactified on a Riemann surface $C$ of genus $g$ with $n$ punctures (complex codimension-1 defect operators) located at points $y_i\in C$, $i=1,\dots, n$ \cite{Gaiotto:2009we}. The Seiberg-Witten curve of the 4d low-energy theory is given by an equation of the form $\lambda^N=\sum_{k=2}^N \lambda^{N-k}\phi_k$, where $\lambda$ is the Seiberg-Witten differential, and the $\phi_k(y)$ ($k=2,\dots, N$) are $k$-differentials. The $\phi_k$ are allowed to have poles of various orders at the punctures $y_i$.

When there are no mass-deformation parameters, the Coulomb branch is a graded vector space,

\begin{equation}
V=\bigoplus_{k=2}^N V_k,
\label{linspace}\end{equation}
where $V_k = H^0\Bigl(C, K^k\left(\sum_{i=1}^n p_k^{(i)}y_i\right)\Bigr)$ is the space of meromorphic of $k$-differentials, $\phi_k$, with poles of order at most $p^{i}_k$ at the punctures $y_i$. The graded dimension of $V$ is given by

\begin{displaymath}
d_k = (2k-1)(g-1) + \sum_{i=1}^{n} p^{(i)}_k,
\end{displaymath}
provided that

\begin{displaymath}
\begin{cases}
g\gt 1\\
\sum_i p^{(i)}_k \gt 0 & g=1 \\
\sum_i p^{(i)}_k \geq (2k-1) & g=0
\end{cases}
\end{displaymath}
At a generic point on the Coulomb branch, we have an Abelian gauge theory (whose infrared dynamics is governed by the Seiberg-Witten solution). At the origin of the Coulomb branch sits a non-Abelian SCFT.

As we vary the moduli (including the locations of the punctures) of the curve $C$, the $V_k$ fit together to form a graded vector bundle over the moduli space $\mathcal{M}_{g,n}$ of $C$.

At the boundary of the moduli space, the curve $C$ degenerates into a collection of 3-punctured spheres (``{}fixtures''{}) connected by cylinders. The plumbing parameter\footnote{In the limit that the other gauge couplings are turned off,

\begin{displaymath}
s = \frac{\theta_{1 0}^4(\tau)}{\theta_{0 0}^4(\tau)}
\end{displaymath}} , $s\sim 16 q^{1/2} +\dots$, with $q=e^{2\pi i \tau}$, for each cylinder controls the strength of the gauge coupling for that factor of the gauge group,

\begin{displaymath}
\tau = \frac{\theta}{\pi} + \frac{8\pi i}{g^2}.
\end{displaymath}
Since there are various different degeneration limits of $C$, there are various different gauge theory presentations of the same underlying SCFT. These are related by S-duality.

What we would like to do is understand the taxonomy of gauge theory presentations which arise in this way from compactifying a given (2,0) SCFT. To do this, we need a catalogue of what the allowed fixtures (3-punctured spheres) are, and what cylinders (gauge groups) connect them. We can then build up the surface $C$, in a degeneration limit, as a ``{}tinkertoy''{}, by connecting fixtures together with cylinders, according to the allowed rules.

We will restrict ourselves to the $A_{N-1}$ series of (2,0) theories, for small $N$. For the most part, we will also work at genus, $g=0$, so that the only degenerations come from the collisions of (multiple) punctures.

\hypertarget{regular}{}\subsection{{Regular Punctures}}\label{regular}

For the 6D $A_{N-1}$ theory, the defects on $C$ can be characterized by the structure of poles $\{p_k\}=\{p_2,\dots,p_N\}$ of the $\phi_k$ ($k=2,\dots, N$) at each puncture on $C$ in the 4D theory. Gaiotto \cite{Gaiotto:2009we} studied the Seiberg-Witten curves of known 4D linear superconformal quivers to deduce what pole structures are actually allowed. Let us briefly review his arguments.

The Seiberg-Witten curve $\lambda^N(y)=\sum_{k=2}^N \lambda^{N-k}(y)\phi_k(y)$, where $\lambda=x d y$ is the Seiberg-Witten differential, can be understood as a polynomial equation of degree $N$ for the fiber coordinate $x$ on the cotangent bundle, $T^*C$. Gaiotto showed that at each of the $N$ solutions for $x$, $\lambda$ has a simple pole at every puncture, with residues being linear combinations of the mass deformations of the theory. In particular, the Seiberg-Witten differential of a theory with no mass deformations has no simple poles at the punctures for any of the $N$ solutions of the Seiberg-Witten equation. For the mass-deformed theory, one can associate to each puncture its set of $N$ corresponding residues. Notice that because of the absence of a $x^{N-1}$ term, the sum of the $N$ residues at a puncture must be zero. Some of these residues can be equal to each other (but at most $(N-1)$ of them can be all equal to each other, if they are to be non-zero). One can then associate a global symmetry group to the puncture, which rotates these residues into each other. The rank of this group should be equal to the number of linearly independent residues. Gaiotto deduced that this flavor group can be easily found if one first associates an auxiliary $SU(N)$ Young diagram to the puncture. If the $N$ residues at a puncture are divided into groups of equal residues, each group having $h_t$ residues (so that $\sum_t h_t=N$), one can construct a Young diagram by making the $h_t$ be the heights of its columns. If there are now $n_h$ columns of height $h$, the global symmetry group associated to the puncture is $G_{\text{global}} = S\left(\prod_h U(n_h)\right)$.

A valid $SU(N)$ Young diagram has at most $N-1$ rows. Correspondly, if we tried to construct the Young diagram with $N$ rows of one box each, it would not correspond to a puncture. The residues of the mass-deformed Seiberg-Witten differential would all be necessarily zero, and so it would have no global symmetry group associated to it.

From the Young diagram, one can easily reconstruct the pole structure $\{p_k\}$ of the $k$-differentials $\phi_k$. Given a Young diagram with rows of lengths decreasing from top to bottom, we construct its corresponding pole structure $\{p_k\}$ by the rules:

\begin{itemize}%
\item We label the $N$ boxes of the Young diagram with integers. We start with the longest row, and we assign to its leftmost box the label 0.
\item We increase the label by one as we move to the right along a row.
\item Once we have finished labelling a row, we move on to the row immediately below it, and we assign to its leftmost box the same label as the rightmost box in the previous row. We then repeat the labelling procedure along the current row.
\item We call the resulting sequence of $N$ labels (again, read from left to right within a row, and then moving to the row below) $\{p_1,p_2,p_3,\dots,p_{N}\}$. Discarding the auxiliary $p_1=0$, we obtain the pole structure corresponding to the Young diagram. (Notice that we have necessarily $p_2=1$.)

\end{itemize}
This construction of the $\{p_k\}$ is found by computing the leading order of the $N$ solutions of the Seiberg-Witten equation with no mass deformations, seen as a polynomial equation for $x$, by the method of the Newton polygon. Indeed, plotting the $(k,p_k)$ ($k=1,\dots, N$, $p_1=0$) in a plane, the convex envelope that starts at $(0,0)$ has sides with slopes of the form $(q-1)/q$, for some positive integer $q$, and this guarantees that the Seiberg-Witten differential $\lambda$ of the theory with no mass deformations will have no simple poles at the punctures.

For example, for $N=6$, the Young diagram with two columns of height 3 corresponds to the pole structure $\{1,1,2,2,3\}$ and global symmetry group $SU(2)$. In general, for even $N$, the Young diagram with two columns of the same height will correspond to the pole structure $\{1,1,2,2,3,3,\dots, N-1, N-1, N\}$ and global symmetry group $SU(2)$.

We call the pole structures associated to a Young diagram ``{}regular''{}, and by extension we refer to a puncture with such a structure as a ``{}regular''{} puncture.

Given a regular pole structure, $\{p_k\}$, it is trivial to invert the above construction and obtain the corresponding Young diagram.

\begin{itemize}%
\item Start with a Young diagram with two boxes in a row.

\item For each $k=3,\dots,N$:

\begin{itemize}%
\item If $p_k-p_{k-1}=1$, add a box to the current row.
\item If $p_k-p_{k-1}=0$, start a new row, with one box.

\end{itemize}

\end{itemize}
At least for regular punctures, we can thus think of the pole structure, and the Young diagram, interchangeably.

There are two regular punctures that deserve special names. The regular puncture with $p_k=k-1$, for $k=2,\dots, N$, will be called a ``{}maximal puncture''{}. It corresponds to the situation with $N$ different residues of the mass-deformed Seiberg-Witten differential, so its associated Young diagram consists of one row with $N$ boxes, and its associated global symmetry group is $SU(N)$. On the other hand, the regular puncture with $p_k=1,\, \forall k$, will be called ``{}minimal''{}; it corresponds to having $(N-1)$ equal residues of the mass-deformed Seiberg-Witten differential, its Young Diagram consists of one row with two boxes, and $N-2$ rows with one box, and its associated global symmetry group is $U(1)$.

For the $A_{N-1}$ theory, there are $P(N)-1$ regular punctures, according to these rules. A colliding pair of regular punctures will give rise to a fixture connected by a cylinder to the rest of the surface. Our job will be to characterize the fixtures that arise as well as the cylinders that connect them.

\hypertarget{irregular}{}\subsection{{Irregular punctures}}\label{irregular}

As we will see below, when two regular punctures collide, the resulting fixture will correspond to one of three possibilities: a number of free hypermultiplets, an interacting SCFT, or an interacting SCFT accompanied by a number of free hypermultiplets. The case with only free hypermultiplets corresponds to a fixture with no Coulomb branch, while the two cases involving an interacting SCFT correspond to a fixture with a (positive-dimensional) Coulomb branch.

We want the graded dimension of the Coulomb branch of the degenerate surface (defined as the sum of the graded dimensions of the Coulomb branch of the fixture, the Coulomb branch of gauge theory on the attaching cylinder and the Coulomb branch of the rest of the surface) to agree with the graded dimension of the Coulomb branch of the original surface, $C$. To achieve this, we would like ---{} as a bookkeeping device ---{} for the (graded) virtual dimension and the actual (graded) dimension of the Coulomb branch of the fixture to agree. This determines, uniquely, the pole structure at the attaching puncture (the third puncture of the 3-punctured sphere).

For a fixture corresponding to free hypermultiplets, the Coulomb branch is zero-dimensional. To achieve this, we are forced in most cases (the exception being the collision of a minimal and a maximal puncture) to introduce punctures with pole structures that are not regular, i.e., that do not arise from the construction detailed in \S\ref{regular}. We call punctures with such pole structures ``{}irregular''{}.

Irregular punctures will also appear in some fixtures associated to interacting SCFTs. There, too, they will be determined by requiring that, when certain $d_k$ are supposed to vanish, the actual and virtual value of $d_k$ agree (and are zero).

Irregular punctures will turn out to satisfy the properties:

\begin{itemize}%
\item $p_2=1$.

\item $\operatorname{max}(k-1,p_{k-1})\leq p_k \leq \operatorname{min}(2k-3, p_{k-1}+2)$.

\item $\sum_{k=2}^{N} p_k \gt N(N-1)/2$.

\item The subset of $k$ such $p_k=k-1$ form the exponents of a simple Lie group, $G_F$, (We work, throughout, with the convention that the exponents of $SU(N)$ are $2,3,\dots, N$.)

\item Define the conjugate pole structure $\{p'_k\}$, via

\begin{equation}
p'_k = \begin{cases}
            p_k   & p_k  =  k-1\\
         2k-1-p_k & p_k \gt k-1
        \end{cases}\quad .
\label{conjpole}\end{equation}
The $\{p'_k\}$ must be a regular pole structure, in the sense of the \S\ref{regular}.

\end{itemize}
As with regular punctures, we will associate a global symmetry group with an irregular puncture; in this case, it is the group $G_F$.

\hypertarget{fixtures}{}\subsection{{Fixtures}}\label{fixtures}

For a Riemann surface of genus $g=0$, with $n$ punctures, the dimension of the Coulomb branch subspace $V_k$ is

\begin{equation}
d_k=1-2k+\left(\sum_{i=1}^{n} p^{(i)}_k\right),
\label{dims}\end{equation}
where $p^{(i)}_k$ ($k=2,\dots, N$) represents the pole structure of the $i$-th puncture, $i=1,\dots, n$. Moreover, we \emph{require} that the RHS of \eqref{dims} be non-negative, for each $k$, i.e. that the virtual dimension and the actual dimension agree. Having done this, our bookkeeping rules will ensure that, when $C$ degenerates, the same is true of the $d_k$ of the Coulomb branches associated to each of component pieces.

For a 3-punctured sphere (a ``{}fixture''{}), we will require, for each $k$, that

\begin{itemize}%
\item if $d_k\gt 0$, then $p^{i}_k\leq k-1$, for $i=1,2,3$.

\end{itemize}
As a simple consequence any fixture has at most one irregular puncture.

If $d_k=0$ for all $k$, we have a free-field fixture. If the three punctures are regular, then necessarily one of them is minimal and the other two are maximal. On the other hand, an interacting SCFT fixture (which could also have free hypermultiplets) consists of three punctures such that $d_k\gt 0$ for at least one $k$.

\hypertarget{cylinders}{}\subsection{{Cylinders}}\label{cylinders}

When two or more punctures on a Riemann surface collide, the surface degenerates, and a long cylinder connecting the two pieces appears (which could still be attached somewhere else). When the cylinder becomes infinitely long and thin, a new puncture appears at each of the two pieces of Riemann surface where the ends of the cylinder were. The long, thin cylinder corresponds to a weakly-coupled gauge group. When the gauge coupling is infinitely weak, we are left with flavor symmetries at each end of the cylinder, corresponding to the two new punctures. Similarly, two punctures on a Riemann surface (or on two initially disconnected Riemann surfaces) can sometimes be glued to each other by a cylinder. In both cases the gauge group corresponding to the cylinder is a subgroup of the flavor groups associated to the punctures. Given two (regular or irregular) punctures, we want to see when they can be connected to each other, and what the arising gauge group is.

We will denote a cylinder connecting a puncture of pole structure $\{p_k\}$ with a puncture of pole structure $\{p'_k\}$ by

\begin{displaymath}
\{p_k\}\xleftrightarrow{\qquad G\qquad}\{p'_k\},
\end{displaymath}
where $G$ denotes a gauged subgroup of the flavor symmetry groups of the two theories connected by the cylinder.

Let $q_k=\operatorname{min}(p_k,p'_k)$. For the cylinder to be valid, $G$, $\{p_k\}$ and $\{p'_k\}$ must satisfy the following requirements:

\begin{itemize}%
\item $q_k$ is a regular pole structure.
\item $G$ is a subgroup of the global symmetry group, $G_{q}$, where $G_{q}$ is the symmetry group associated to $\{q_k\}$, following the Young diagram prescription.
\item $\operatorname{rank}(G)=N^2 -1 - \sum_{k=2}^N (p_k + p'_k)$.
\item For each $k$, we have either $\begin{cases}p_k=p'_k=k-1\\ p_k+p'_k = 2k-1\end{cases}$.
\item The exponents of $G$ are the set of $k$ such that $p_k = p'_k = k-1$. (Notice there cannot be repeated exponents.)

\end{itemize}
In particular, for the $A_{N-1}$ theories, $G=SU(n)$ or $Sp([n/2])$, for some $n\leq N$.

Since we must have $1\leq\operatorname{rank}(G)\leq N-1$, two regular punctures can be connected by a cylinder if and only if they are maximal, in which case the gauge group is $G=SU(N)$. The vast majority of cylinders will connect a regular and an irregular puncture.

Occasionally, though, cylinders connecting two irregular punctures will appear (see the case \hyperlink{A3}{$A_3$} below). These are rare, as the tension between the rank condition and the condition on the exponents is quite restrictive.

We can now explain how the irregular punctures serve as a useful bookkeeping device. Consider the collision of two punctures $\{p_k\}$ and $\{p'_k\}$ on a Riemann surface $C$. They bubble off a sphere $S$, which is attached by a cylinder $T$ to the rest of $C$. Let the pole structure of the new puncture to which $S$ is attached by $T$ be $\{p''_k\}$. Before the collision, the contribution of $\{p_k\}$, $\{p'_k\}$ to the total dimension of the Coulomb branch of the theory on $C$ was

\begin{equation}
\sum_{k=2}^N p_k +p'_k.
\label{before}\end{equation}
After the collision, such contribution becomes

\begin{equation}
d_S + \operatorname{rank}(G_T) + \sum_{k=2}^N p''_k,
\label{after}\end{equation}
where $d_S\geq 0$ is the dimension of the Coulomb branch associated to the fixture $S$, and $G_T$ is the gauge group associated to the cylinder $T$. The requirements on the cylinder that we listed above ensure that \eqref{before} and \eqref{after} agree.

The rules above actually guarantee that the agreement is finer than that. Recall that the Coulomb branch \eqref{linspace} is not just a vector space, but a \emph{graded} vector space (with grading given by $k$). We want to ensure that the \emph{graded} dimensions, $d_k=\operatorname{dim}(V_k)$, agree. In the degeneration limit, certain of the $\phi_k$ (precisely the ones satisfying the $p_k=p'_k=k-1$ condition) are allowed to have a $k$-th order pole at the node, with the residues agreeing on the two sides. The residue is the Coulomb-branch parameter for the gauge theory on the cylinder. The degrees of these Coulomb-branch parameters are precisely the exponents of $G$. In other words, when $p_k=p'_k=k-1$, the dimension of that graded component of the Coulomb branch of $G$ is 1. When $p_k+p'_k=2k-1$, the dimension (and virtual dimension) of that graded component of the Coulomb branch of $G$ vanishes.

\hypertarget{symmetries}{}\section{{Symmetries and Central Charges}}\label{symmetries}

The $\mathcal{N}=2$ SCFT, obtained by compactifying the $A_{N-1}$ (2,0) SCFT on $C$, has a number of obvious S-duality invariant properties; we will use these to support our identification of the various fixtures which arise.

The most obvious ones are the graded dimension of the Coulomb branch, and the global symmetry group, $G_{\text{global}}$. More subtle is the central charge, $k_{G_i}$, of each nonabelian factor $G_i\subset G_{\text{global}}$, defined via the current algebra \cite{Argyres:2007cn, Osborn:1993cr}

\begin{equation}
J_\mu^a(x)J_\nu^b(0) = \frac{3 k_{G}}{4\pi^4} \delta^{a b} \frac{g_{\mu\nu} x^2 -2 x_\mu x_\nu}{(x^2)^4} + \frac{2}{\pi^2} \tensor{f}{^a^b_c} \frac{x_\mu x_\nu x\cdot J^c}{(x^2)^3},
\label{CurrentAlg}\end{equation}
and the conformal anomaly coefficients, $a$, $c$, defined via

\begin{equation}
\tensor{T}{_{\mu}_{}^{\mu}} = \frac{c}{16\pi^2} {(\text{Weyl})}^2 -\frac{a}{16\pi^2} (\text{Euler}),
\label{ConfAnom}\end{equation}
where

\begin{displaymath}
\begin{aligned}
{(\text{Weyl})}^2 &= R_{\mu\nu\lambda\rho}^2 -2 R_{\mu\nu}^2 +\tfrac{1}{3} R^2,\\
(\text{Euler}) &= R_{\mu\nu\lambda\rho}^2 -4 R_{\mu\nu}^2 + R^2.
\end{aligned}
\end{displaymath}
These are straightforwardly calculable in an $\mathcal{N}=2$ gauge theory with a Lagrangian description, and are invariant under marginal deformations \cite{Kuzenko:1999pi}. Using S-duality, one obtains predictions for $G_{\text{global}}$, and the coefficients $k, a, c$ of the intrinsically-interacting SCFT fixtures. These have been computed before \cite{Aharony:2007dj} for the exceptional SCFTs of Minahan and Nemeschansky \cite{Minahan:1996cj}, and more recently for the $T_N$ interacting SCFTs \cite{Gaiotto:2009gz, Benini:2009gi}.

For a Lagrangian field theory, with $n_v$ vector multiplets and $n_h$ hypermultiplets,

\begin{equation}
\begin{gathered}
a=\frac{5n_v+n_h}{24},\\
c=\frac{2n_v+n_h}{12}.
\end{gathered}
\label{acvals}\end{equation}
For an interacting SCFT fixture, coupled to some gauge fields and, perhaps, other fixtures corresponding to free hypermultiplets (or other fixtures for which we have already computed these coefficients), we use S-duality to relate it to another gauge theory where $(a,c)$ are easy to calculate, either because it'{}s a Lagrangian theory, or because it involves SCFTs whose central charges $(a,c)$ have been previously calculated.

As suggested by \cite{Gaiotto:2009gz}, it proves convenient to use \eqref{acvals} to define an ``{}effective''{} $n_v$ and $n_h$ for these interacting SCFTs. One then has the quaternionic dimension of the Higgs branch  \cite{Gaiotto:2008nz}, $\mathcal{H}$, given by

\begin{displaymath}
\operatorname{dim}_{\mathbf{H}}\mathcal{H}=n_h-n_v = 24(c-a).
\end{displaymath}
These obey some simple formul\ae.

Let $\mathfrak{g}$ be a semisimple Lie algebra, of rank $r_\mathfrak{g}$, and let $m_\alpha$, $\alpha=1,\dots r_\mathfrak{g}$ be its exponents. It'{}s well known\footnote{The result follows, for instance, from \cite{Kostant1959}, which introduces the notion of a ``{}principal''{} $su(2)\subset\mathfrak{g}$, such that the $m_\alpha-1$, $\alpha =1,\dots,r$ are the $su(2)$ highest weights which occur in the decomposition of $\mathfrak{g}$ with respect to this embedding.}  that

\begin{displaymath}
\operatorname{dim}(\mathfrak{g}) = \sum_\alpha (2 m_\alpha -1).
\end{displaymath}
As a trivial consequence, for any of our theories, possessing an S-duality frame with a Lagrangian interpretation,

\begin{equation}
n_v = \sum_{k=2}^N (2k-1)d_k.
\label{nvdef}\end{equation}
Moreover, the RHS of \eqref{nvdef} is determined entirely in terms of the genus, $g$, and the pole structures, $p_k^{(i)}$, at the punctures.

As will be clear from our analysis, \eqref{nvdef} will provide the correct definition for the effective $n_v$, even in cases where there is no weakly-coupled Lagrangian dual.

For the effective number of hypermultiplets, we combine the above with a result of Nanopoulos and Xie \cite{Nanopoulos:2010ga}, to obtain

\begin{equation}
n_h = - (1-g) \frac{4N(N^2-1)}{3} +\sum_{i=1}^n f^{(i)},
\label{nhdef}\end{equation}
where $f^{(i)}$ is the contribution of the $i^{\text{th}}$ puncture. For a regular puncture,

\begin{equation}
f^{(\text{reg})} = \tfrac{1}{2}\left(- N + \sum_{r} l_r^2  \right) + \sum_{k=2}^{N} (2k-1) p_k,
\label{fregdef}\end{equation}
where $l_r$ is the length of the $r^{\text{th}}$ row of the Young diagram. For an irregular puncture, define the conjugate pole structure $\{p'_k\}$, via \eqref{conjpole}. $\{p'_k\}$ is, by definition, a regular pole structure. The contribution of an irregular puncture, then, is\footnote{The origin of this formula is clear. The irregular puncture, $\{p_k\}$, can be attached to the rest of the surface via a cylinder $\{p_k\}\xleftrightarrow{\qquad\qquad}\{p'_k\}$. Cylinders do not contribute any hypermultiplets, and \eqref{firregdef} is simply the embodiment of the requirement that $n_h$ should be the same, before and after sewing.} 

\begin{equation}
f^{(\text{irreg})} = \frac{4(N^2 - 1)N}{3} -
\tfrac{1}{2}\left(- N + \sum_{r} (l'_r)^2  \right) - \sum_{k=2}^{N} (2k-1) p'_k\quad.
\label{firregdef}\end{equation}

Applying this to the case of a sphere with three maximal punctures, one recovers the result for the $T_N$ theories \cite{Benini:2009gi, Gaiotto:2009gz},

\begin{displaymath}
\begin{gathered}
k=2N\\
a=\frac{N^3}{6}-\frac{5N^2}{16}-\frac{N}{16}+\frac{5}{24}\\
c=\frac{N^3}{6}-\frac{N^2}{4}-\frac{N}{12}+\frac{1}{6}.
\end{gathered}
\end{displaymath}
We will check these results explicitly for the cases up to $N=5$, as well as identify a host of new theories.

\hypertarget{identifying}{}\section{{Identifying fixtures}}\label{identifying}

We take as our starting point that

\begin{enumerate}%
\item The $A_{N-1}$ fixture arising from the collision of a minimal puncture and a maximal puncture corresponds to $N^2$ free hypermultiplets. (The third puncture in this fixture is then also maximal.)
\item The $A_{N-1}$ fixture arising from the collision of two minimal punctures corresponds to $2$ free hypermultiplets. (The third puncture in this fixture is then irregular, of the form $\{1,3,\dots,2k-3,\dots, 2N-3\}$.)
\item Fixtures corresponding to $n_h$ free hypermultiplets (with $n_h$ given by \eqref{nhdef}) will have $n_v=0$, according to \eqref{nvdef} and have zero-dimensional Coulomb branches.

\end{enumerate}
By studying collisions of more regular punctures, we can bootstrap the properties above to identify further fixtures. Consider, for instance, the collision of several minimal punctures. When two of them collide, the resulting fixture

\begin{displaymath}
 \includegraphics[width=181px]{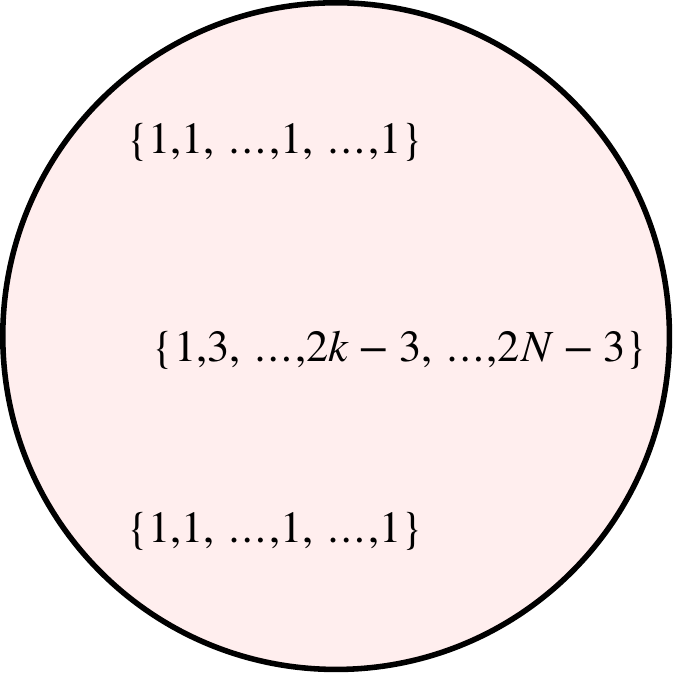}
\end{displaymath}
is attached to the rest of the surface with the cylinder

\begin{displaymath}
\{1,3,\dots,2k-3,\dots,2N-3\}\xleftrightarrow{\qquad SU(2)\qquad}\{1,2,\dots,2,\dots,2\}.
\end{displaymath}
Colliding the $\{1,2,\dots,2,\dots,2\}$ puncture with another minimal puncture produces a free-field fixture

\begin{displaymath}
 \includegraphics[width=181px]{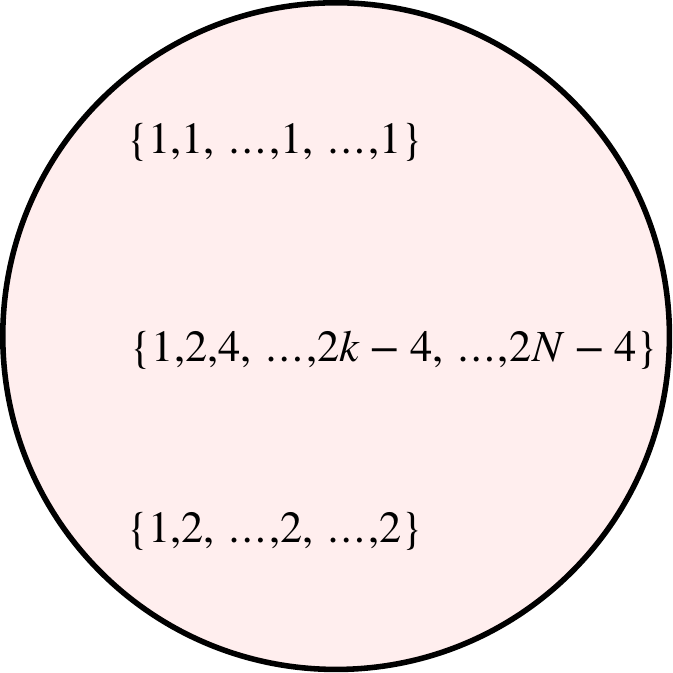}
\end{displaymath}
By conformality of the $SU(2)$, this consists of 6 hypermultiplets (transforming as 3 copies of the 2). This fixture, in turn, is attached to the rest of the surface by the cylinder

\begin{displaymath}
\{1,2,4,\dots,2k-4,\dots,2N-4\}\xleftrightarrow{\qquad SU(3)\qquad}\{1,2,3,\dots,3,\dots,3\}.
\end{displaymath}
Colliding the $\{1,2,3,\dots,3\}$ puncture with another minimal puncture produces a fixture which (by conformality of the $SU(3)$) consists of 12 hypermultiplets, transforming as 4 copies of the 3.

Repeating the process, we deduce a series of fixtures of the form

\begin{displaymath}
 \includegraphics[width=271px]{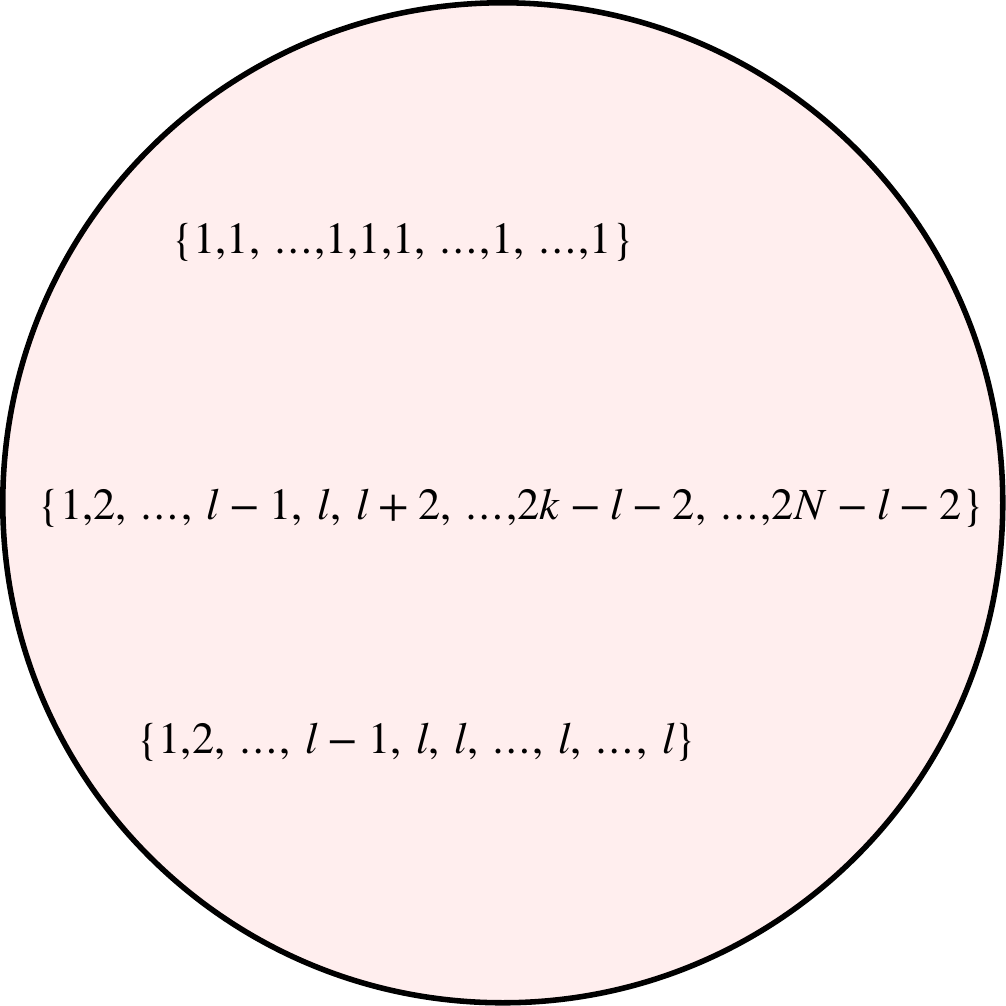}
\end{displaymath}
consisting of $l(l+1)$ hypermultiplets, transforming as the bifundamental of $SU(l)\times SU(l+1)$.

The next simplest puncture has pole structure, $\{1,1,2,\dots,2,\dots,2\}$, corresponding to the Young diagram with two rows of length 2, and the rest of length 1.

Colliding this with a minimal puncture, we produce the fixture

\begin{equation}
 \includegraphics[width=181px]{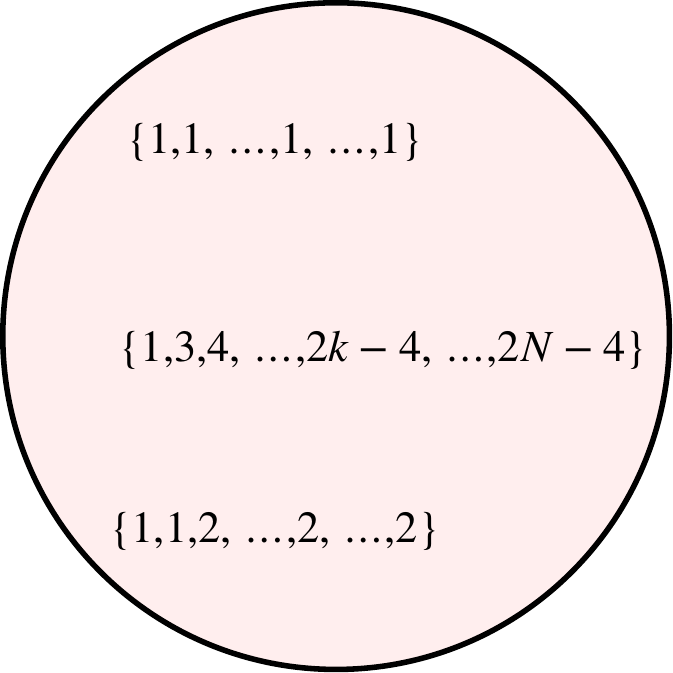}
\label{zeroHypers}\end{equation}
This attaches to the rest of the surface via the cylinder

\begin{displaymath}
\{1,3,4,\dots,2k-4,\dots,2N-4\}\xleftrightarrow{\qquad SU(2)\qquad}\{1,2,3,\dots,3,\dots,3\}.
\end{displaymath}
If we collide that puncture with another minimal puncture, we obtain a fixture we'{}ve seen before, namely

\begin{displaymath}
 \includegraphics[width=181px]{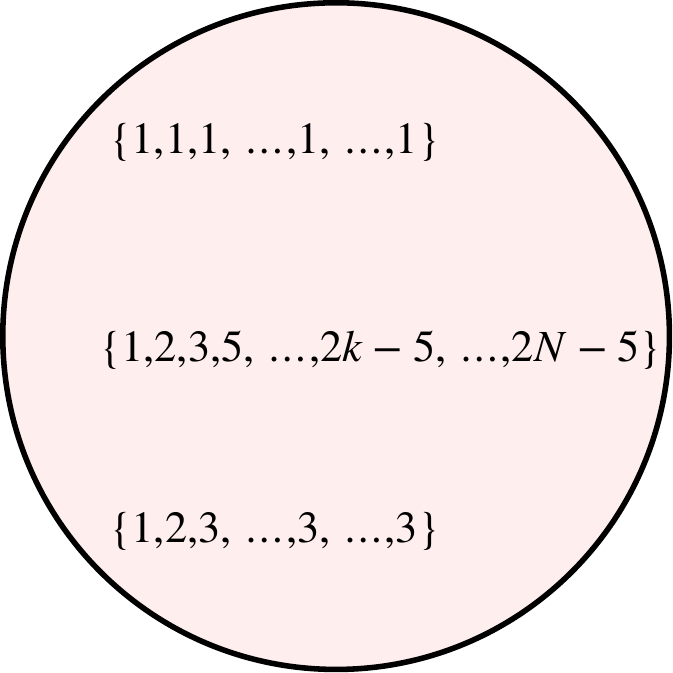}
\end{displaymath}
which consisted of 12 hypermultiplets, transforming as the $(3,4)$ of $SU(3)\times SU(4)$. Here, we'{}re gauging an $SU(2)\subset SU(3)$. This fixture, by itself, provides enough matter to make the $SU(2)$ conformal. Thus, the fixture \eqref{zeroHypers} consists of zero hypermultiplets.

In similar fashion, we can proceed to identify the free-field fixtures corresponding to the collision of \emph{any} regular puncture with a minimal puncture.

We can then go on to identify other fixtures, which arise as collisions of punctures we'{}ve studied already. For instance, colliding two $\{1,1,2,2,\dots,2,\dots,2\}$ punctures, we obtain the fixture

\begin{equation}
 \includegraphics[width=181px]{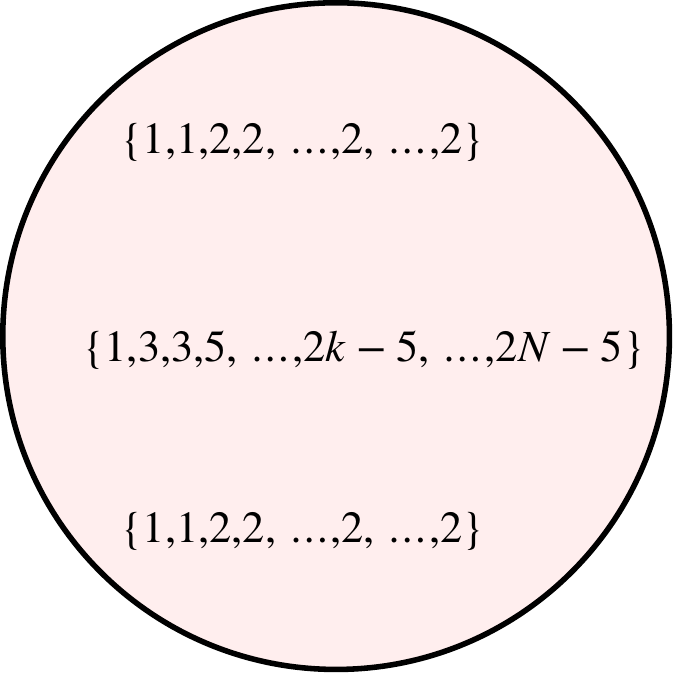}
\label{oneSp2}\end{equation}
This attaches to the rest of the surface via the cylinder

\begin{displaymath}
\{1,3,3,5,,\dots,2k-5,\dots,2N-5\}\xleftrightarrow{\qquad Sp(2)\qquad}\{1,2,3,4,4,\dots,4,\dots,4\}.
\end{displaymath}
If we collide that puncture with another minimal puncture, we again obtain a fixture we'{}ve seen before: this time, 20 hypermultiplets transforming as the $(4,5)$ of $SU(4)\times SU(5)$. Here we'{}re gauging $Sp(2)\subset SU(4)$, so conformality of the $Sp(2)$ requires that the fixture \eqref{oneSp2} consists of 4 hypermultiplets, transforming as the fundamental of $Sp(2)$.

Colliding a $\{1,1,2,2,\dots,2,\dots,2\}$ puncture with a $\{1,2,2,2,\dots,2,\dots,2\}$ puncture, we obtain the free-field fixture

\begin{displaymath}
 \includegraphics[width=181px]{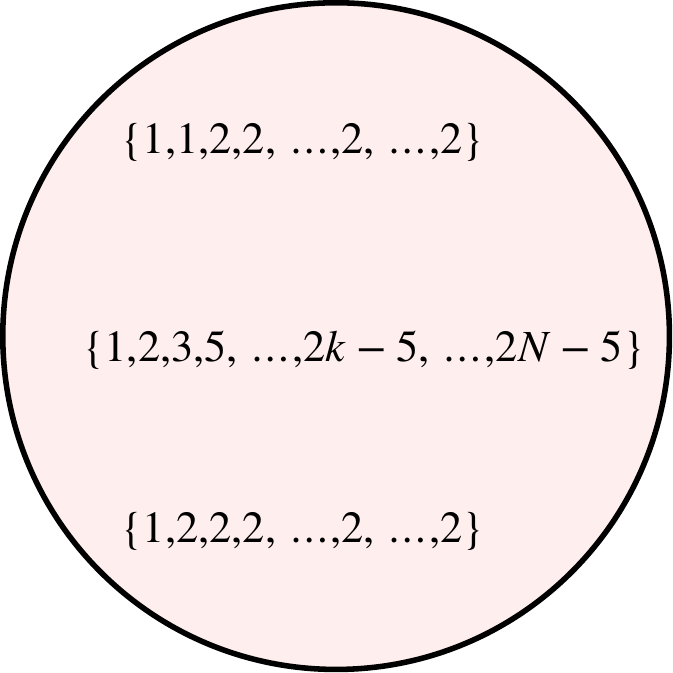}
\end{displaymath}
On the one hand, we can gauge this fixture by attaching a

\begin{displaymath}
\{1,3,5,\dots,2k-3,\dots,2N-3\}\xleftrightarrow{\qquad SU(2)\qquad}\{1,2,2,\dots,2,\dots,2\}
\end{displaymath}
cylinder. On the other, we can attach a

\begin{displaymath}
\{1,2,3,5,\dots,2k-5,\dots,2N-5\}\xleftrightarrow{\qquad SU(4)\qquad}\{1,2,3,4,\dots,4,\dots,4\}.
\end{displaymath}
To ensure conformality of both the $SU(2)$ and the $SU(4)$, we conclude that this fixture consists of 10 hypermultiplets, transforming as the $(1,4)+\tfrac{1}{2}(2,6)$ of $SU(2)\times SU(4)$. (Note that the $(2,6)$ representation is pseudo-real, so we can have matter in a half-hypermultiplet, in that representation. Also, $\ell_6=2$, which ensures conformality of the $SU(4)$.)

Having proceeded as far as we can, in this fashion, we can then use these ``{}known''{} fixtures, plus S-duality, to deduce the identity of other fixtures (including the interacting SCFTs). To see how that works, it is perhaps best to proceed by example.

\section{Taxonomy}\label{taxonomy}
\hypertarget{A1}{}\subsection{{$A_1$}}\label{A1}

For $A_1$, there'{}s just one type of regular puncture, $\{1\}$, where the quadratic differential, $\phi_2$ is allowed to have a simple pole, and there are no irregular punctures. Correspondingly, there is one type of cylinder, which has gauge group $SU(2)$. Similarly, there is only one fixture, with three $\{1\}$ punctures, which is the free theory of four hypermultiplets, or, in a language more appropriate for the $A_1$ case, eight half-hypermultiplets, which transform as a $(2,2,2)$ representation of the $SU(2)\times SU(2)\times SU(2)$ flavor subgroup of this fixture. As before, half-hypermultiplets are allowed because the fundamental representation of $SU(2)$ is pseudo-real.

\hypertarget{A2}{}\subsection{{$A_2$}}\label{A2}

There are now two regular punctures: the maximal puncture $\{1,2\}$, and the minimal puncture, $\{1,1\}$. And there are three distinct types of collisions giving rise to three different fixtures: the collision of two minimal punctures, a minimal and a maximal puncture, and two maximal punctures. The first two cases yield free-field fixtures. The third yields a fixture with a one-dimensional Coulomb branch, the interacting $E_6$ SCFT of Minahan and Nemeschansky \cite{Minahan:1996cj}.

There is one type of irregular puncture, $\{1,3\}$, and two types of cylinders:

\begin{center}
\begin{tabular}{|c|c|}
\hline
Cylinder&Gauge Group\\
\hline 
$\{1,2\}\xleftrightarrow{\qquad\qquad}\{1,2\}$&$SU(3)$\\
$\{1,3\}\xleftrightarrow{\qquad\qquad}\{1,2\}$&$SU(2)$\\
\hline
\end{tabular}
\end{center}

The free-field fixtures are:

\begin{center}
\begin{tabular}{|l|c|c|}
\hline
Fixture&Number of Hypers&Representation\\
\hline 
$ \includegraphics[width=75px]{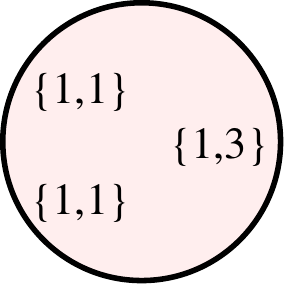}$&2&2\\
\hline
$ \includegraphics[width=75px]{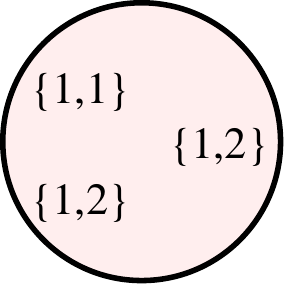}$&9&$(3,3)$\\
\hline
\end{tabular}
\end{center}

Here we have listed the matter representation of the (non-Abelian) global symmetry group of each puncture (or, in the case of an irregular puncture, of the gauge group of the attaching cylinder).

The interacting fixture is

\begin{center}
\begin{tabular}{|c|c|c|c|}
\hline
Fixture&$(d_2,d_3)$&$(a,c)$&${(G_{\text{global}})}_k$\\
\hline 
$ \includegraphics[width=75px]{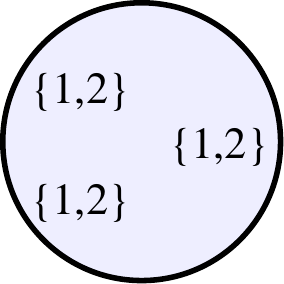}$&$(0,1)$&$\left(\tfrac{41}{24},\tfrac{13}{6}\right)$&${(E_6)}_6$\\
\hline
\end{tabular}
\end{center}

Here we have listed the graded dimensions $d_k$ of the Coulomb branch (the total dimension is $d=\sum_k d_k$), the central charges, $(a,c)$, the global symmetry group $G_{\text{global}}$ of the SCFT, and the central charge $k$ of the $G_{\text{global}}$ current algebra.

The basic S-duality of the $A_2$ theory (discovered by Argyres and Seiberg \cite{Argyres:2007cn}), can be seen by studying the various degenerations of the 4-punctured sphere.

\begin{displaymath}
 \includegraphics[width=410px]{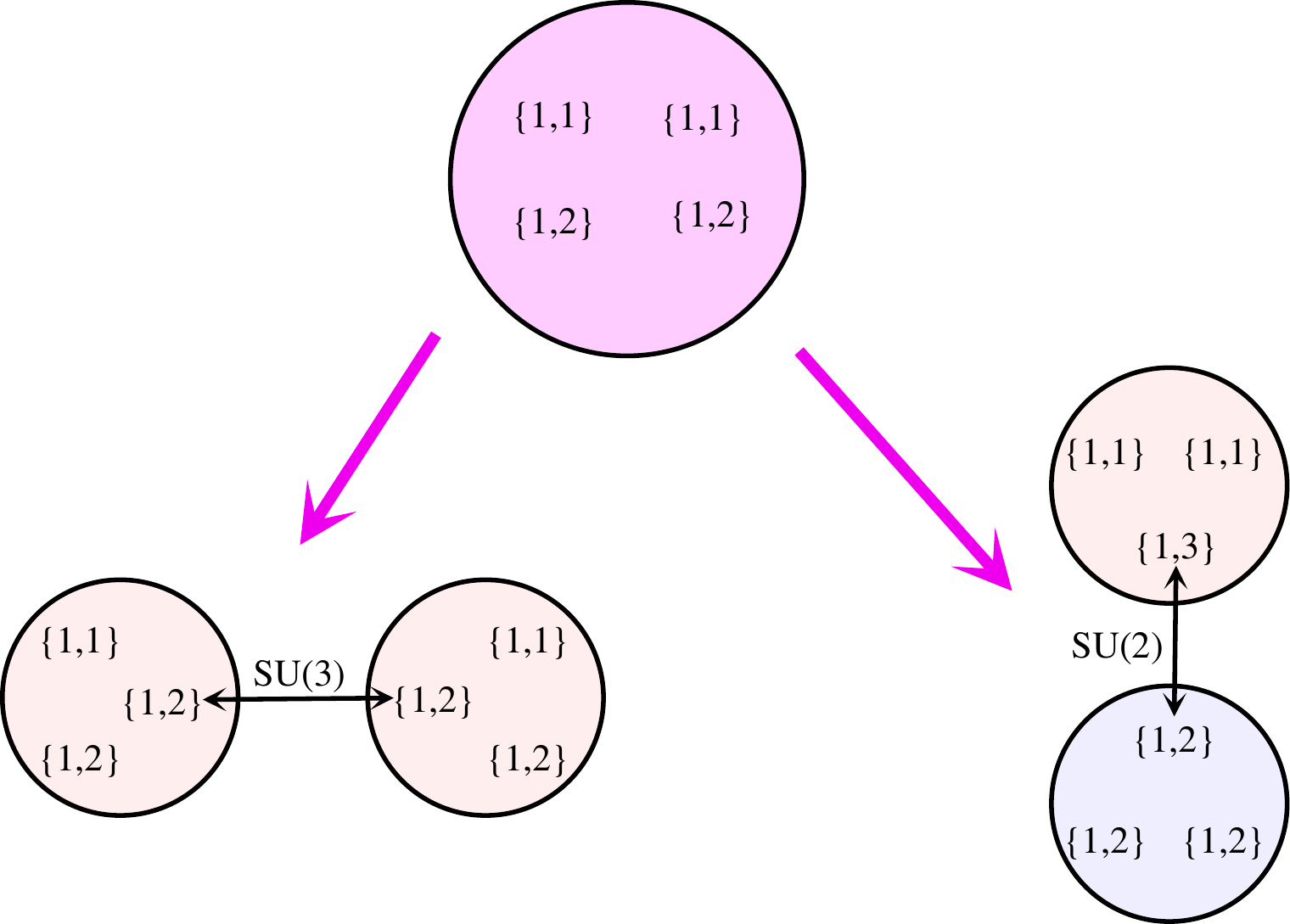}
\end{displaymath}
On one side we have an $SU(3)$ gauge theory with 6 hypermultiplets in the fundamental (3 from each fixture). On the other, we have an $SU(2)$ gauge theory coupled to one fundamental hypermultiplet, where the $SU(2)$ is a gauged subgroup of the original $\subset E_6$ flavor symmetry of the interacting $E_6$ SCFT. The central charge of the $E_6$ current algebra is such that the $\beta$-function of the $SU(2)$ vanishes. In both cases, the global symmetry group is $SU(6)\times U(1)$. In the $SU(2)$ gauge theory, the $SU(6)$ global symmetry arises as the commutant of $SU(2)\subset E_6$.

We can use this example of S-duality to compute the $(a,c)$ central charges of the $E_6$ SCFT. The effective number of vector multiplets and hypermultiplets of the $SU(3)$ $N_f=6$ theory are $n_v=8$ and $n_h=18$, respectively. In the S-dual theory, the $SU(2)$ gauge group and the fundamental hypermultiplet contribute $n_v=3$ and $n_h=2$, so the difference gives $n_v=5$ and $n_h=16$ for the $E_6$ theory. From these numbers we compute $a=\frac{41}{24}$ and $c=\frac{13}{6}$. The results, of course, agree with our explicit formul\ae, \eqref{nhdef} and \eqref{nvdef}.

\hypertarget{A3}{}\subsection{{$A_3$}}\label{A3}

Now we turn to the $A_3$ theory. There are four regular punctures:

\begin{center}
\begin{tabular}{|c|c|c|}
\hline
Pole Structure&Young Diagram&Global Symmetry\\
\hline 
$\{1,2,3\}$&$ \includegraphics[width=31px]{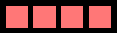}$&$SU(4)$\\
\hline
$\{1,2,2\}$&$ \includegraphics[width=24px]{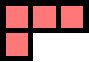}$&$SU(2)\times U(1)$\\
\hline
$\{1,1,2\}$&$ \includegraphics[width=16px]{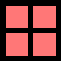}$&$SU(2)$\\
\hline
$\{1,1,1\}$&$ \includegraphics[width=16px]{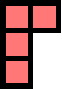}$&$U(1)$\\
\hline
\end{tabular}
\end{center}

and four irregular punctures:

\begin{displaymath}
\begin{gathered}
\{1,3,5\}\\
\{1,3,4\}\\
\{1,3,3\}\\
\{1,2,4\}\\
\end{gathered}
\end{displaymath}
The cylinders are:

\begin{center}
\begin{tabular}{|c|c|}
\hline
Cylinder&Gauge Group\\
\hline 
$\{1,3,5\}\xleftrightarrow{\qquad\qquad}\{1,2,2\}$&$SU(2)$\\
\hline
$\{1,3,4\}\xleftrightarrow{\qquad\qquad}\{1,2,3\}$&$SU(2)$\\
\hline
$\{1,3,3\}\xleftrightarrow{\qquad\qquad}\{1,2,4\}$&$SU(2)$\\
\hline
$\{1,3,3\}\xleftrightarrow{\qquad\qquad}\{1,2,3\}$&$Sp(2)$\\
\hline
$\{1,2,4\}\xleftrightarrow{\qquad\qquad}\{1,2,3\}$&$SU(3)$\\
\hline
$\{1,2,3\}\xleftrightarrow{\qquad\qquad}\{1,2,3\}$&$SU(4)$\\
\hline
\end{tabular}
\end{center}

To determine the fixtures, we need to consider all possible collisions of pairs of regular punctures. There are ten such collisions; six lead to free-field fixtures, and four to interacting SCFT fixtures. The ones which lead to free-field fixtures are (we draw the pair of punctures that collide on the left):

\begin{center}
\begin{tabular}{|l|c|c|}
\hline
Fixture&Number of Hypers&Representation\\
\hline 
$ \includegraphics[width=75px]{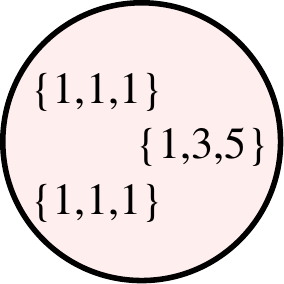}$&2&2\\
\hline
$ \includegraphics[width=75px]{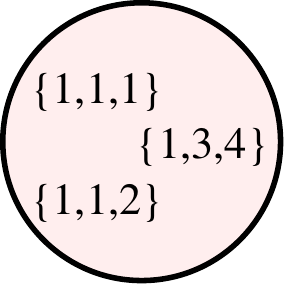}$&0&$-$\\
\hline
$ \includegraphics[width=75px]{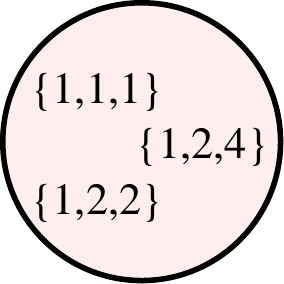}$&6&$(2,3)$\\
\hline
$ \includegraphics[width=75px]{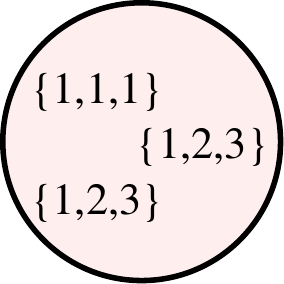}$&16&$(4,4)$\\
\hline
$ \includegraphics[width=75px]{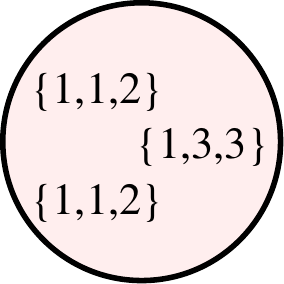}$&8&$\tfrac{1}{2}(2,2,4)$\\
\hline
$ \includegraphics[width=75px]{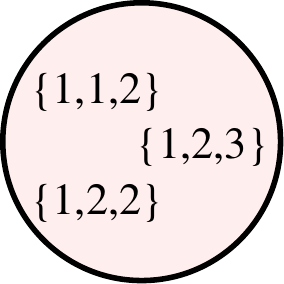}$&14&$(2,1,4)+\tfrac{1}{2}(1,2,6)$\\
\hline
\end{tabular}
\end{center}

The interacting fixtures are:
\begin{center}
\begin{tabular}{|c|c|c|c|l|}
\hline
Fixture&$(d_2,d_3,d_4)$&$(a,c)$&${(G_{\text{global}})}_k$&Theory\\
\hline 
$ \includegraphics[width=75px]{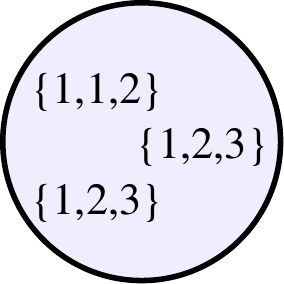}$&$(0,0,1)$&$\left(\tfrac{59}{24},\tfrac{19}{6}\right)$&${(E_7)}_8$&\parbox{1.75in}{\raggedright The $E_7$ SCFT of Minahan-Nemeschansky}\\
\hline
$ \includegraphics[width=75px]{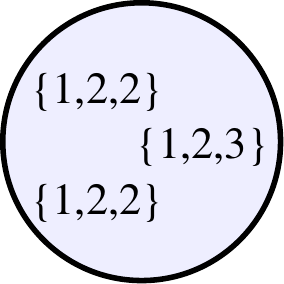}$&$(0,1,0)$&$\left(\tfrac{15}{8},\tfrac{5}{2}\right)$&${(E_6)}_6$&\parbox{1.35in}{The $E_6$ SCFT plus 4 free hypers}\\
\hline
$ \includegraphics[width=75px]{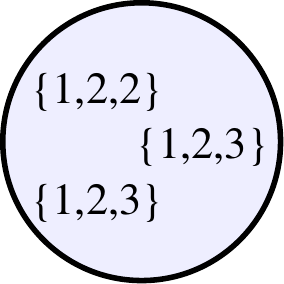}$&$(0,1,1)$&$\left(\tfrac{15}{4},\tfrac{9}{2}\right)$&${SU(2)}_6\times{SU(8)}_8$&New. ``$R_{0,4}$".\\
\hline
$ \includegraphics[width=75px]{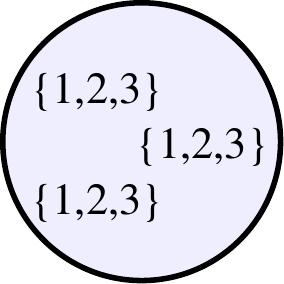}$&$(0,1,2)$&$\left(\tfrac{45}{8},\tfrac{13}{2}\right)$&${SU(4)}_8^3$&``New." $T_4$.\\
\hline
\end{tabular}
\end{center}

To understand the free-field fixtures, it is helpful to repeat the analysis that Gaiotto did, of ``{}the ends of linear quivers''{} \cite{Gaiotto:2009we}. In the present notation, we have a set of punctures colliding, in hierarchical fashion, producing a chain of fixtures, connected to the rest of $C$.

Consider the following chain, obtained as the collision of four minimal ($\{1,1,1\}$) punctures on $C$.

\begin{displaymath}
 \includegraphics[width=384px]{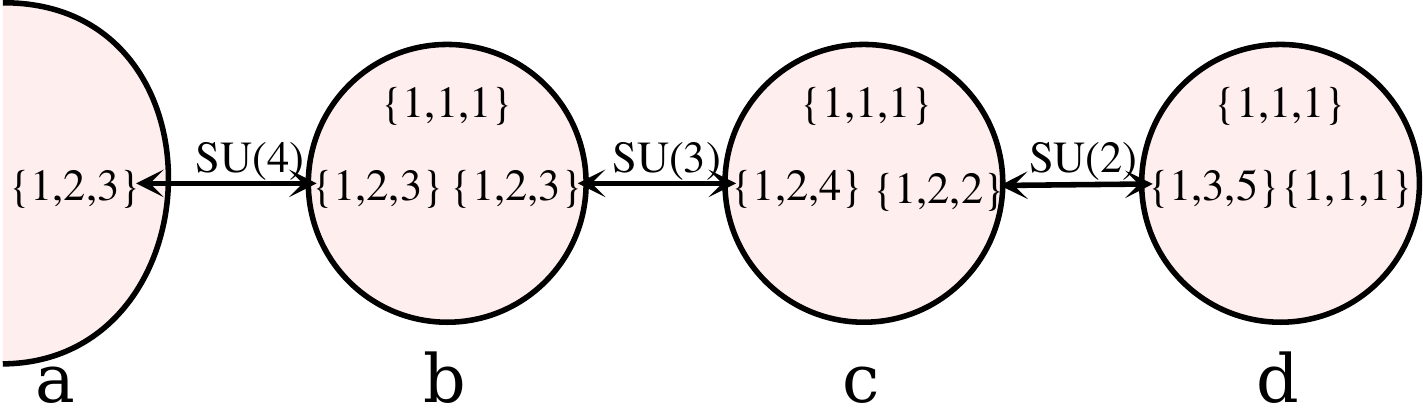}
\end{displaymath}
Representing the rest of $C$ as four hypermultiplets in the fundamental of $SU(4)$, the matter content of this theory is

\begin{center}
\begin{tabular}{|r|c|c|c|c|}
\hline
~{}&$\#$ hypers&$SU(4)$&$SU(3)$&$SU(2)$\\
\hline 
a&4&4&1&1\\
\hline
b&1&4&1&1\\
~{}&1&4&3&1\\
\hline
c&1&1&3&2\\
\hline
d&1&1&1&2\\
\hline
\end{tabular}
\end{center}

Each gauge group factor has vanishing $\beta$-function. We can obtain the gauge theories correponding to other, related, collisions by lopping fixtures off of the end of the picture. For instance, the gauge theory corresponding to the collision of two minimal punctures and a $\{1,2,2\}$ puncture is obtained by omitting fixture ``{}d''{} and the $SU(2)$ gauge group factor.

The collision of two minimal punctures and a $\{1,1,2\}$ puncture gives rise to

\begin{displaymath}
 \includegraphics[width=271px]{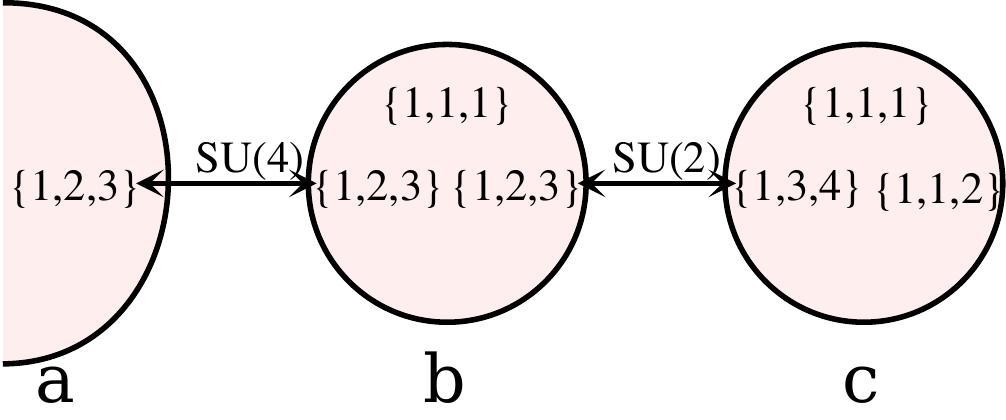}
\end{displaymath}
with matter content

\begin{center}
\begin{tabular}{|r|c|c|c|}
\hline
~{}&$\#$ hypers&$SU(4)$&$SU(2)$\\
\hline 
a&4&4&1\\
\hline
b&1&4&2\\
~{}&2&4&1\\
\hline
c&$-$&$-$&$-$\\
\hline
\end{tabular}
\end{center}

This theory is S-dual to

\begin{displaymath}
 \includegraphics[width=271px]{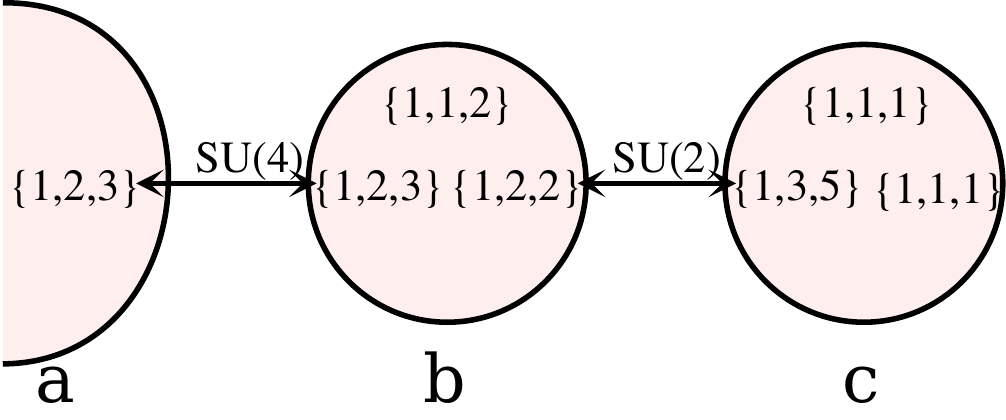}
\end{displaymath}
with matter content

\begin{center}
\begin{tabular}{|r|c|c|c|}
\hline
~{}&$\#$ hypers&$SU(4)$&$SU(2)$\\
\hline 
a&4&4&1\\
\hline
b&2&4&1\\
~{}&$\frac{1}{2}$&6&2\\
\hline
c&1&1&2\\
\hline
\end{tabular}
\end{center}

The collision of one minimal puncture and two $\{1,1,2\}$ punctures gives

\begin{displaymath}
 \includegraphics[width=271px]{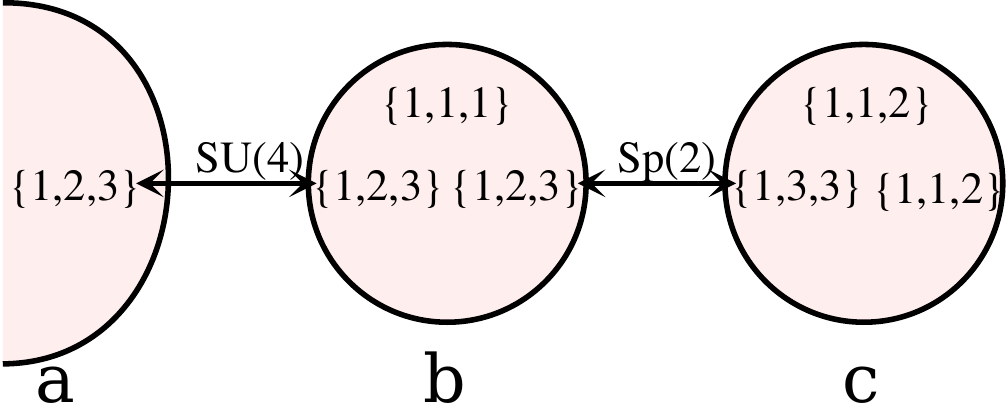}
\end{displaymath}
with matter content

\begin{center}
\begin{tabular}{|r|c|c|c|}
\hline
~{}&$\#$ hypers&$SU(4)$&$Sp(2)$\\
\hline 
a&4&4&1\\
\hline
b&1&4&4\\
\hline
c&2&1&4\\
\hline
\end{tabular}
\end{center}

If we S-dualize this, we end up with an interacting SCFT fixture. To study that, in its simplest context, let'{}s turn off the $SU(4)$, and consider the simpler theory

\begin{displaymath}
 \includegraphics[width=189px]{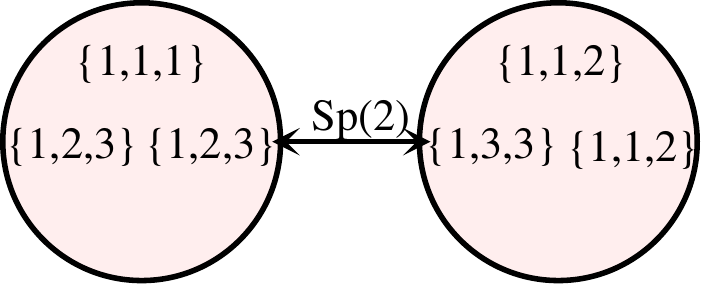}
\end{displaymath}
which is an $Sp(2)$ gauge theory with 6 hypers in the fundamental (4 from the fixture on the left, and 2 from the fixture on the right). The global symmetry group is $SO(12)$. The Seiberg-Witten solution can be found in  \cite{Argyres:1995fw}.

S-dualizing, we obtain

\begin{displaymath}
 \includegraphics[width=189px]{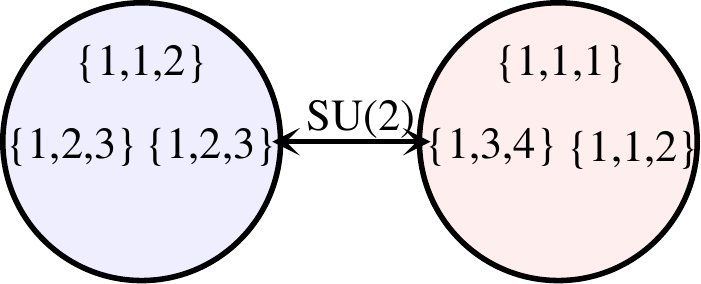}
\end{displaymath}
The fixture on the right contains no matter, so the theory is an $SU(2)$ gauging of the interacting fixture on the left. The commutant of $SU(2)\subset G$ must be $SO(12)$, and conformality implies that the central charge $k=8$. Exactly these considerations led Argyres and Seiberg \cite{Argyres:2007cn} to identify the SCFT corresponding to this fixture as the $E_7$ SCFT of Minahan and Nemeschansky \cite{Minahan:1996cj}. We can use this example to find $n_v=7$ and $n_h=24$ for the $E_7$ SCFT, from which we compute $a=\frac{59}{24}$ and $c=\frac{19}{6}$ (which, again, agree with our explicit formul\ae, \eqref{nvdef},\eqref{nhdef}).

We can use our rules to find the $E_7$ theory in a different example, as the strong coupling point of a Lagrangian theory with $SU(4)$ gauge group. Consider

\begin{displaymath}
 \includegraphics[width=189px]{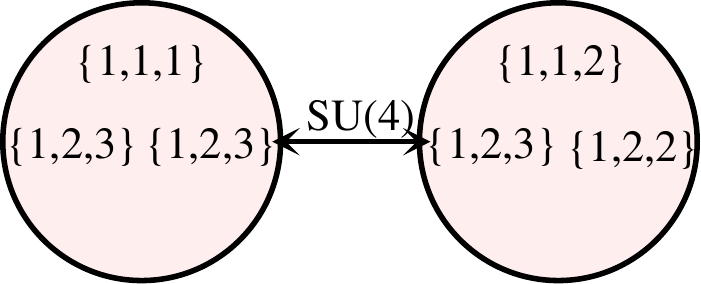}
\end{displaymath}
This is a $SU(4)$ gauge theory with 6 fundamental hypermultiplets, and 1 hypermultiplet in the 6 of $SU(4)$. The S-dual frame contaning the $E_7$ theory is

\begin{displaymath}
 \includegraphics[width=189px]{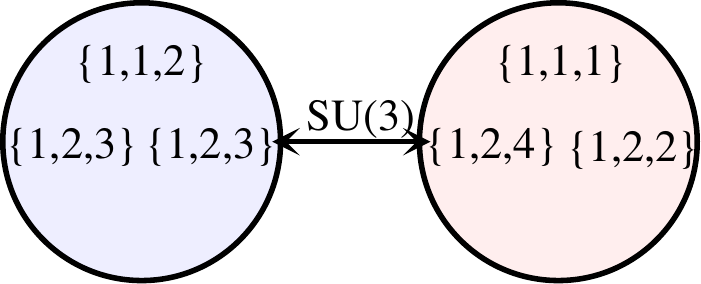}
\end{displaymath}
This is an $SU(2)$ gauging of the $E_7$ theory, coupled to 2 fundamental hypermultiplets. One can also compute $n_v=7$ and $n_h=24$ for the $E_7$ theory from this example, which agrees with what we obtained previously.

Let us study the next in the list of interacting SCFT fixtures. Start with

\begin{displaymath}
 \includegraphics[width=189px]{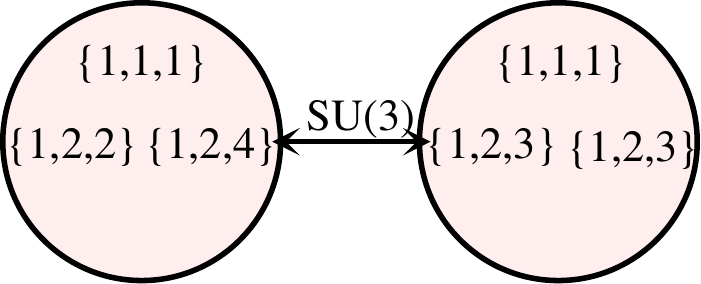}
\end{displaymath}
This is $SU(3)$ with 6 fundamental hypers, and 4 free hypers. S-dualizing, we obtain

\begin{displaymath}
 \includegraphics[width=189px]{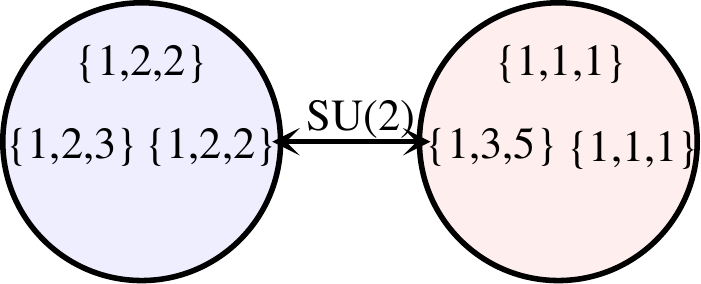}
\end{displaymath}
But we'{}ve seen this S-duality before (without the 4 free hypers) when we studied the $A_2$ theory. The fixture on the right is two hypers (one fundamental of $SU(2)$). So the fixture on the left must be the $E_6$ SCFT plus 4 free hypers. Indeed, one finds $n_v=5$ and $n_h=20$ (and so $a=\frac{15}{8}$ and $c=\frac{5}{2}$) for this fixture, which is what we expected, given the values $n_v=5$ and $n_h=16$ for the $E_6$ SCFT alone.

As a further check on this identification, consider

\begin{displaymath}
 \includegraphics[width=189px]{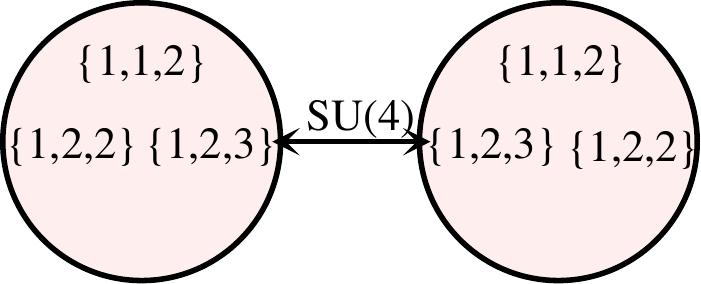}
\end{displaymath}
This is an $SU(4)$ gauge theory with 4 hypermutiplets in the fundamental, and 2 hypermultiplets in the 6. The global symmetry group is

\begin{equation}
G_{\text{global}}= SU(4)_8\times Sp(2)_6\times U(1).
\label{su4sp2u1}\end{equation}
S-dualizing, we obtain

\begin{displaymath}
 \includegraphics[width=189px]{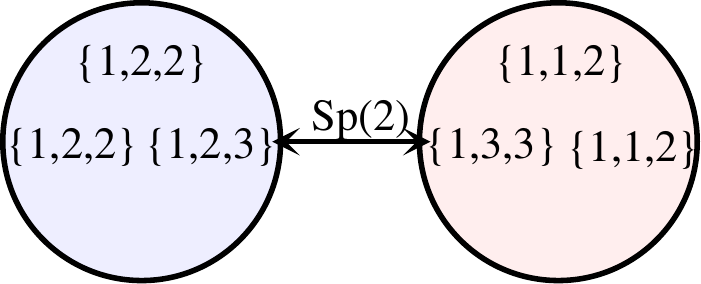}
\end{displaymath}
This is an $Sp(2)$ gauge theory. The fixture on the right supplies two hypermultiplets in the fundamental. According to our identification, the fixture on the left provides one more fundamental hypermultiplet, making a total of 3 fundamental hypers. Gauging an $Sp(2)\subset E_6$, with $k=6$, ensures conformality. The global symmetry group associated to the 3 fundamental hypers is $SO(6)\sim SU(4)$. The commutant of $Sp(2)\subset E_6$ is $Sp(2)\times U(1)$, giving an overall global symmetry group which agrees with \eqref{su4sp2u1}.

Next we turn to

\begin{displaymath}
 \includegraphics[width=189px]{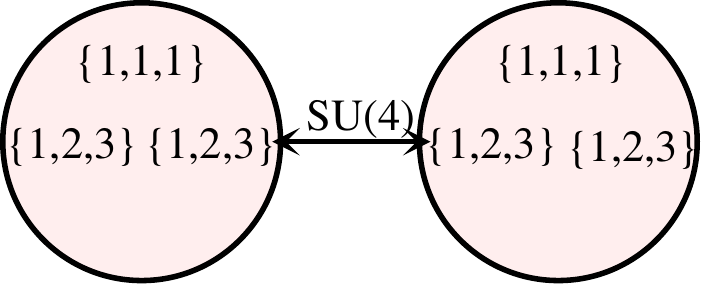}
\end{displaymath}
This is $SU(4)$ with 8 fundamental hypers. It is conformal, and has an ${SU(8)}_8\times U(1)$ global symmetry. S-dualizing, we obtain

\begin{displaymath}
 \includegraphics[width=189px]{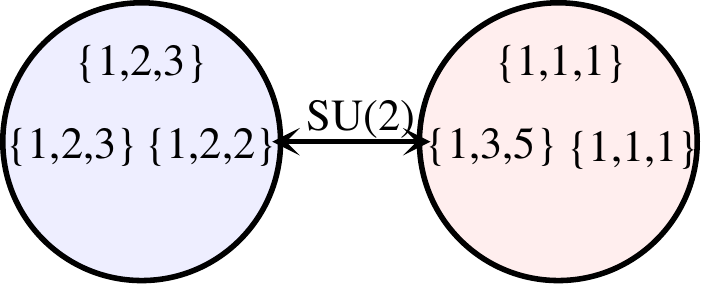}
\end{displaymath}
This is $SU(2)$ with one fundamental hyper (from the fixture on the right), coupled to an $SU(2)$ subgroup of the global symmetry group of the interacting SCFT fixture on the left. The commutant of $SU(2)$ must be $SU(8)$, and the central charge of the $SU(2)$ current algebra must be $k=6$.

To gain more information, consider

\begin{displaymath}
 \includegraphics[width=189px]{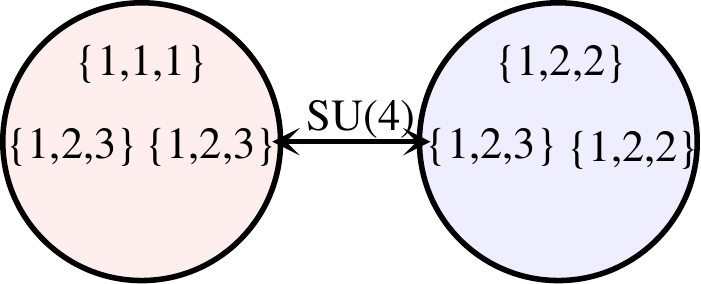}
\end{displaymath}
This is an $SU(4)$ gauge theory. The fixture on the left provides 4 hypermultipets in the fundamental. The free hypers from the fixture on the right provide one more fundamental (making a total of 5 fundamental hypers). Gauging an $SU(4)\subset E_6$, at $k=6$ makes the theory conformal. The commutant of $SU(4)\subset E_6$ is $SU(2)\times SU(2)\times U(1)$, so the global symmetry group of this gauge theory is

\begin{displaymath}
G_{\text{global}} = {SU(5)}_8\times {SU(2)}^2_6\times {U(1)}^2.
\end{displaymath}
S-dualizing, we obtain

\begin{displaymath}
 \includegraphics[width=189px]{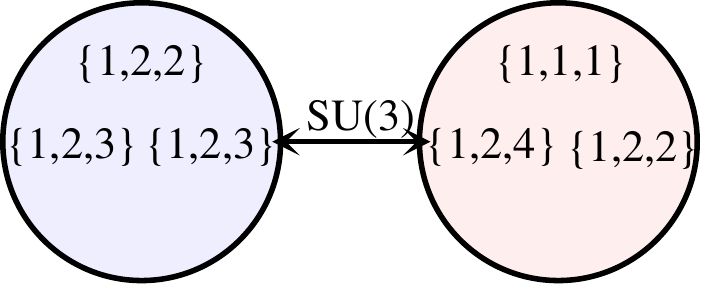}
\end{displaymath}
The fixture on the right supplies 2 hypermultiplets in the fundamental. These supply an $SU(2)\times U(1)$ subgroup of the global symmetry group.

If we gauge an $SU(3)\subset SU(8)$ of the fixture on the right, we obtain conformality for $k=8$. Moreover, the commutant of $SU(3)\subset SU(8)$ is $SU(5)\times U(1)$. So we obtain conformality and the correct global symmetry groups for our two examples if

\begin{displaymath}
G_{\text{SCFT}} = SU(2)_{k=6}\times SU(8)_{k=8}.
\end{displaymath}
From either of the above two gaugings of this $SU(2)_{k=6}\times SU(8)_{k=8}$ SCFT we can compute $n_v=12$ and $n_h=30$, and so its central charges are $a=\frac{15}{4}$ and $c=\frac{9}{2}$.

This SCFT with global symmetry $SU(2)_{k=6}\times SU(8)_{k=8}$ belongs to a series, $R_{0,N}$, of $A_{N-1}$ ($N\geq 3$) interacting SCFTs with global symmetry

\begin{displaymath}
G_{\text{global}} = SU(2)_{k=6}\times SU(2N)_{k=2N}\; ,
\end{displaymath}
which we will discuss in \S\ref{series}.

Finally, let us pass to the last of the interacting fixtures on our list. Consider

\begin{displaymath}
 \includegraphics[width=189px]{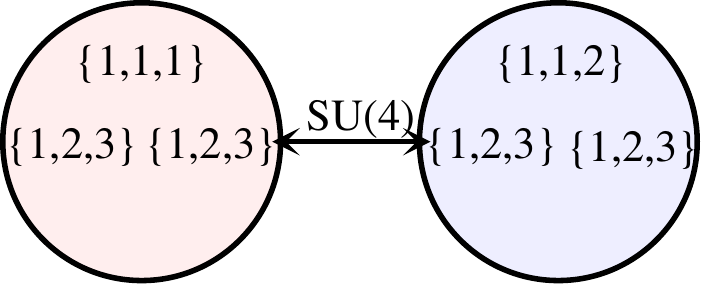}
\end{displaymath}
The fixture on the left provides 4 hypermultiplets in fundamental. Gauging an $SU(4)\subset E_7$ at $k=8$ achieves conformality. The commutant of $SU(4)\subset E_7$ is $SU(4)\times SU(2)$. So, overall, the global symmetry group is

\begin{equation}
G_{\text{global}}= {SU(4)}_8^2\times {SU(2)}_8\times U(1).
\label{su4su4su2u1}\end{equation}
S-dualizing, we obtain

\begin{displaymath}
 \includegraphics[width=189px]{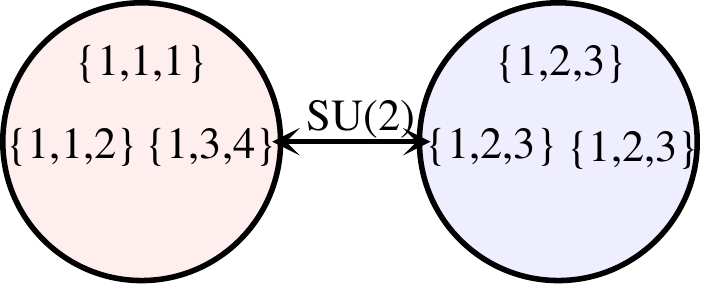}
\end{displaymath}
The fixture on the left supplies no matter. To achieve conformality, gauging an $SU(2)$ subgroup of $G_{\text{SCFT}}$, we must have $k=8$. For the global symmetries to agree with \eqref{su4su4su2u1}, the commutant of $SU(2)$ must be $SU(4)^2\times SU(2)\times U(1)$, which suggests that

\begin{displaymath}
G_{\text{SCFT}}= {SU(4)}^3_{k=8}.
\end{displaymath}
As another check, consider

\begin{displaymath}
 \includegraphics[width=189px]{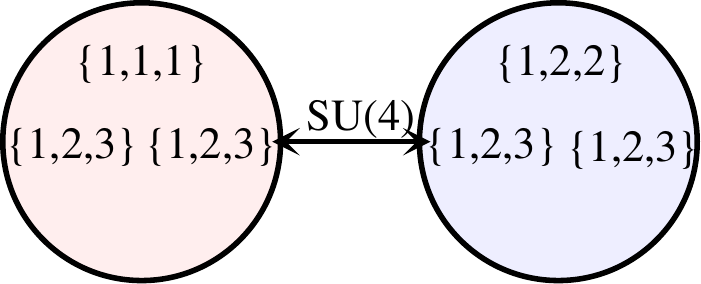}
\end{displaymath}
The fixture on the left supplies 4 hypermultiplets in the fundamental of $SU(4)$, which contribute an $SU(4)\times U(1)$ to $G_{\text{global}}$. On the right, we gauge an $SU(4)\subset {SU(2)}_{k=6}\times {SU(8)}_{k=8}$. The commutant is $SU(2)\times SU(4)\times U(1)$. So, overall,

\begin{equation}
G_{\text{global}}= {SU(4)}_8^2\times {SU(2)}_6\times U(1)^2\; .
\label{su4su4su2u1u1}\end{equation}
S-dualizing, we obtain

\begin{displaymath}
 \includegraphics[width=189px]{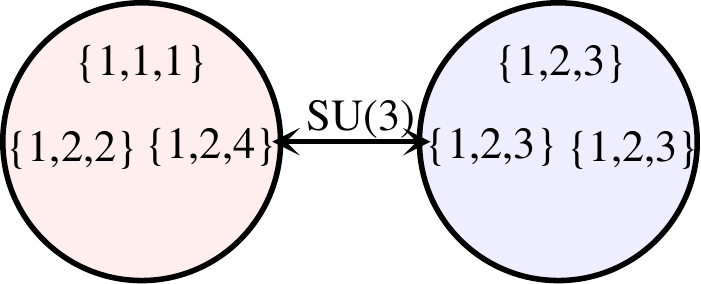}
\end{displaymath}
The fixture on the left supplies 2 hypermultiplets in the fundamental of $SU(3)$ (contributing an ${SU(2)}_6\times U(1)$ factor to $G_{\text{global}}$). On the right, we gauge an $SU(3)\subset {SU(4)}^3_{k=8}$, which yields a conformal theory. And the commutant, ${SU(4)}_8^2\times U(1)$, combines to give \eqref{su4su4su2u1u1}.

Using any of these gaugings we find $n_v=19$ and $n_h=40$, and so $a=\frac{45}{8}$ and $c=\frac{13}{2}$, for the ${SU(4)}^3_{k=8}$ SCFT. This SCFT is part of the $T_N$ series \cite{Gaiotto:2009we, Gaiotto:2009gz, Benini:2009gi}, which for $N\geq 4$ has ${SU(N)}^3_{k=2N}$ global symmetry.

Finally, let us note that the cylinder between the pair of irregular punctures is crucial to understanding certain S-duality frames. For instance, consider the 5-punctured sphere

\begin{displaymath}
 \includegraphics[width=425px]{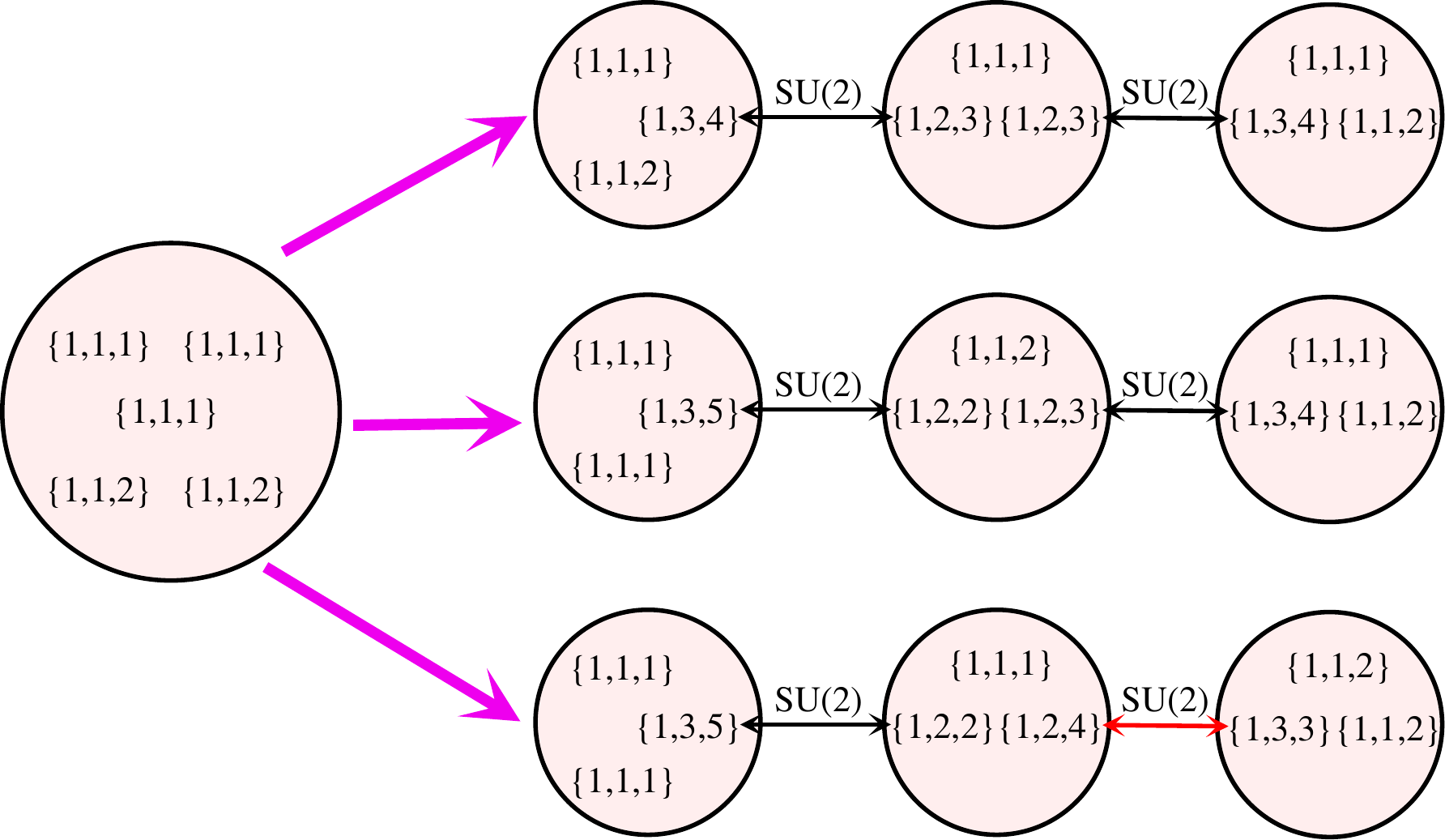}
\end{displaymath}
Note that, for each degeneration, we have an $SU(2)\times SU(2)$ gauge theory, with matter in the $(2,2)+2(2,1)+2(1,2)+4(1,1)$, so that

\begin{displaymath}
G_{\text{global}} = {SU(2)}^2\times {U(1)}^3 + \text{4 free hypers}.
\end{displaymath}
But, to make sense of the last degeneration, we crucially need the cylinder between two irregular punctures.

\hypertarget{A4}{}\subsection{{$A_4$}}\label{A4}

Now we turn to the $A_4$ theory. There are six regular punctures:

\begin{center}
\begin{tabular}{|c|c|c|}
\hline
Pole Structure&Young Diagram&Global Symmetry\\
\hline 
$\{1,2,3,4\}$&$ \includegraphics[width=39px]{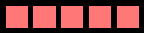}$&$SU(5)$\\
\hline
$\{1,2,3,3\}$&$ \includegraphics[width=39px]{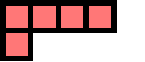}$&$SU(3)\times U(1)$\\
\hline
$\{1,2,2,3\}$&$ \includegraphics[width=24px]{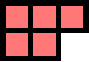}$&$SU(2)\times U(1)$\\
\hline
$\{1,2,2,2\}$&$ \includegraphics[width=24px]{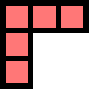}$&$SU(2)\times U(1)$\\
\hline
$\{1,1,2,2\}$&$ \includegraphics[width=16px]{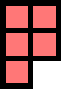}$&$U(1)$\\
\hline
$\{1,1,1,1\}$&$ \includegraphics[width=16px]{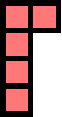}$&$U(1)$\\
\hline
\end{tabular}
\end{center}

\noindent
and six irregular punctures:
\begin{displaymath}
\begin{gathered}
\{1,3,5,7\}\\
\{1,3,4,6\}\\
\{1,3,3,5\}\\
\{1,2,4,6\}\\
\{1,2,4,5\}\\
\{1,2,3,5\}
\end{gathered}
\end{displaymath}
The cylinders are:
\begin{center}
\begin{tabular}{|c|c|}
\hline
Cylinder&Gauge Group\\
\hline 
$\{1,3,5,7\}\xleftrightarrow{\qquad\qquad}\{1,2,2,2\}$&$SU(2)$\\
\hline
$\{1,3,4,6\}\xleftrightarrow{\qquad\qquad}\{1,2,3,3\}$&$SU(2)$\\
\hline
$\{1,3,3,5\}\xleftrightarrow{\qquad\qquad}\{1,2,3,4\}$&$Sp(2)$\\
\hline
$\{1,2,4,6\}\xleftrightarrow{\qquad\qquad}\{1,2,3,3\}$&$SU(3)$\\
\hline
$\{1,2,4,5\}\xleftrightarrow{\qquad\qquad}\{1,2,3,4\}$&$SU(3)$\\
\hline
$\{1,2,3,5\}\xleftrightarrow{\qquad\qquad}\{1,2,3,4\}$&$SU(4)$\\
\hline
$\{1,2,3,4\}\xleftrightarrow{\qquad\qquad}\{1,2,3,4\}$&$SU(5)$\\
\hline
\end{tabular}
\end{center}

The free-field fixtures are

\begin{center}
\begin{tabular}{|l|c|c|}
\hline
Fixture&Number of Hypers&Representation\\
\hline 
$ \includegraphics[width=91px]{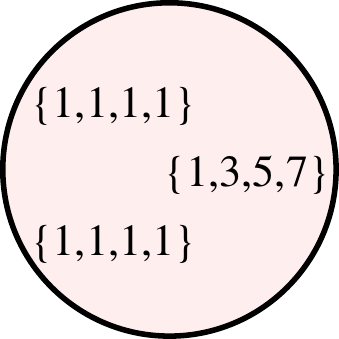}$&2&2\\
\hline
$ \includegraphics[width=91px]{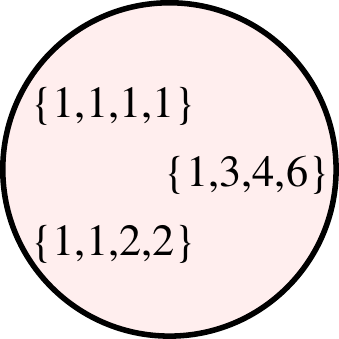}$&0&$-$\\
\hline
\end{tabular}
\end{center}

\begin{center}
\begin{tabular}{|l|c|c|}
\hline
Fixture&Number of Hypers&Representation\\
\hline
$ \includegraphics[width=91px]{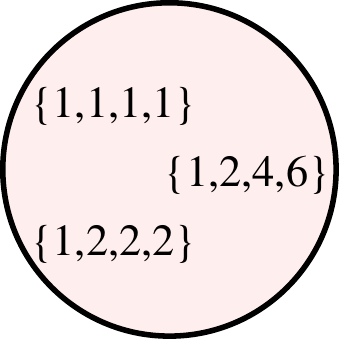}$&6&$(2,3)$\\
\hline
$ \includegraphics[width=91px]{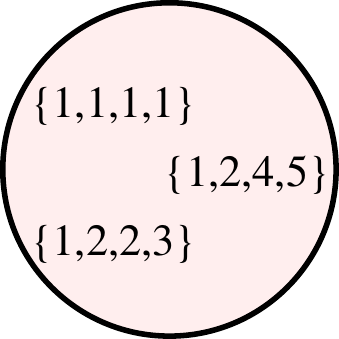}$&3&$(1,3)$\\
\hline
$ \includegraphics[width=91px]{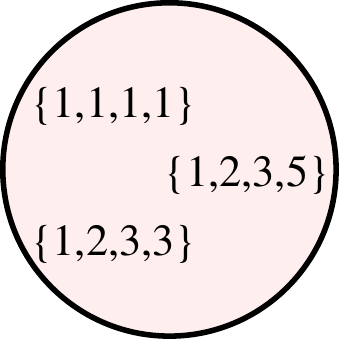}$&12&$(3,4)$\\
\hline
$ \includegraphics[width=91px]{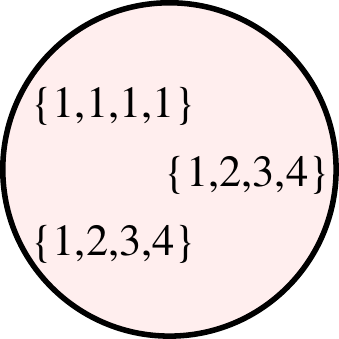}$&25&$(5,5)$\\
\hline
$ \includegraphics[width=91px]{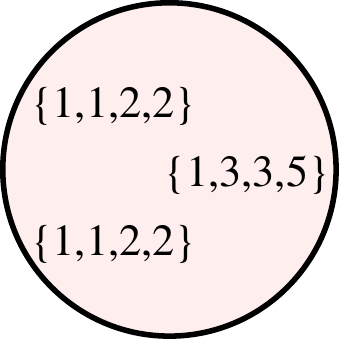}$&4&4\\
\hline
$ \includegraphics[width=91px]{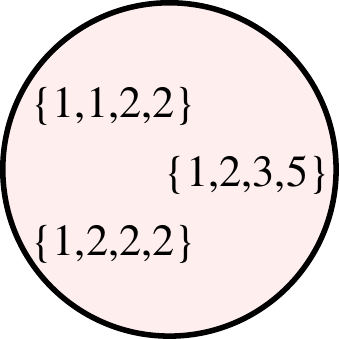}$&10&$(1,4)+\tfrac{1}{2}(2,6)$\\
\hline
\end{tabular}
\end{center}

\begin{center}
\begin{tabular}{|l|c|c|}
\hline
Fixture&Number of Hypers&Representation\\
\hline
$ \includegraphics[width=91px]{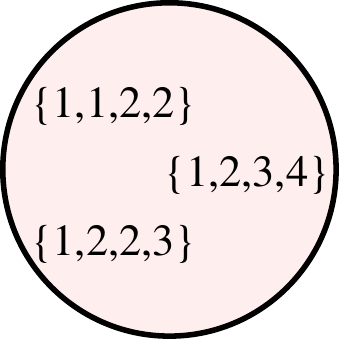}$&20&$(2,5)+(1,10)$\\
\hline
\end{tabular}
\end{center}

The interacting fixtures are:

\begin{center}
\begin{tabular}{|c|c|c|c|l|}
\hline
Fixture&$(d_2,d_3,d_4,d_5)$&$(a,c)$&${(G_{\text{global}})}_k$&Theory\\
\hline 
$ \includegraphics[width=91px]{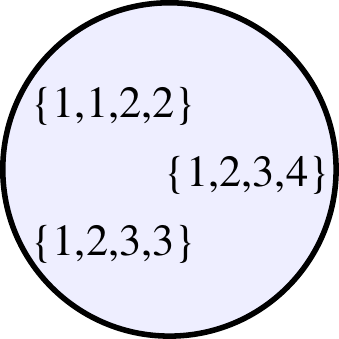}$&$(0,0,1,0)$&$\left(\tfrac{8}{3},\tfrac{43}{12}\right)$&${(E_7)}_8$&\parbox{1.25in}{\raggedright The $E_7$ SCFT plus 5 hypers}\\
\hline
\includegraphics[width=91px]{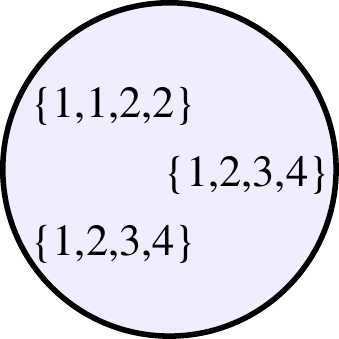}&$(0,0,1,1)$&\parbox{.5in}{$\left(\tfrac{61}{12},\tfrac{37}{6}\right)$}&${SU(10)}_{10}$&New. ``$S_5$".\\
\hline
\includegraphics[width=91px]{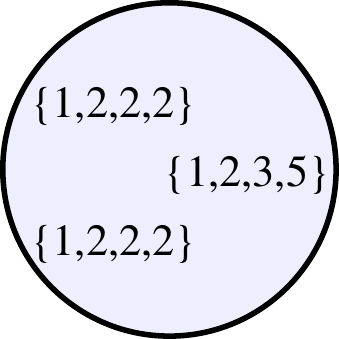}&$(0,1,0,0)$&$\left(\tfrac{41}{24},\tfrac{13}{6}\right)$&${(E_6)}_6$&The $E_6$ SCFT\\
\hline 
$ \includegraphics[width=91px]{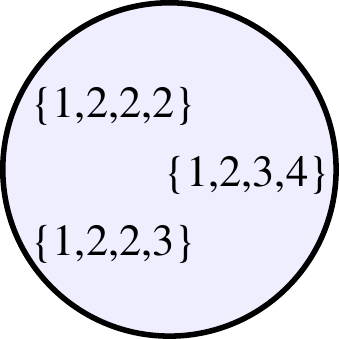}$&$(0,1,0,0)$&$\left(\tfrac{17}{8},3\right)$&${(E_6)}_6$&\parbox{1.5in}{\raggedright The $E_6$ SCFT plus 10 hypers, in the $(1,2,5)$}\\
\hline
\end{tabular}
\end{center}

\begin{center}
\begin{tabular}{|c|c|c|c|l|}
\hline
Fixture&$(d_2,d_3,d_4,d_5)$&$(a,c)$&${(G_{\text{global}})}_k$&Theory\\
\hline
$ \includegraphics[width=91px]{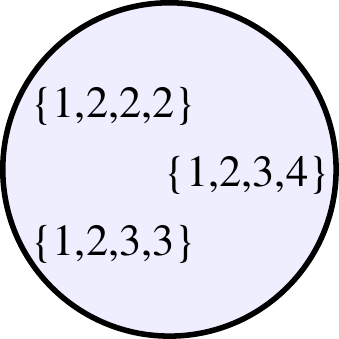}$&$(0,1,1,0)$&$\left(\tfrac{95}{24},\tfrac{59}{12}\right)$&${SU(2)}_6\times{SU(8)}_8$&\parbox{1.75in}{\raggedright The ${SU(2)}_6\times{SU(8)}_8$ SCFT plus 5 hypers}\\
\hline
$ \includegraphics[width=91px]{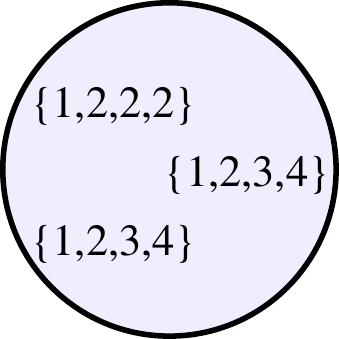}$&$(0,1,1,1)$&$\left(\tfrac{51}{8},\tfrac{15}{2}\right)$&${SU(2)}_6\times{SU(10)}_{10}$&``{}New.''{} $R_{0,5}$.\\
\hline
$ \includegraphics[width=91px]{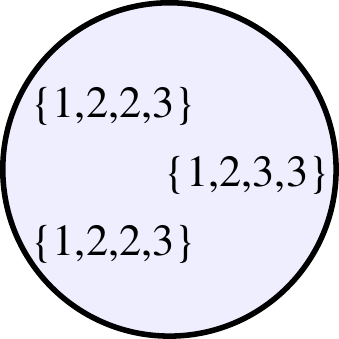}$&$(0,1,0,0)$&$\left(2,\tfrac{11}{4}\right)$&${(E_6)}_6$&\parbox{1.5in}{\raggedright The $E_6$ SCFT plus 7 hypers in the $(2,2,1)+(1,1,3)$}\\
\hline
$ \includegraphics[width=91px]{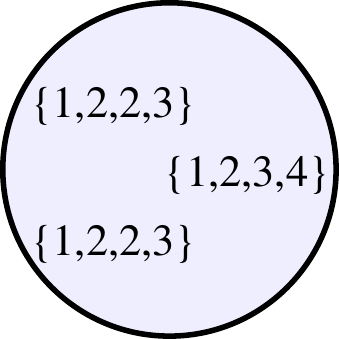}$&$(0,1,0,1)$&$\left(\tfrac{53}{12},\tfrac{16}{3}\right)$&${SO(14)}_{10}\times {U(1)}$&New. ``$R_{2,5}$".\\
\hline
$ \includegraphics[width=91px]{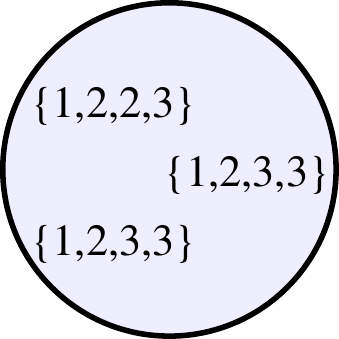}$&$(0,1,1,0)$&$\left(\tfrac{23}{6},\tfrac{14}{3}\right)$&${SU(2)}_{6}\times {SU(8)}_8$&\parbox{1.65in}{\raggedright The ${SU(2)}_{6}\times {SU(8)}_8$ SCFT plus 2 hypers}\\
\hline
\end{tabular}
\end{center}

\begin{center}
\begin{tabular}{|c|c|c|c|l|}
\hline
Fixture&$(d_2,d_3,d_4,d_5)$&$(a,c)$&${(G_{\text{global}})}_k$&Theory\\
\hline
$ \includegraphics[width=91px]{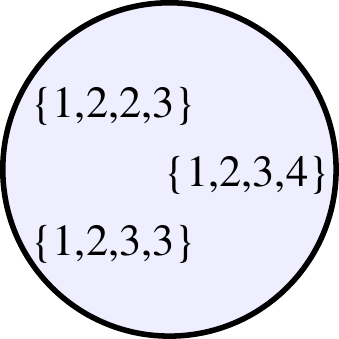}$&$(0,1,1,1)$&$\left(\tfrac{25}{4},\tfrac{29}{4}\right)$&${SU(3)}_8\times {SU(7)}_{10}\times U(1)$&New. ``$R_{1,5}$".\\
\hline
$ \includegraphics[width=91px]{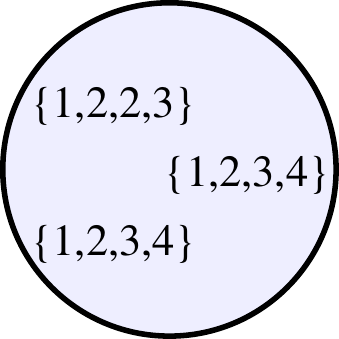}$&$(0,1,1,2)$&$\left(\tfrac{26}{3},\tfrac{59}{6}\right)$&${SU(5)}_{10}^2\times {SU(2)}_{10}\times U(1)$&New. ``$V_N$"\\
\hline
$ \includegraphics[width=91px]{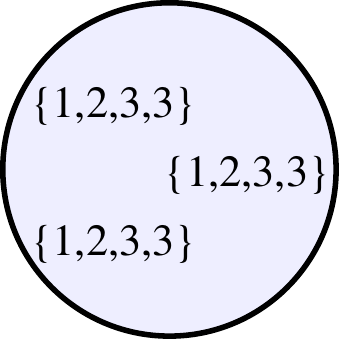}$&$(0,1,2,0)$&$\left(\tfrac{17}{3},\tfrac{79}{12}\right)$&${SU(4)}_{8}^3$&\parbox{1.5in}{\raggedright The $SU(4)_{8}^3$ SCFT plus 1 hyper}\\
\hline
$ \includegraphics[width=91px]{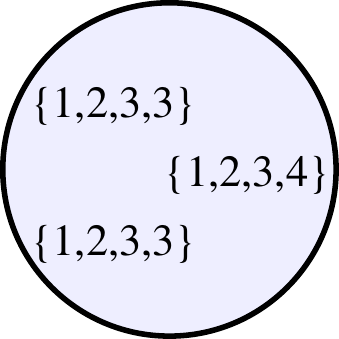}$&$(0,1,2,1)$&$\left(\tfrac{97}{12},\tfrac{55}{6}\right)$&${SU(6)}_{10}\times {SU(3)}_{8}^2\times U(1)$&New\\
\hline
$ \includegraphics[width=91px]{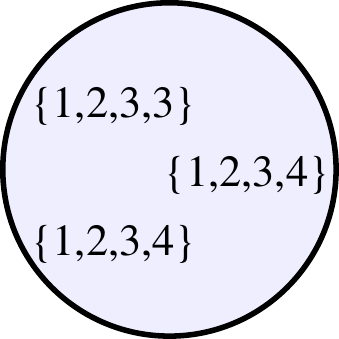}$&$(0,1,2,2)$&$\left(\tfrac{53}{6},\tfrac{47}{4}\right)$&${SU(5)}_{10}^2\times {SU(3)}_{8}\times U(1)$&New. ``$U_5$".\\
\hline
$ \includegraphics[width=91px]{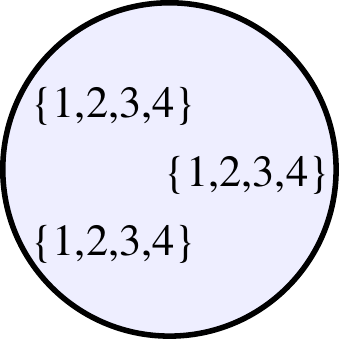}$&$(0,1,2,3)$&$\left(\tfrac{155}{12},\tfrac{43}{3}\right)$&${SU(5)}_{10}^3$&``{}New.''{} $T_5$.\\
\hline
\end{tabular}
\end{center}

Since our procedures should by now be more or less straightforward, let us simply present the $A_4$ interacting SCFTs as strong coupling points of linear quivers of special unitary groups.

For the $SU(10)$ theory, we study the following theory

\begin{displaymath}
 \includegraphics[width=189px]{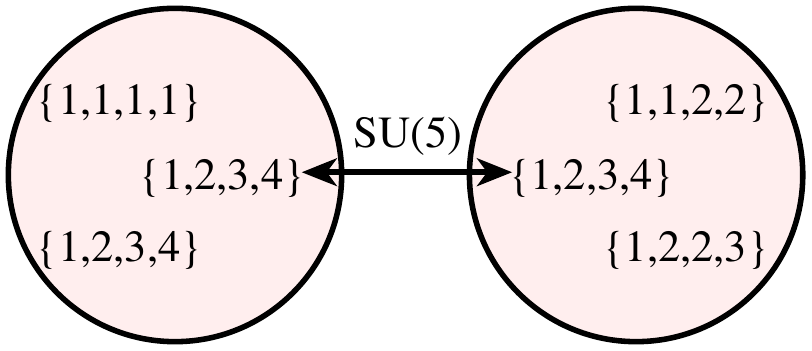}
\end{displaymath}
which is a $SU(5)$ gauge theory with 7 fundamental hypermultiplets and one hypermultiplet in the 10 of $SU(5)$. The S-dual frame in which we are interested is

\begin{displaymath}
 \includegraphics[width=189px]{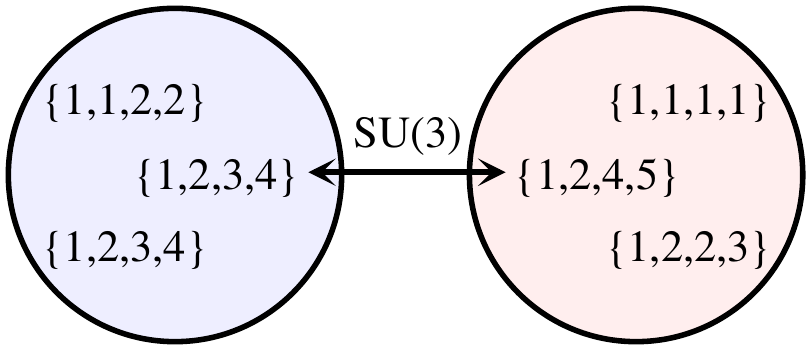}
\end{displaymath}
which is a $SU(3)$ gauging of the $SU(10)$ theory coupled to one fundamental hypermultiplet. The $SU(10)$ theory is the first in a series of interacting SCFTs, $S_N$ ($N\geq 5$), which we discuss in \S\ref{series}.

For the $SU(2)\times SU(10)$ theory, consider the Lagrangian theory

\begin{displaymath}
 \includegraphics[width=189px]{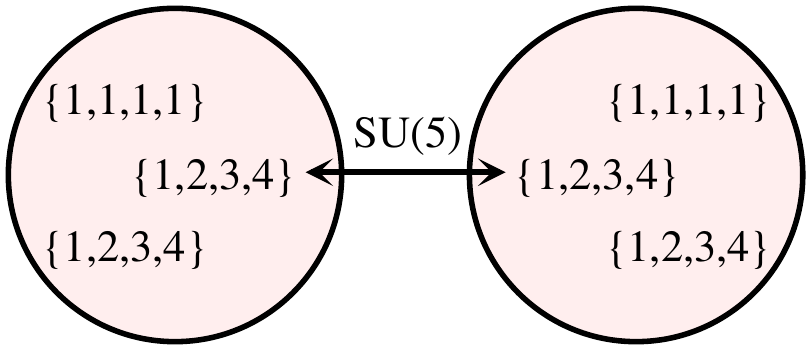}
\end{displaymath}
which is the $SU(5)$ $N_f=10$ gauge theory. The S-dual theory, which we are interested in, is

\begin{displaymath}
 \includegraphics[width=189px]{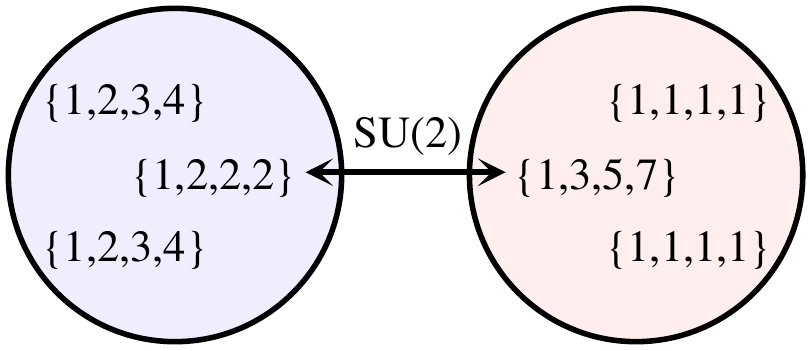}
\end{displaymath}
which is a $SU(2)$ gauging of the $SU(2)\times SU(10)$ theory, coupled to one fundamental hypermultiplet.

For the $SO(14)\times U(1)$ theory we consider the Lagrangian theory

\begin{displaymath}
 \includegraphics[width=189px]{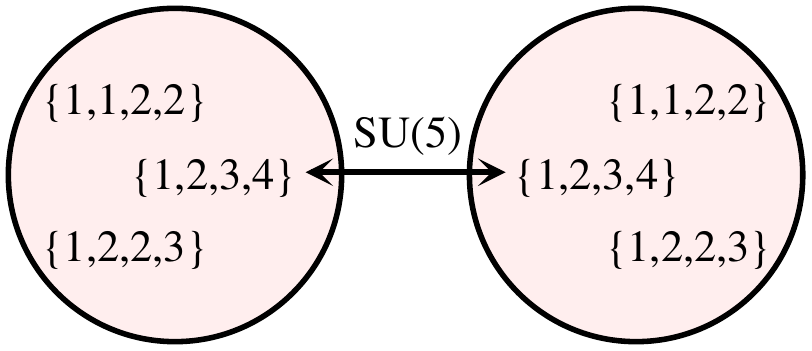}
\end{displaymath}
which is a $SU(5)$ gauge theory with 4 fundamental hypermultiplets and 2 hypermultiplets in the $10$ representation. The S-dual frame in which we are interested is

\begin{displaymath}
 \includegraphics[width=189px]{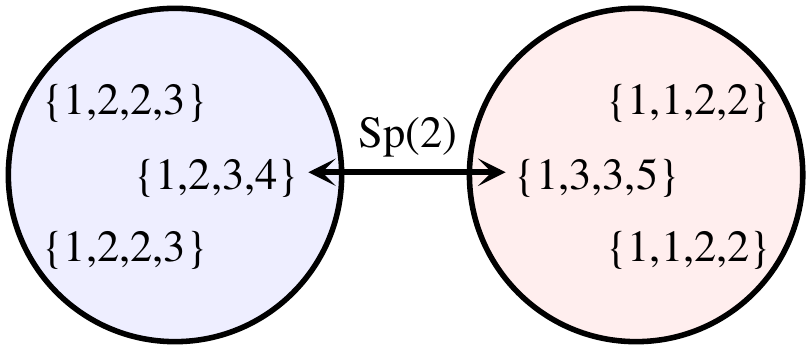}
\end{displaymath}
which is an $Sp(2)$ gauging of the $SO(14)\times U(1)$ theory with 1 fundamental hypermultiplet. The $SO(14)\times U(1)$ theory is part of an infinite series of interacting SCFTs we call $R_{2,N}$, for $N$ odd, with global symmetry group

\begin{displaymath}
G_{\text{global}}= SO(2N+4)_{k=2N}\times U(1).
\end{displaymath}
For $N=3$, the $SO(10)_{6}\times U(1)$ is enhanced to ${(E_6)}_{6}$, and we identify $R_{2,3}\equiv T_3$.

For the $SU(3)\times SU(7)\times U(1)$ theory, we consider the Lagrangian theory

\begin{displaymath}
 \includegraphics[width=189px]{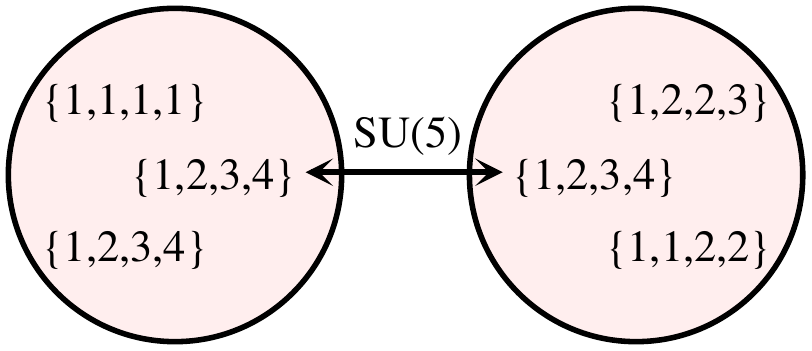}
\end{displaymath}
which is a $SU(5)$ gauge theory with 7 fundamental hypermultiplets and 1 hypermultiplet in the $10$ of $SU(5)$. The S-dual frame in which we are interested is

\begin{displaymath}
 \includegraphics[width=189px]{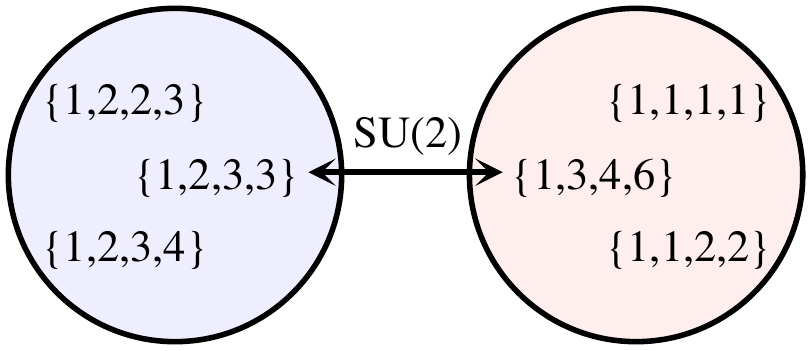}
\end{displaymath}
which is an $SU(2)$ gauging of the $SU(3)\times SU(7)\times U(1)$ SCFT.

As discussed in \S\ref{series}, this theory, too, is part of an infinite series of interacting SCFTs, $R_{1,N}$, for odd $N$, with global symmetry group

\begin{displaymath}
G_{\text{global}} = SU(3)_{k=8}\times {SU(N+2)}_{k=2N}\times U(1).
\end{displaymath}
For the $SU(5)^2\times SU(2) \times U(1)$ theory, consider the Lagrangian theory

\begin{displaymath}
 \includegraphics[width=359px]{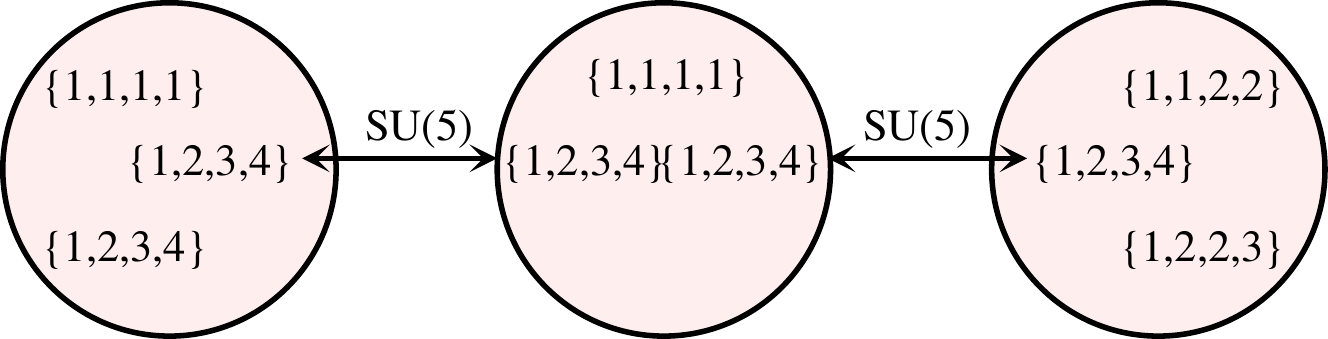}
\end{displaymath}
which is an $SU(5)\times SU(5)$ gauge theory with matter in the $5(5,1)+(5,5)+2(1,5)+(1,10)$. The S-dual frame in which we are interested is

\begin{displaymath}
 \includegraphics[width=359px]{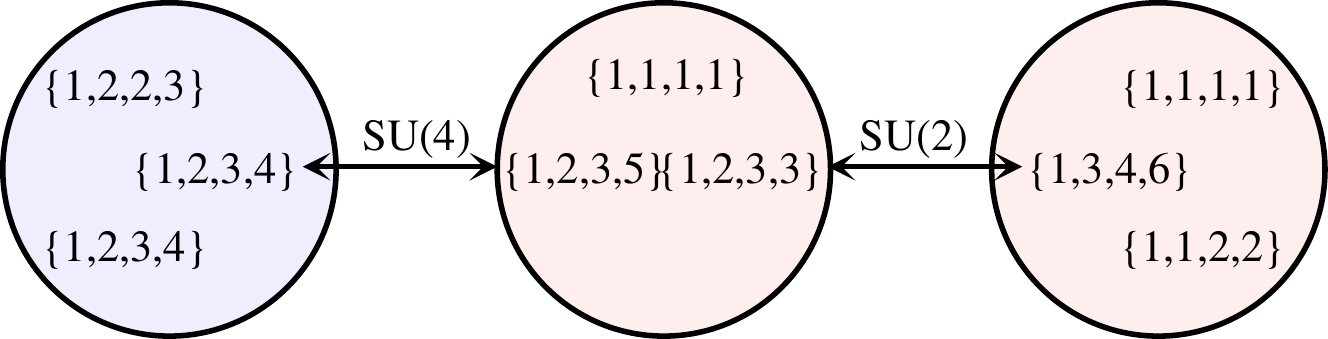}
\end{displaymath}
which is an $SU(4)$ gauging of the $SU(5)^2\times SU(2)\times U(1)$ SCFT coupled to a $SU(2)$ gauge theory with matter in the $(4,2)$ of $SU(4)\times SU(2)$.

For the $SU(6)\times SU(3)^2\times U(1)$ theory, consider the following Lagrangian theory

\begin{displaymath}
 \includegraphics[width=359px]{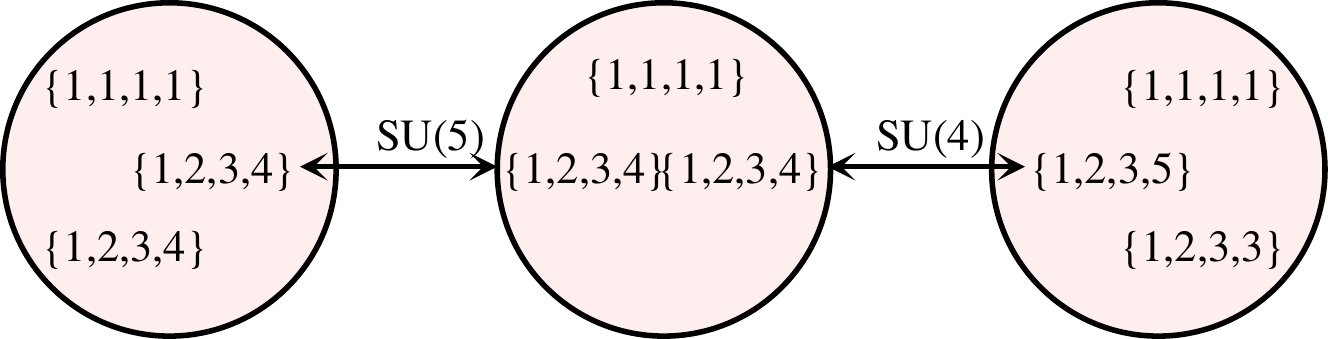}
\end{displaymath}
which is an $SU(5)\times SU(4)$ gauge theory with matter in the $6(5,1)+(5,4)+3(1,4)$ representation of $SU(5)\times SU(4)$. The S-dual frame in which we are interested is

\begin{displaymath}
 \includegraphics[width=359px]{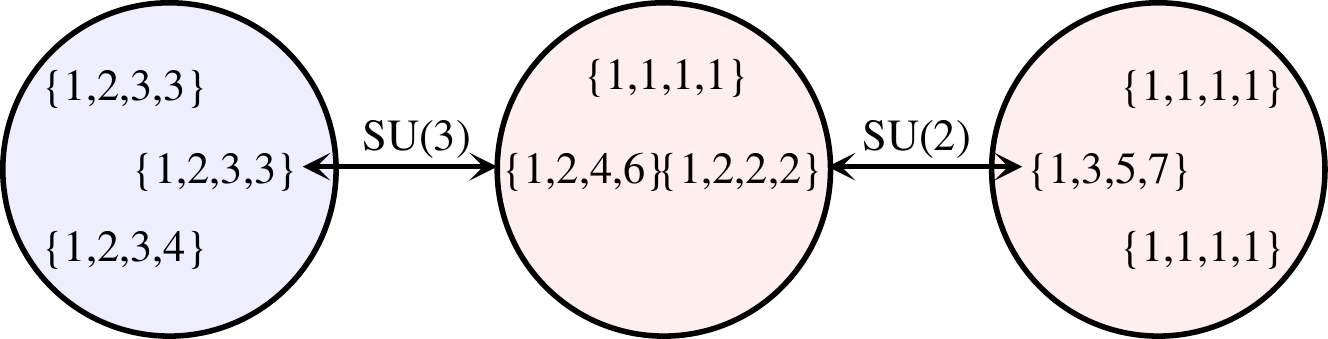}
\end{displaymath}
which is an $SU(3)$ gauging of the $SU(6)\times SU(3)^2\times U(1)$ SCFT coupled to a $SU(2)$ gauge theory with matter in the $(3,2)+(1,2)$ representation of $SU(3)\times SU(2)$.

For the $SU(5)^2\times SU(3)\times U(1)$ theory, we consider the following Lagrangian theory,

\begin{displaymath}
 \includegraphics[width=359px]{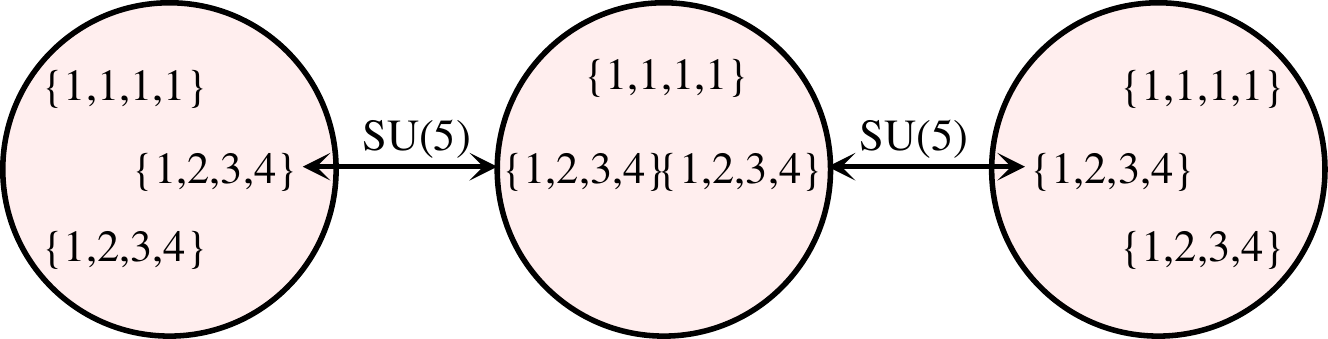}
\end{displaymath}
which is an $SU(5)\times SU(5)$ gauge theory with matter in the $5(5,1)+(5,5)+5(1,5)$ representation. The S-dual frame in which we are interested is

\begin{displaymath}
 \includegraphics[width=359px]{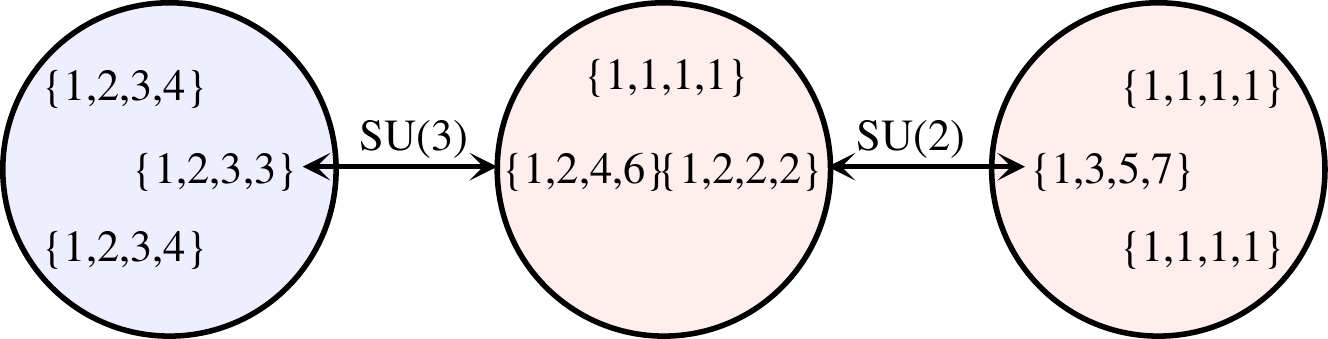}
\end{displaymath}
which is an $SU(3)$ gauging of the $SU(5)^2\times SU(3)\times U(1)$ SCFT coupled to a $SU(2)$ gauge theory with matter in the $(3,2)+(1,2)$ representation of $SU(3)\times SU(2)$. This interacting fixture is, again, the first of an infinite series we call $U_N$.

\hypertarget{spheres}{}\subsubsection*{{4-Punctured Spheres}}\label{spheres}

As a concrete test that our enumeration of fixtures and cylinders, in the $A_4$ theory, didn'{}t miss anything, we decided to systematically study \emph{all} 4-punctured spheres --{} that is, all theories with a single gauge group factor --{} which arise from the $A_4$ theory. There are 90 such spheres, consisting of 4 regular punctures and a positive (graded) dimensional Coulomb branch.

\begin{itemize}%
\item Three are spheres with 4 identical punctures.
\item Twenty-one are spheres with 3 identical punctures.

\end{itemize}
In each of these cases, the gauge theory is self-dual, and so does not yield much of an interesting check on our predictions.

\begin{itemize}%
\item Fifty-four are spheres with two identical punctures. These lead to \emph{pairs} of distinct gauge theories, which are related by S-duality.
\item Twelve are spheres with four distinct punctures. These lead to \emph{triples} of distinct gauge theories, related by S-duality.

\end{itemize}
We have checked that our rules reproduce the correct global symmetry groups, Coulomb branch dimension and conformal anomaly coefficients for all 66 theories. Since each fixture, and each cylinder appears multiple times among the 144 distinct degenerations, this provides a powerful check on our methods. We give a brief summary of the results in the \hyperlink{appendix}{Appendix}.

\hypertarget{mirrors}{}\section{{3D Mirrors}}\label{mirrors}

To bolster our identification of the global symmetry groups of the interacting SCFTs that we have found, we will use an approach described by Benini, Tachikawa and Xie \cite{Benini:2010uu}.

They compactify from four down to three dimension, and construct the mirror of the 3D SCFT. The 3D mirror of the $A_{N-1}$ theory on an $n$-punctured sphere ($\times S^1$) is a star-shaped quiver gauge theory, with $n$ arms, whose central node is $U(N)$. We will be interested in the case $n=3$. The other $U(k)$ gauge groups, in each arm of the quiver, are dictated by the Young diagram associated to the puncture. Starting at the central node, we reduce the rank of each successive $U(k)$ gauge group by the \emph{height} of each successive column of the Young diagram. Since all of the matter is in bifundamental hypermultiplets, the mirror gauge group is $\left(\prod_i U(k_i)\right)/{U(1)}_{\text{diag}}$.

Having constructed the quiver, Gaiotto and Witten \cite{Gaiotto:2008ak} tell you how to extract the global symmetry group (by construction, all of our quivers are ``{}good quivers''{}, in the sense of Gaiotto and Witten):

\begin{itemize}%
\item Mark each ``{}balanced''{} node of the quiver (one for which $\sum k_i$ for the adjacent nodes is equal to $2k$).
\item If all of the nodes of the quiver are balanced, remove one of the $U(1)$ nodes (since we are modding out by the diagonal $U(1)$.
\item The marked nodes form the Dynkin diagram of the semi-simple part of $G_{\text{global}}$. The abelian part is $U(1)^{p-1}$, where $p$ is the number of unmarked nodes.

\end{itemize}
For the $A_2$ theory, there'{}s just one interacting SCFT, and the quiver corresponding to its 3D mirror has the shape of the $E_6$ extended Dynkin diagram.

\begin{displaymath}
\begin{matrix} \includegraphics[width=75px]{fig82}\end{matrix}
\qquad\xRightarrow{\qquad}\qquad
\begin{matrix} \includegraphics[width=89px]{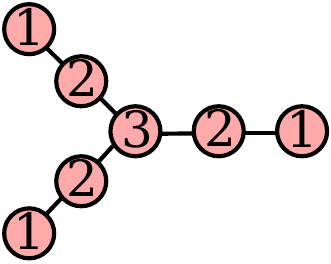}\end{matrix}
\end{displaymath}
After modding out by the diagonal $U(1)$, we reproduce the global symmetry group, $E_6$.

In the $A_3$ theory, there are three ``{}new''{} interacting SCFTs.

The first has a mirror quiver in the shape of the extended Dynkin diagram of $E_7$.

\begin{displaymath}
\begin{matrix} \includegraphics[width=75px]{fig15}\end{matrix}
\qquad\xRightarrow{\qquad}\qquad
\begin{matrix} \includegraphics[width=125px]{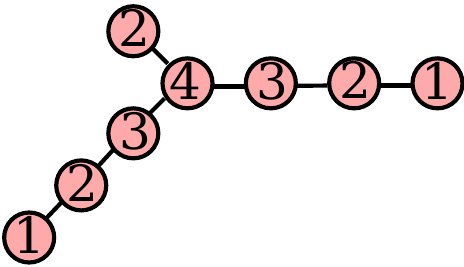}\end{matrix}
\end{displaymath}
After modding out by the diagonal $U(1)$, this yields the flavour symmetry $E_7$.

In the 3D mirror of the second SCFT

\begin{displaymath}
\begin{matrix} \includegraphics[width=75px]{fig17}\end{matrix}
\qquad\xRightarrow{\qquad}\qquad
\begin{matrix} \includegraphics[width=125px]{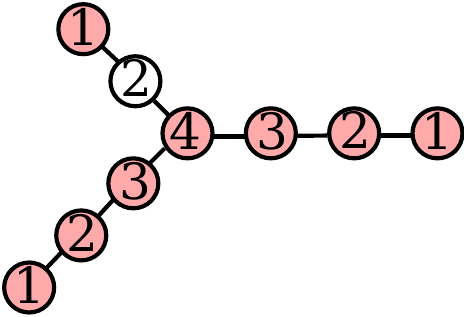}\end{matrix}
\end{displaymath}
not all the nodes of the quiver are superconformal. Modding out by the diagonal $U(1)$ kills one of the non-superconformal nodes (in this case, there'{}s only one), leaving $SU(2)\times SU(8)$ as the global symmetry group.

Finally, the $T_4$ theory has an $SU(4)^3$ global symmetry group.

\begin{displaymath}
\begin{matrix}
 \includegraphics[width=75px]{fig18}
\end{matrix}
\qquad\xRightarrow{\qquad}\qquad
\begin{matrix} \includegraphics[width=125px]{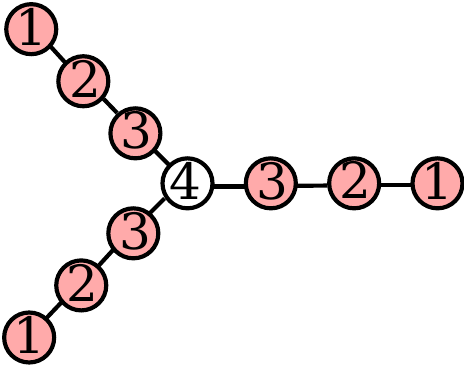}
\end{matrix}
\end{displaymath}
Turning to the $A_4$ theory, there are 8 new interacting SCFTs which arise. The 3D dual theories each have $l\gt 0$ nodes of the quiver which are non-superconformal. Modding out by the diagonal $U(1)$ yields a ${U(1)}^{l-1}$ factor in the global symmetry group.

\begin{center}
\begin{tabular}{|c|c|l|}
\hline
SCFT&3D Mirror&$G_k$\\
\hline 
$\begin{matrix} \includegraphics[width=91px]{fig55}\end{matrix}$&$\begin{matrix} \includegraphics[width=162px]{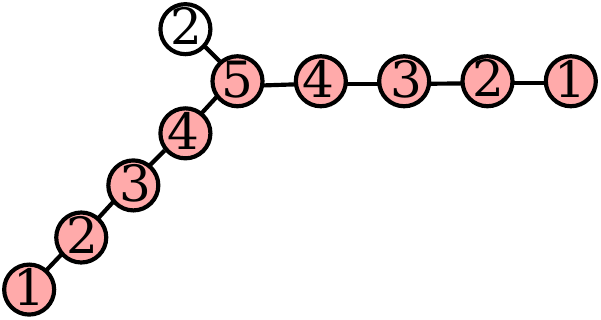}\end{matrix}$&${SU(10)}_{10}$\\
\hline
$\begin{matrix} \includegraphics[width=91px]{fig59}\end{matrix}$&$\begin{matrix} \includegraphics[width=161px]{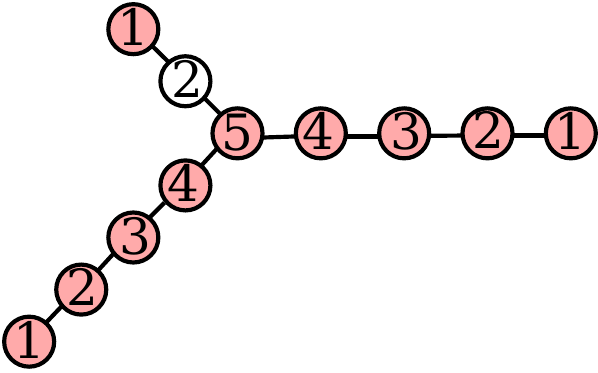}\end{matrix}$&${SU(2)}_6\times {SU(10)}_{10}$\\
\hline
\end{tabular}
\end{center}

\begin{center}
\begin{tabular}{|c|c|l|}
\hline
SCFT&3D Mirror&$G_k$\\
\hline
$\begin{matrix} \includegraphics[width=91px]{fig83}\end{matrix}$&$\begin{matrix} \includegraphics[width=134px]{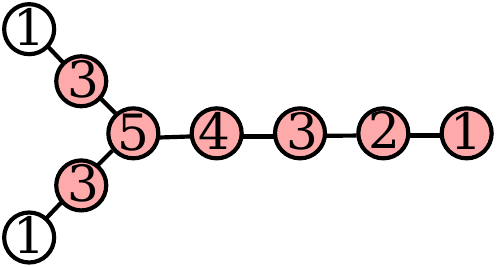}\end{matrix}$&${SO(14)}_{10}\times U(1)$\\
\hline
$\begin{matrix} \includegraphics[width=91px]{fig62}\end{matrix}$&$\begin{matrix} \includegraphics[width=148px]{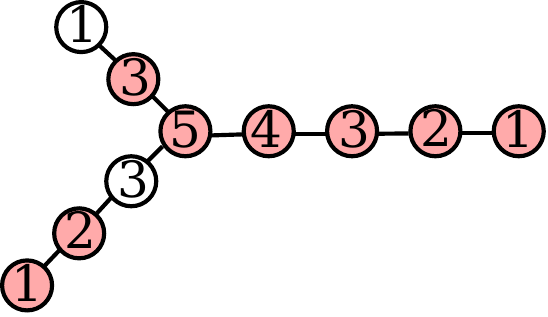}\end{matrix}$&${SU(3)}_8\times{SU(7)}_{10}\times U(1)$\\
\hline
$\begin{matrix} \includegraphics[width=91px]{fig63}\end{matrix}$&$\begin{matrix} \includegraphics[width=161px]{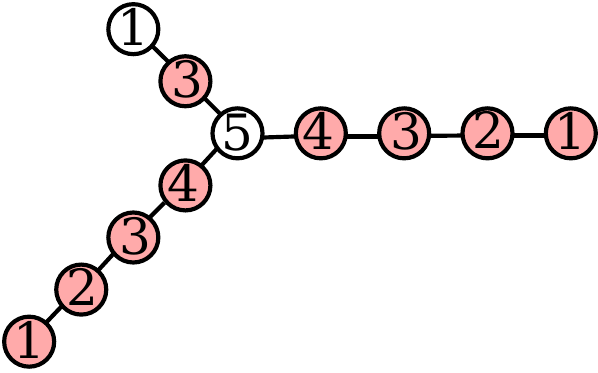}\end{matrix}$&${SU(5)}_{10}^2\times {SU(2)}_{10}\times U(1)$\\
\hline
$\begin{matrix} \includegraphics[width=91px]{fig65}\end{matrix}$&$\begin{matrix} \includegraphics[width=147px]{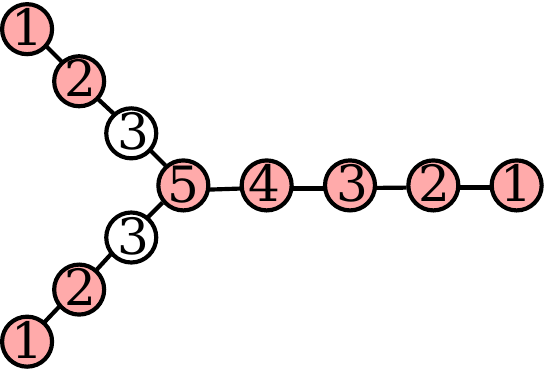}\end{matrix}$&${SU(6)}_{10}\times{SU(3)}_{8}^2\times U(1)$\\
\hline
$\begin{matrix} \includegraphics[width=91px]{fig66}\end{matrix}$&$\begin{matrix} \includegraphics[width=161px]{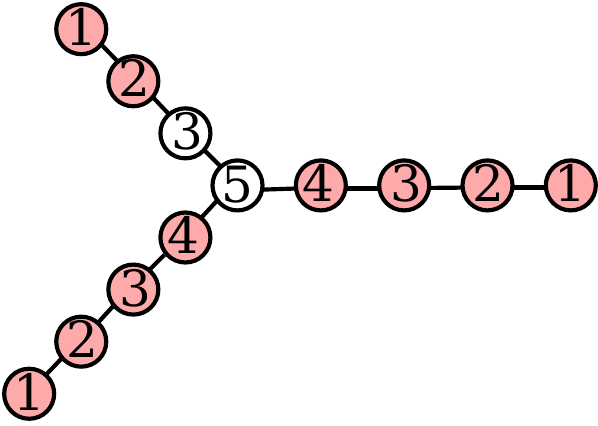}\end{matrix}$&${SU(5)}_{10}^2\times {SU(3)}_8\times U(1)$\\
\hline
\end{tabular}
\end{center}

\begin{center}
\begin{tabular}{|c|c|l|}
\hline
SCFT&3D Mirror&$G_k$\\
\hline
$\begin{matrix} \includegraphics[width=91px]{fig67}\end{matrix}$&$\begin{matrix} \includegraphics[width=161px]{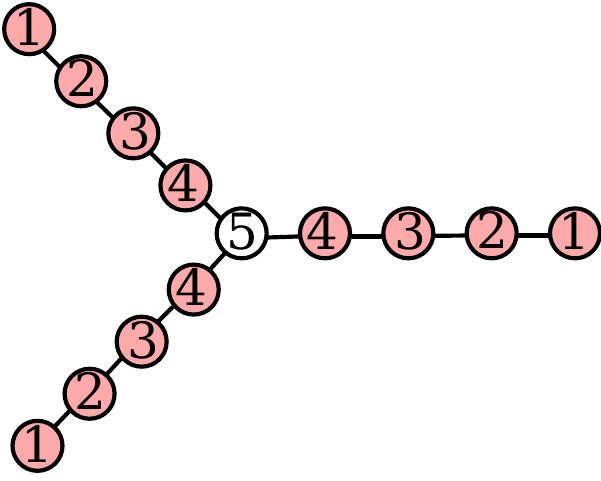}\end{matrix}$&${SU(5)}_{10}^3$\\
\hline
\end{tabular}
\end{center}

\hypertarget{series}{}\section{{Infinite Series}}\label{series}

\paragraph*{\includegraphics[width=136px]{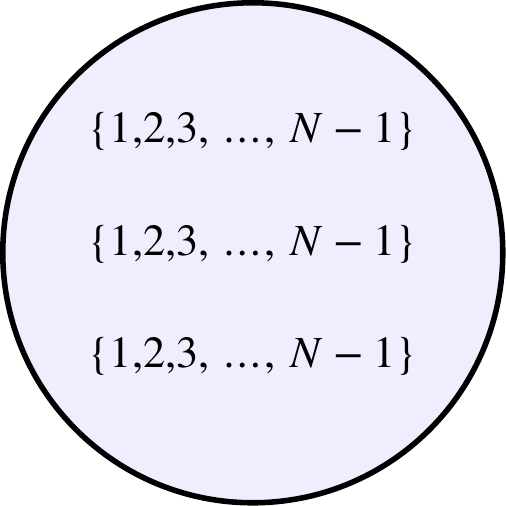}}

We are already familiar with the $T_N$ series of interacting SCFTs, introduced by Gaiotto, whose fixture consists of three maximal punctures. The global symmetry group is

\begin{displaymath}
G_{\text{global}}= {SU(N)}^3_{k=2N}.
\end{displaymath}
The graded dimension of the Coulomb branch is

\begin{displaymath}
(d_2,d_3,d_4,\dots,d_N)= (0,1,2,3,\dots,N-2),
\end{displaymath}
and conformal anomaly coefficients are

\begin{displaymath}
\begin{gathered}
a=\frac{N^3}{6}-\frac{5N^2}{16}-\frac{N}{16}+\frac{5}{24},\\
c=\frac{N^3}{6}-\frac{N^2}{4}-\frac{N}{12}+\frac{1}{6}.
\end{gathered}
\end{displaymath}
For $N=3$, $G_{\text{global}}$ is enhanced to ${E_6}_{k=6}$.

In our investigations, we have come across several new series of interacting SCFTs. Below, we will discuss seven of them.
\bigskip

\paragraph*{\includegraphics[width=129px]{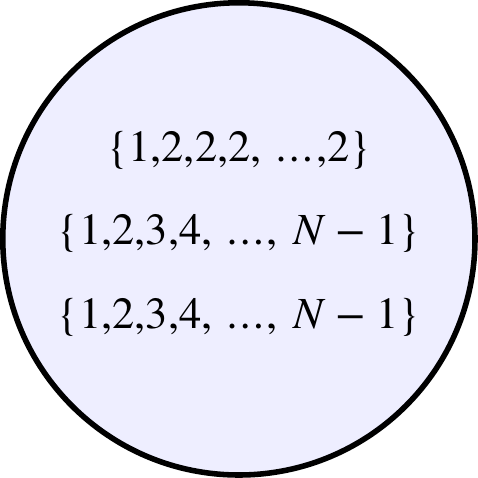}}

The $R_{0,N}$ series of interacting SCFTs has global symmetry

\begin{displaymath}
G_{\text{global}} = SU(2)_{k=6}\times SU(2N)_{k=2N},
\end{displaymath}
and has a Coulomb branch of graded dimension

\begin{displaymath}
(d_2,d_3,d_4,\dots,d_N)= (0,1,1,\dots,1).
\end{displaymath}
The strong coupling cusp of $SU(N)$, $N_f=2N$ gauge theory \cite{Argyres:1995wt, Argyres:2007cn} is S-dual to an $SU(2)$ gauging of the $SU(2)_{k=6}\subset G_{\text{global}}$ coupled to a fundamental hypermultiplet.

\begin{displaymath}
 \includegraphics[width=286px]{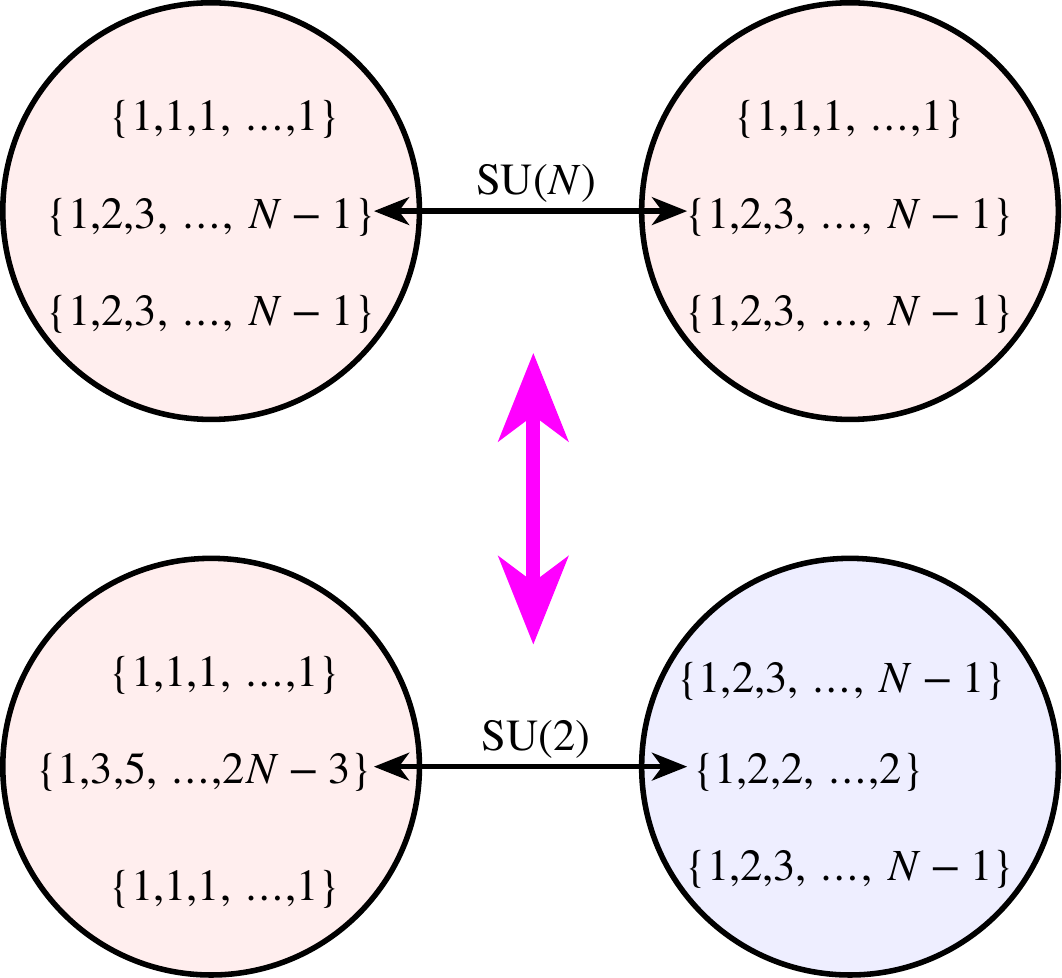}
\end{displaymath}
For $R_{0,3}\, (\equiv T_3)$, the ${SU(2)}_6\times {SU(6)}_6$ global symmetry is enhanced to $(E_6)_6$, and we get back the classic example of Argyres-Seiberg duality.) The conformal anomaly coefficients for the $R_{0,N}$ series are

\begin{displaymath}
\begin{gathered}
a = \frac{7N^2-22}{24},\\
c = \frac{2 N^2 -5}{6}.
\end{gathered}
\end{displaymath}

\paragraph*{\includegraphics[width=104px]{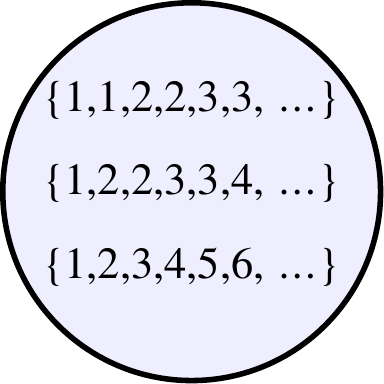}}

The fixture for the $R_{1,N}$ ($N\geq5$) series has one maximal puncture, and two other punctures, corresponding to Young diagrams of the form

\begin{displaymath}
\begin{matrix} \includegraphics[width=24px]{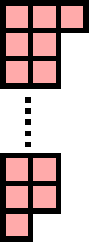}\end{matrix},\, 
\begin{matrix} \includegraphics[width=31px]{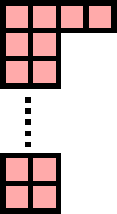}\end{matrix}\, \text{for}\, N\, \text{even, or}\,
\begin{matrix} \includegraphics[width=24px]{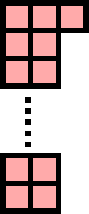}\end{matrix}, 
\begin{matrix} \includegraphics[width=31px]{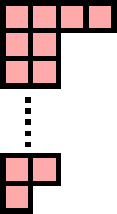}\end{matrix}\, \text{for}\, N\, \text{odd.}
\end{displaymath}
The Coulomb branch has graded dimension

\begin{displaymath}
(d_2,d_3,d_4,\dots,d_N)=(0,1,1,1,\dots,1),
\end{displaymath}
and the conformal anomaly coefficients are

\begin{displaymath}
\begin{gathered}
a=\frac{13N^2+3N-40}{48},\\
c=\frac{7N^2+3N-16}{24}.
\end{gathered}
\end{displaymath}
$R_{1,N}$ has global symmetry group

\begin{displaymath}
G_{\text{global}} = SU(2)_{k=8}\times {SU(N+2)}_{k=2N}\times {U(1)}^2
\end{displaymath}
(enhanced to $SU(3)_{k=8}\times SU(7)_{k=10}\times U(1)$ for $N=5$).

However, the realization differs slightly in the $N$ even versus $N$ odd cases. This is easily seen by examining the 3D mirrors

\begin{displaymath}
N\, \text{even}: \begin{matrix} \includegraphics[width=283px]{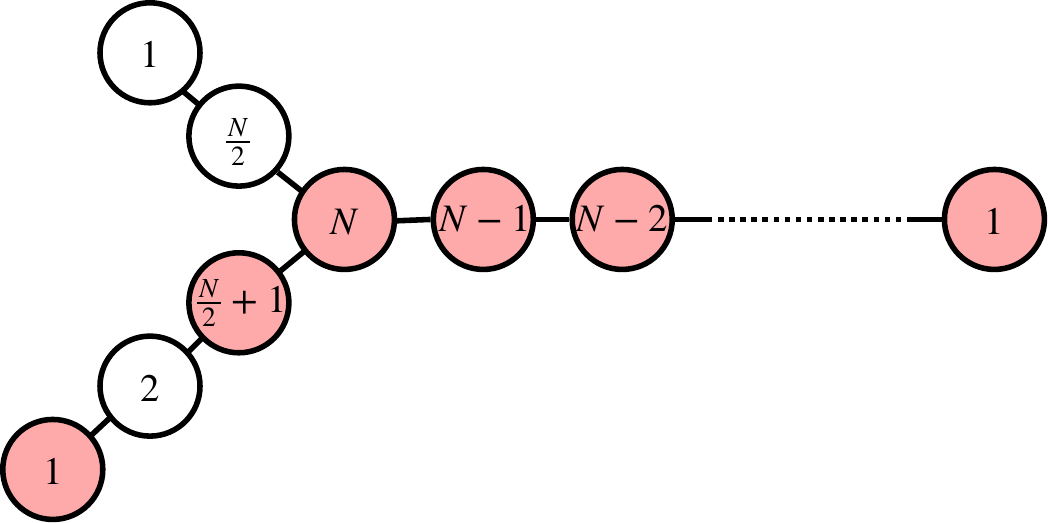}\end{matrix}
\end{displaymath}
\begin{displaymath}
N\, \text{odd}: \begin{matrix} \includegraphics[width=283px]{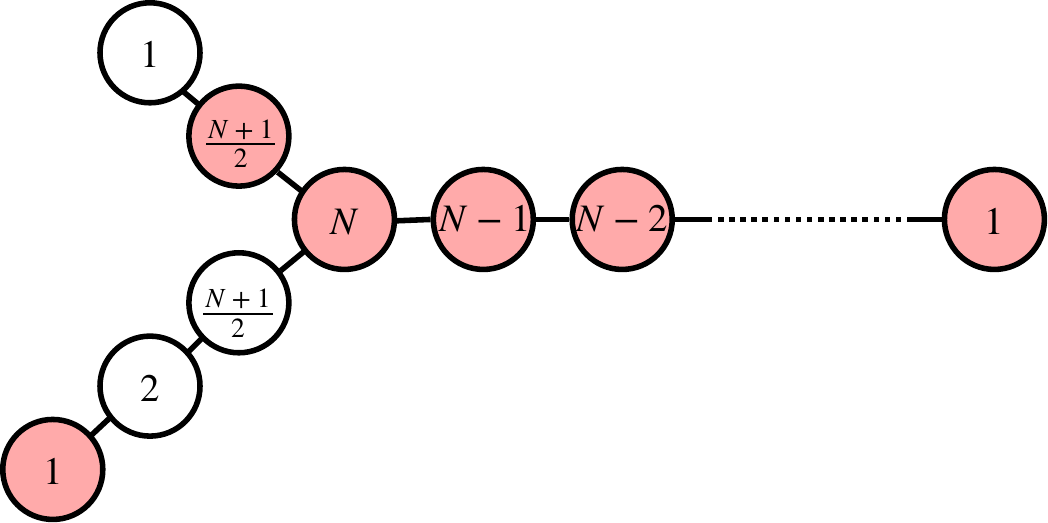}\end{matrix}
\end{displaymath}
From the 3D mirrors, one also readily sees the enhancement in $G_{\text{global}}$ for $N=5$.

(One of) the S-duals of $SU(N)$ with matter in the $(N+2)(\begin{matrix} \includegraphics[width=8px]{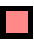}\end{matrix})+ \begin{matrix} \includegraphics[width=8px]{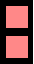}\end{matrix}$ is a gauging of the $SU(2)_8\subset G_{\text{global}}$ symmetry of $R_{1,N}$.

\begin{displaymath}
 \includegraphics[width=361px]{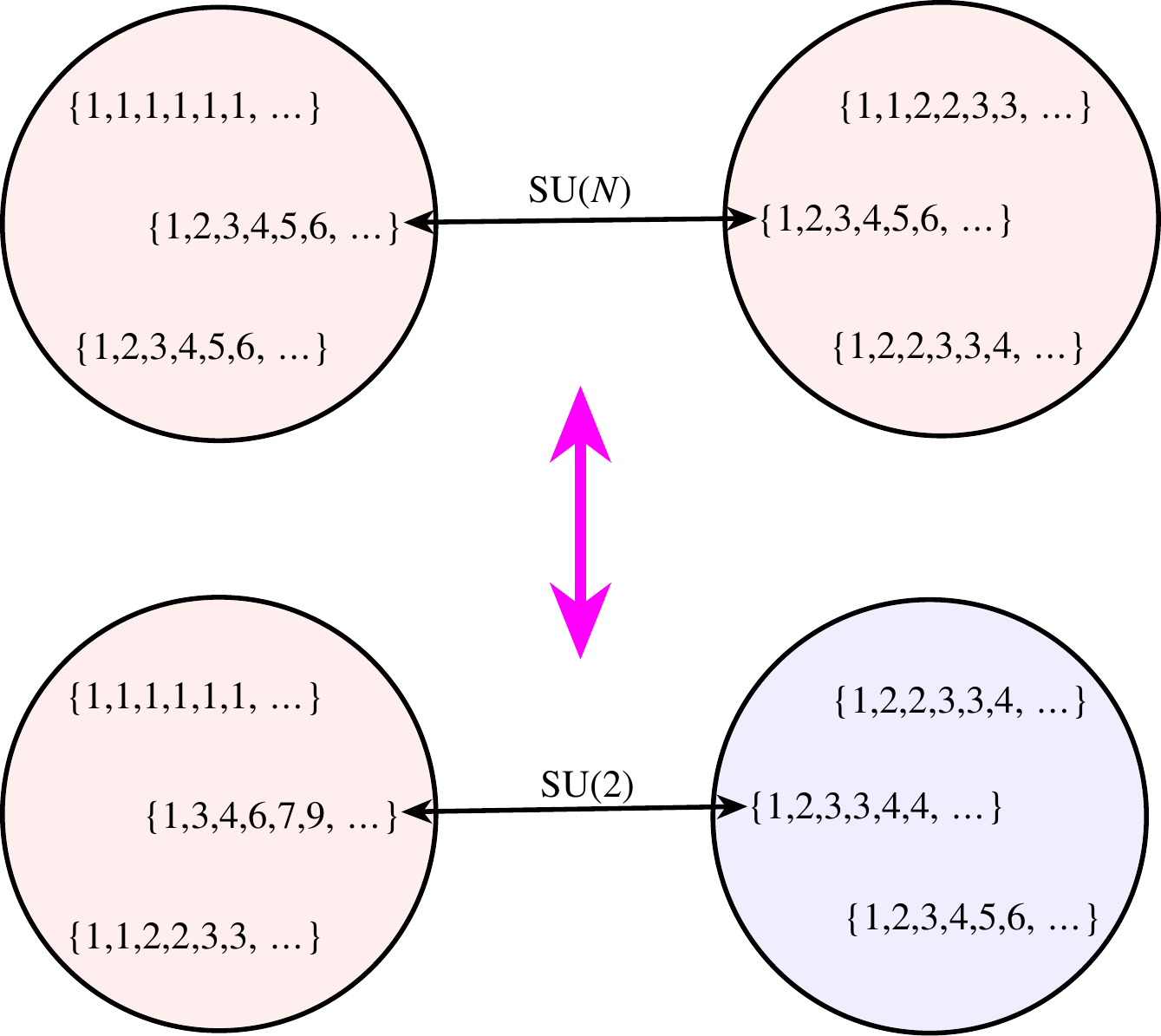}
\end{displaymath}
In the upper figure, the fixture on the left contributes $N$ fundamentals; the fixture on the right contributes 2 fundamental and one $\begin{matrix} \includegraphics[width=8px]{fig100}\end{matrix}$. In the lower figure, the fixture on the left contributes nothing; the fixture on the right is $R_{1,N}$.

Of course, the above 4-punctured sphere has another degeneration, which leads us to our fourth series of interacting SCFTs

\paragraph*{\includegraphics[width=121px]{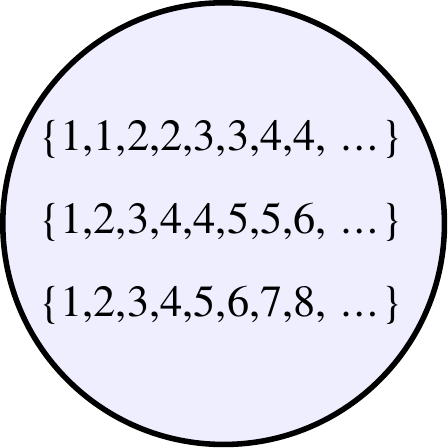}}

The $S_N$ series has global symmetry

\begin{displaymath}
G_{\text{global}}= SU(N+2)_{k=2N}\times SU(3)_{k=10}\times U(1)
\end{displaymath}
(enhanced to $SU(10)_{10}$, for $N=5$). Its Coulomb branch has graded dimension

\begin{displaymath}
(d_2,d_3,d_4,d_5,\dots) = (0,0,1,1,1,\dots,1).
\end{displaymath}
The conformal anomaly coefficients are

\begin{displaymath}
\begin{gathered}
a=\frac{13N^2+3N-96}{48},\\
c=\frac{7N^2+3N-42}{24}.
\end{gathered}
\end{displaymath}
The third S-duality frame of the $SU(N)$ gauge theory we have been discussing is

\begin{displaymath}
 \includegraphics[width=332px]{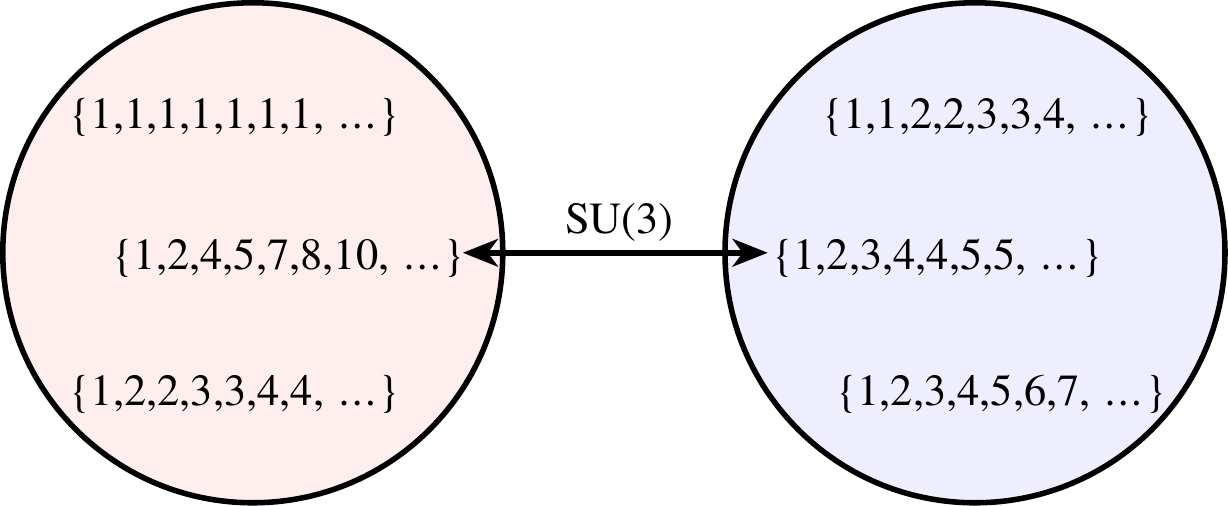}
\end{displaymath}
an $SU(3)$ gauging of the $S_N$ theory, coupled to a single fundamental hypermultiplet.

\paragraph*{\includegraphics[width=151px]{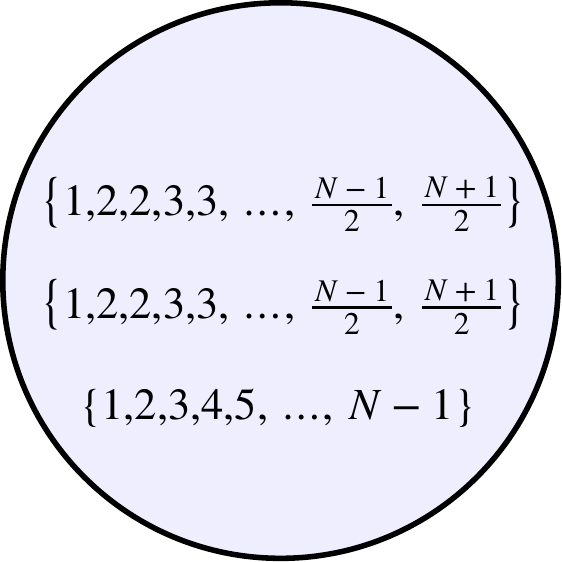}}

Next, we turn to the $R_{2,N}$ theory, which appears, for $N$ odd, as a fixture in the (unique) S-dual of $SU(N)$, with matter in the $4(\begin{matrix} \includegraphics[width=8px]{fig99}\end{matrix})+ 2\Bigl(\begin{matrix} \includegraphics[width=8px]{fig100}\end{matrix}\Bigr)$.

The global symmetry group of $R_{2,N}$ is

\begin{displaymath}
G_{\text{global}}={SO(2N+4)}_{k=2N}\times U(1)
\end{displaymath}
(enhanced to ${(E_6)}_6$ for $N=3$, where there is no distinction between a fundamental hypermultiplet and an antisymmetric tensor). The graded dimension of the Coulomb branch is

\begin{displaymath}
(d_2,d_3,d_4,d_5,d_6,\dots,d_N) = (0,1,0,1,0,\dots,1).
\end{displaymath}
The conformal anomaly coefficients for the $R_{2,N}$ series are

\begin{displaymath}
\begin{gathered}
a = \frac{7N^2+9N-8}{48},\\
c = \frac{2N^2+3N-1}{12}.
\end{gathered}
\end{displaymath}
The strong coupling S-dual of $SU(N)$ ($N$ odd), with matter in the $4(\begin{matrix} \includegraphics[width=8px]{fig99}\end{matrix})+ 2\Bigl(\begin{matrix} \includegraphics[width=8px]{fig100}\end{matrix}\Bigr)$ is an $Sp\left(\tfrac{N-1}{2}\right)$ gauge theory coupled to one fundamental hypermultiplet and gauging an $Sp\left(\tfrac{N-1}{2}\right)\subset SO(2N+4)_{2N}$ of the $R_{2,N}$ theory.

\begin{displaymath}
 \includegraphics[width=402px]{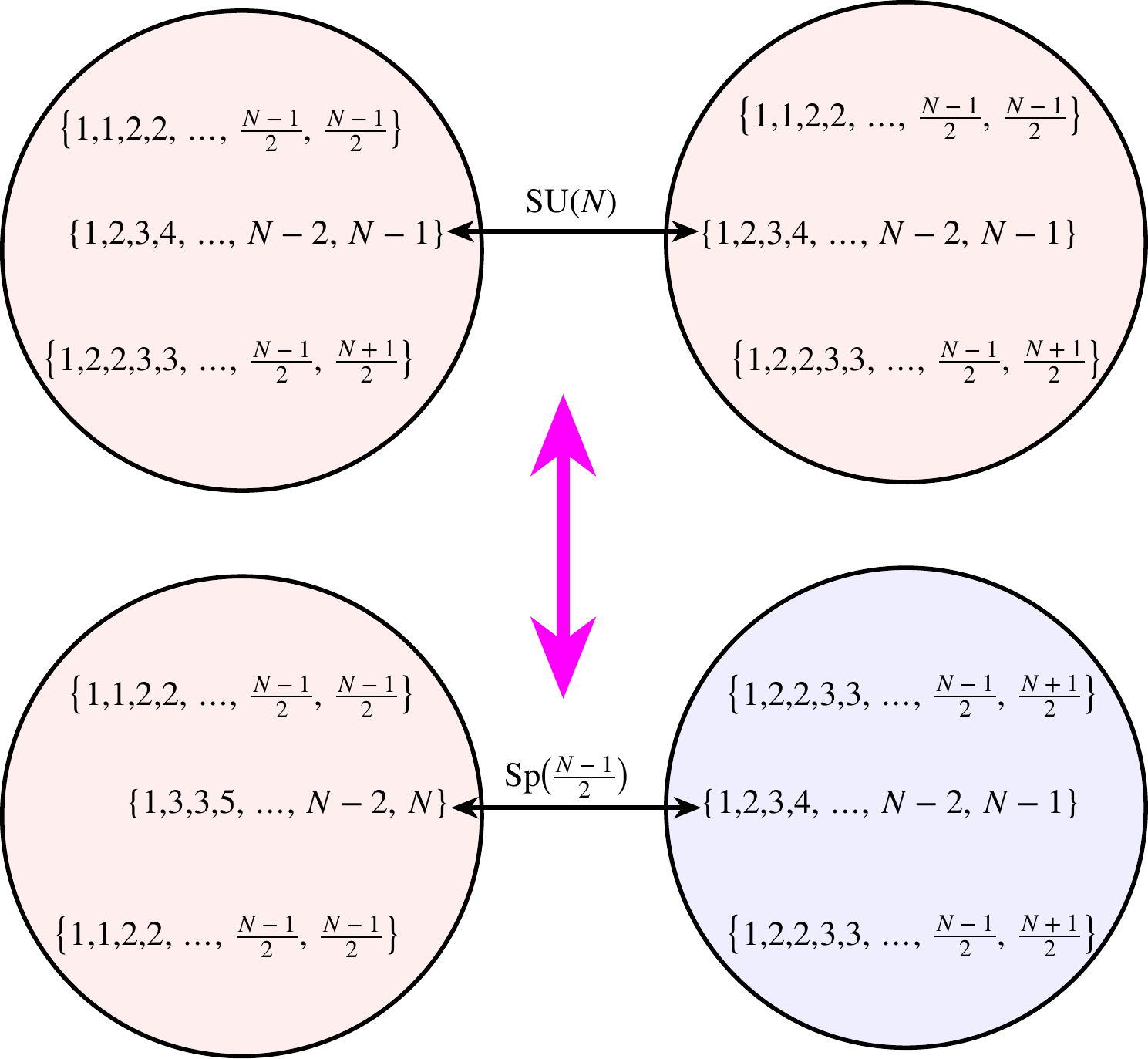}
\end{displaymath}
For $N$ even, the S-duality of $SU(N)$, with matter in the $4(\begin{matrix} \includegraphics[width=8px]{fig99}\end{matrix})+ 2\Bigl(\begin{matrix} \includegraphics[width=8px]{fig100}\end{matrix}\Bigr)$, looks almost identical to the picture above. The S-dual gauge group is $Sp(N/2)$. The fixture on the left contributes $2N$ hypermultiplets, transforming as 2 fundamentals of $Sp(N/2)$ (instead of $N-1$ hypermultiplets, transforming as one fundamental of $Sp((N-1)/2)$, as it did, for $N$ odd). The fixture on the right is $R_{2,N-1}$ plus $N$ hypermultiplets (which contribute another fundamental of $Sp(N/2)$).

All together, the S-dual of $SU(N)$ ($N$ even), with matter in the $4(\begin{matrix} \includegraphics[width=8px]{fig99}\end{matrix})+ 2\Bigl(\begin{matrix} \includegraphics[width=8px]{fig100}\end{matrix}\Bigr)$, is $Sp(N/2)$ with 3 hypermultiplets in the fundamental, gauging the $R_{2,N-1}$ theory.

The fixture

\begin{displaymath}
 \includegraphics[width=104px]{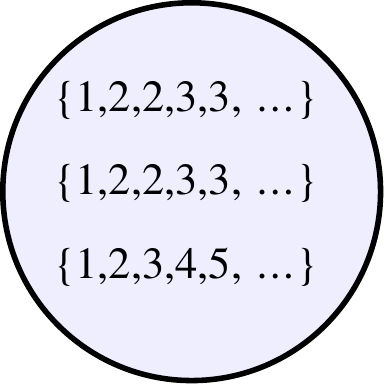}
\end{displaymath}
is $R_{2,N}$, for $N$ odd, and $R_{2,N-1}$ plus $N$ hypermultiplets, for $N$ even.

\paragraph*{\includegraphics[width=114px]{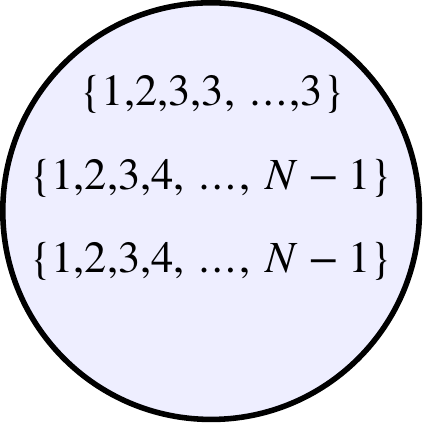}}

The $U_N$ series has global symmetry

\begin{displaymath}
G_{\text{global}}= {SU(N)}^2_{k=2N}\times {SU(3)}_{k=8}\times {U(1)}
\end{displaymath}
(enhanced to ${SU(4)}^3_{8}$ for $S_4\equiv T_4$). The Coulomb branch has graded dimension

\begin{displaymath}
(d_2,d_3,d_4,d_5,\dots)=(0,1,2,2,2,\dots,2).
\end{displaymath}
and the conformal anomaly coefficients are

\begin{displaymath}
\begin{gathered}
a=\frac{13 N^2-73}{24},\\
c=\frac{7N^2-34}{12}.
\end{gathered}
\end{displaymath}
Consider an ${SU(N)}^2$ gauge theory, with matter in the $N(N,1)+(N,N)+N(1,N)$. One S-dual frame is, of course, an $SU(2)\times SU(N)$ gauge theory, with matter in the $(2,1)+(1,N)$, gauging an $SU(2)\times SU(N)\subset SU(2)\times SU(2N)_{2N}$ of the $R_{0,N}$ theory. The other S-dual frame is an $SU(2)\times SU(3)$ gauge theory, with matter in the $(2,1)+(2,3)$, where the $SU(3)$ gauges the ${SU(3)}_8\subset G_{\text{global}}$ of $U_N$. 

So far, our infinite series have been fixtures which appear in S-dual descriptions of Lagrangian field theories. In light of recent progress, this seems like a quaint restriction.

Let us turn to a pair of infinite series of interacting SCFT fixtures, consisting of a pair of maximal punctures plus a puncture whose Young diagram's first column has a height that grows like $N$.

\begin{displaymath}
V_N = \begin{matrix} \includegraphics[width=114px]{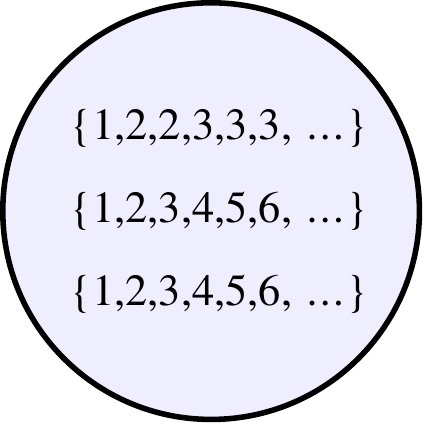}\end{matrix},\qquad
  W_N = \begin{matrix} \includegraphics[width=114px]{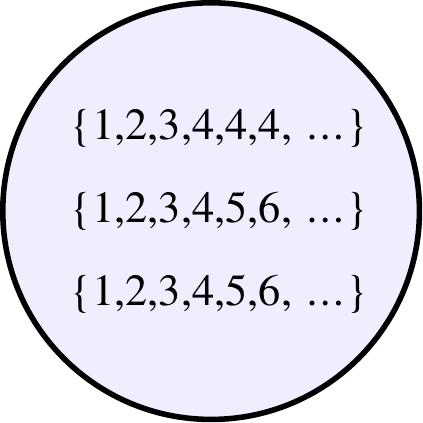}\end{matrix}.
\end{displaymath}
The Coulomb branch of $V_N$ has graded dimension

\begin{displaymath}
(d_2,d_3,d_4,d_5,d_6,d_7,\dots)= (0,1,1,2,2,2,\dots,2).
\end{displaymath}
From the 3D mirror, we find its global symmetry group to be

\begin{displaymath}
G_{\text{global}} = {SU(N)}_{k=2N}^2\times {U(1)}^2
\end{displaymath}
(enhanced to ${SU(5)}_{10}^2\times {SU(2)}_{10}\times U(1)$ for $N=5$). It has $n_v=2N^2-20$, and $n_h= 3N^2-17$, or

\begin{displaymath}
\begin{gathered}
a=\frac{13(N^2-9)}{24},\\
c=\frac{7N^2-57}{12}.
\end{gathered}
\end{displaymath}
The Coulomb branch of $W_N$ has graded dimension

\begin{displaymath}
(d_2,d_3,d_4,d_5,d_6,d_7,\dots)= (0,1,2,3,3,3,\dots,3).
\end{displaymath}
Its global symmetry group is

\begin{displaymath}
G_{\text{global}} = {SU(N)}_{k=2N}^2\times {SU(4)}_{10}\times U(1)
\end{displaymath}
(enhanced to ${SU(5)}_{10}^3$ for $N=5$). It has $n_v=3N^2-29$, and $n_h= 4N^2-20$, or

\begin{displaymath}
\begin{gathered}
a=\frac{19N^2-174}{24},\\
c=\frac{10N^2-87}{12}.
\end{gathered}
\end{displaymath}
Using these interacting fixtures, we construct a family of S-dual theories

\begin{displaymath}
\begin{matrix} \includegraphics[width=301px]{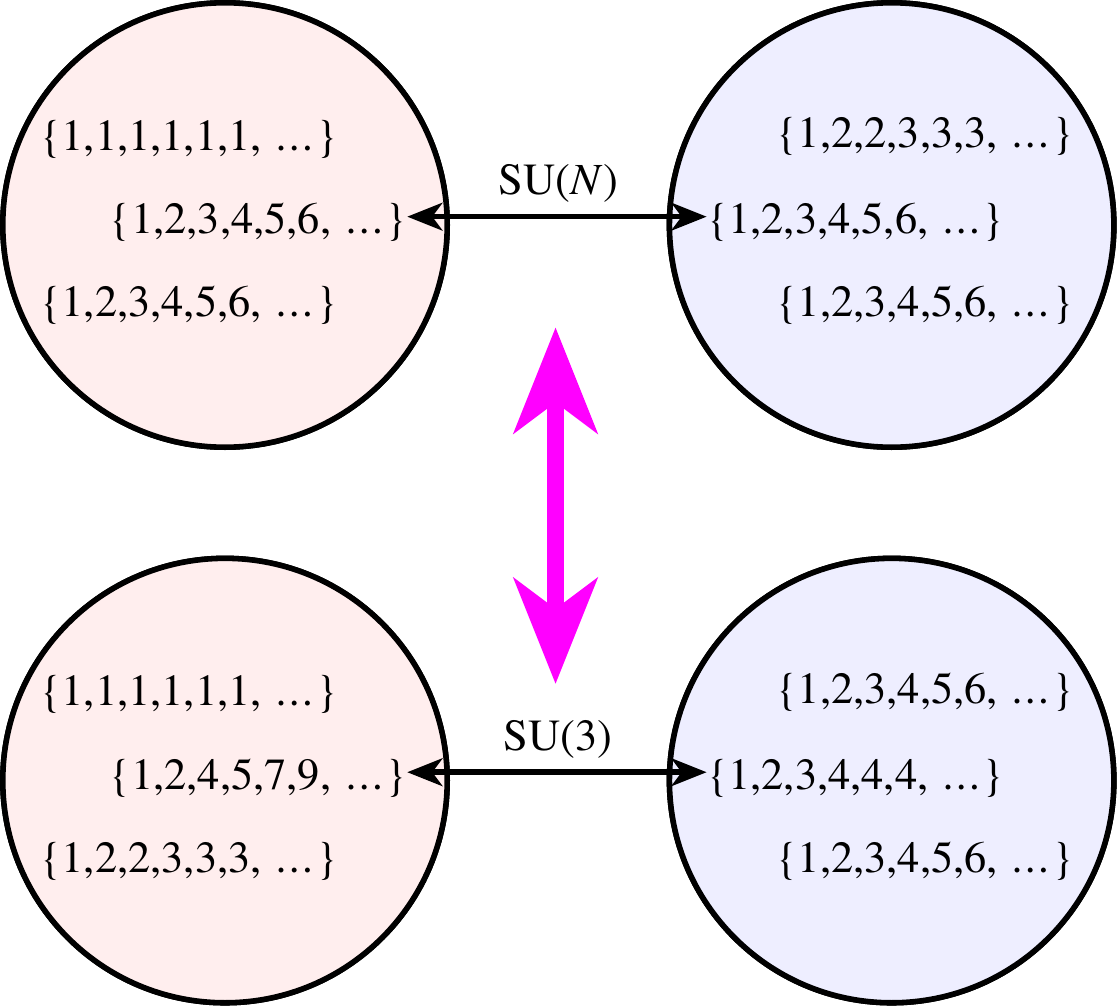}\end{matrix}
\end{displaymath}
The upper theory is an $SU(N)$ gauge theory, with $N$ fundamentals, coupled to $V_N$. The lower theory is an $SU(3)$ gauge theory, with one fundamental, coupled to $W_N$.

Of course, there are an infinite number of arbitrary-$N$ families of Young diagrams, that one can write down, and from there, an infinite number of arbitrary-$N$ families of interacting fixtures. The ones discussed here were those which cropped up in the theories up through $N=5$, and which gave rise to interesting series of S-dualities.

\hypertarget{irregulartheories}{}\section{{Theories with irregular punctures}}\label{irregulartheories}

Having introduced 3-punctured spheres with irregular punctures, we should ask whether ---{} according to our rules ---{} it is possible to construct \emph{connected} curves, $C$, with $g\gt 0$ and/or $n\gt 3$, containing one or more irregular punctures.

It would be most dangerous if we could construct connected surfaces with \emph{two} or more irregular puncture, as we would then have to specify what happens when two irregular punctures collide, and that would take us outside the set of configurations we have allowed.

It is easy to see, however, that this complication does not arise. At least up through $A_4$, we can exhaustively list all the connected surfaces, constructed according to our rules, with one or more irregular punctures. There are a \emph{finite} in number, and contain just one irregular puncture. All have $g=0$.

More generally, we can argue as follows. Assume there exists a connected surface, $C$, with two irregular punctures.

\begin{itemize}%
\item One of the implications of our rules for constructing surfaces is that, for any $k$, if $C$ had $d_k\gt 0$, then, for that value of $k$, $p^{(i)}_k\leq k-1,\, \forall i$.
\item On the other hand, an irregular puncture, by definition, has $p_k\geq k-1,\, \forall k$ and $\gt k-1$ for at least some $k$. Pick one such value of $k$.
\item We demand $0\equiv d_k = -(1-g) (2k-1) + \sum_{i=1}^n p^{(i)}_k$. The second term is manifestly positive, and the two irregular punctures make a contribution $\geq 2k-1$. The only way to satisfy the equality is to set $g=0$, with no other punctures.
\item But, for $g=0$, we must have $n\geq 3$ (otherwise, the virtual dimension $d_2$ is negative).

\end{itemize}
Thus, we reach a contradiction: there can be no connected curves, $C$, with two (or more) irregular punctures.

It remains to list the finite number of $A_{N-1}$ ($N\leq 5$) theories with $g=0$, a single irregular puncture and $n\gt 3$. In the $A_2$ theory, there is only the 3-punctured sphere, listed above. Starting with $A_3$, however, we find a 4-punctured sphere

\begin{displaymath}
 \includegraphics[width=106px]{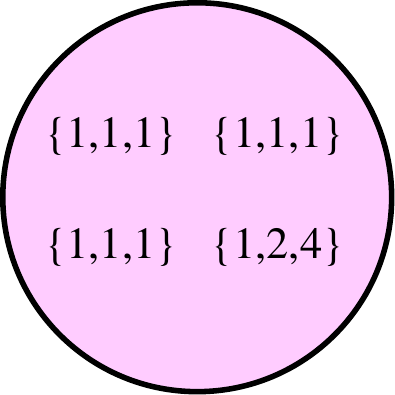}
\end{displaymath}
This is the $SU(2)$, $N_f=4$ theory, as it arises in the $A_3$ theory.

For $A_4$, we find three 4-punctured spheres,

\begin{displaymath}
 \includegraphics[width=384px]{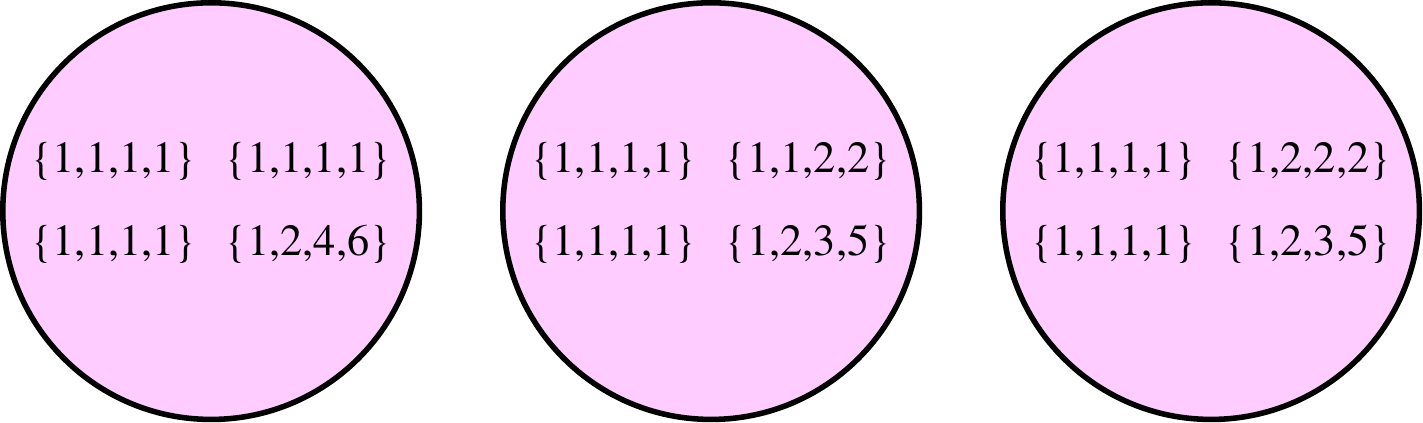}
\end{displaymath}
The first is, again, the $A_4$ expression of the $SU(2)$ $N_f=4$ theory. The second is the $SU(2)$ $N_f=4$ theory plus 4 free hypers. The third is the $SU(3)$ $N_f=6$ theory (or its S-dual).

From the latter, we can construct a 5-punctured sphere

\begin{displaymath}
 \includegraphics[width=109px]{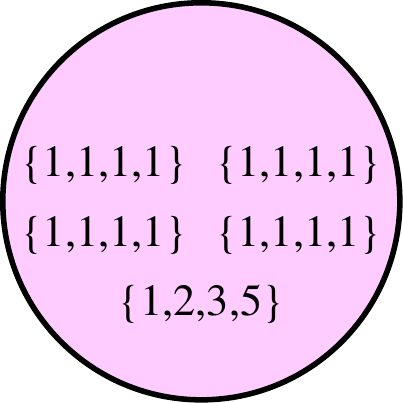}
\end{displaymath}
which is an $SU(2)\times SU(3)$ gauge theory, with matter in the $(2,1)+(2,3)+4(1,3)$.

\hypertarget{conclusions}{}\section{{Conclusions}}\label{conclusions}

In this paper, we have embarked on a systematic classification of the $\mathcal{N}=2$ superconformal field theories which result from compactifying the $A_{N-1}$ (2,0) SCFT on a Riemann surface with punctures.

We gave systematic rules for cataloging the fixtures (3-punctured spheres) and gauge groups (cylinders) that appear, and formul\ae\ for (graded) Coulomb branch dimension, the effective number of vectors multiplets \eqref{nvdef} and hypermultiplets \eqref{nhdef} (equivalently, for the conformal anomaly coefficients, $(a,c)$).

When the Coulomb branch is zero-dimensional, $n_v=0$, and the fixture consists of $n_h$ free hypermultiplets. Otherwise, the fixture is either an intrinsically-interacting SCFT, or a combination of free hypermultiplets and an interacting SCFT (that has already appeared in our catalogue for smaller $N$). In the former case, 3D mirror symmetry gave an easy prescription for reading off the global symmetry group of the SCFT.

Applying these rules, we systematically catalogued \emph{all} of the fixtures and cylinders that appear up through $N=5$. For instance, for $N=4$, there were a total of 10 fixtures --- 6 corresponding to free hypermultiplets, 3 corresponding to interacting SCFTs and 1 corresponding to an interacting SCFT plus some free hypermultiplets --- and 6 cylinders. \emph{All} $\mathcal{N}=2$ superconformal theories which can be constructed by compactifying the $A_{3}$ (2,0) theory on $C_{g,n}$, for \emph{any} $g,n$, are constructed from these basic building blocks.

For $N=2$, there were a mere 3 fixtures and 2 cylinders, whereas, for $N=5$, there are 24 fixtures and 7 cylinders. Clearly, the number grows quite rapidly with $N$, so constructing a complete catalogue, for larger $N$, would be tedious. For larger $N$, we gave some examples of infinite series of interacting SCFTs, $R_{0,N}$, $R_{1,N}$, $S_N$, $T_N$, $U_N$, $V_N$ and $W_N$ (and $R_{2,N}$ for $N$ odd), and examples of the S-dualities that follow from them.

\begin{itemize}%
\item $SU(N)$, with $2N$ fundamentals, is S-dual to $SU(2)$ with one fundamental, gauging the $SU(2)_{k=6}\subset G_{\text{global}}$ of the $R_{0,N}$ theory.

\item $SU(N)$, with $N+2$ fundamentals and an antisymmetric tensor, has two strong-coupling S-duals.

\begin{itemize}%
\item One is the $SU(2)$ gauging of $R_{1,N}$.
\item The other is $SU(3)$, with one fundamental, gauging ${SU(3)}_{k=10}\subset G_{\text{global}}$ of the $S_N$ theory.

\end{itemize}

\item $SU(N)$ with 4 fundamentals and 2 antisymmetric tensors is S-dual to

\begin{itemize}%
\item $Sp\left(\tfrac{N-1}{2}\right)$ with one fundamental, gauging the $R_{2,N}$ theory, for $N$ odd.
\item $Sp\left(\tfrac{N}{2}\right)$ with \emph{three} fundamentals, gauging the $R_{2,N-1}$ theory, for $N$ even.

\end{itemize}

\item $SU(N)^2$, with matter in the $N(N,1)+(N,N)+N(1,N)$ has two strong-coupling S-dual descriptions
\begin{itemize}
\item $SU(2)\times SU(3)$ gauge theory, with matter in the $(2,1)+(2,3)$, with the $SU(3)$ gauging the ${SU(3)}_{k=8}\subset G_{\text{global}}$ of the $U_N$ theory.
\item $SU(2)\times SU(N)$ gauge theory, with matter in the $(2,1)+(1,N)$, gauging an $SU(2)\times SU(N)\subset {SU(2)}_6\times SU(2N)_{2N}$ of the $R_{0,N}$ theory.
\end{itemize}

\end{itemize}
The global symmetry groups and conformal anomaly coefficient are easily seen to agree. It would be interesting to compare the superconformal indices \cite{Gadde:2010te} of the interacting SCFTs, that we have found, to provide a more detailed test of some of these dualities. It would also be nice to interpret certain aspects of the recipe, we have presented, in the language of Liouville theory \cite{Alday:2009aq}.

Even more interesting, would be to extend these results to the $D$- and $E$-series of (2,0) theories (see \cite{Tachikawa:2009rb} for some preliminary results for the $D_N$ series). Work in that direction is in progress.

\section*{Acknowledgements}
The research of the authors is based upon work supported by the National Science Foundation under Grant No. PHY-0455649 and a grant from the United States-Israel Binational Science Foundation. J.~D.~ would like to thank the high energy theory group at Tel Aviv University for their hospitality when this work was initiated, and the Aspen Center for Physics and the Simons Workshop at SUNY Stonybrook, where the bulk of this work was completed. We have benefited trememendously from conversations with Andrew Neitzke and Yuji Tachikawa, as well as well from some useful remarks of David Ben-Zvi.

\vfill\eject
\appendix

\hypertarget{appendix}{}\section*{Appendix}\label{appendix}

Here we list all the sixty-six 4-punctured spheres in the compactification of the $A_4$ theory with more than one distinct S-dual presentation. For brevity, we list the pole structures of the punctures, the gauge groups which arise in each S-duality frame, the graded Coulomb branch dimension and the global symmetry group of each theory.

\begin{center}
Theories with two distinct S-duality frames
\end{center}

\begin{tabular}{|c|c|c|c|l|}
\hline
4-punctured Sphere&$G_{\text{gauge}}$&$G'_{\text{gauge}}$&$(d_2,d_3,d_4,d_5)$&$G_{\text{global}}$\\
\hline 
$\begin{matrix}\{1,1,1,1\}&\{1,2,2,3\}\\ \{1,1,1,1\}&\{1,2,3,4\}\end{matrix}$&$SU(2)$&$SU(3)$&$(1,1,0,0)$&${SU(6)}_6\times U(1)+\text{10 free hypers}$\\
\hline
$\begin{matrix}\{1,1,1,1\}&\{1,2,3,3\}\\ \{1,1,1,1\}&\{1,2,3,4\}\end{matrix}$&$SU(2)$&$SU(4)$&$(1,1,1,0)$&${SU(8)}_8\times U(1)+\text{5 free hypers}$\\
\hline
$\begin{matrix}\{1,1,1,1\}&\{1,2,3,4\}\\ \{1,1,1,1\}&\{1,2,3,4\}\end{matrix}$&$SU(2)$&$SU(5)$&$(1,1,1,1)$&${SU(10)}_{10}\times U(1)$\\
\hline
$\begin{matrix}\{1,1,1,1\}&\{1,1,2,2\}\\ \{1,1,2,2\}&\{1,2,3,4\}\end{matrix}$&$SU(2)$&$Sp(2)$&$(1,0,1,0)$&${SO(12)}_8+\text{5 free hypers}$\\
\hline
$\begin{matrix}\{1,1,1,1\}&\{1,2,2,3\}\\ \{1,1,2,2\}&\{1,2,2,3\}\end{matrix}$&$SU(2)$&$SU(3)$&$(1,1,0,0)$&${SU(6)}_6\times U(1)+\text{5 free hypers}$\\
\hline
$\begin{matrix}\{1,1,1,1\}&\{1,2,3,3\}\\ \{1,1,2,2\}&\{1,2,3,3\}\end{matrix}$&$SU(2)$&$SU(4)$&$(1,1,2,0)$&$\begin{aligned}{SU(4)}_8^2&\times {SU(2)}_8\times U(1)\\ & +\text{1 free hyper}\end{aligned}$\\
\hline
$\begin{matrix}\{1,1,1,1\}&\{1,2,3,4\}\\ \{1,1,2,2\}&\{1,2,3,4\}\end{matrix}$&$SU(2)$&$SU(5)$&$(1,1,2,2)$&${SU(5)}_{10}^2\times {U(1)}^2$\\
\hline
$\begin{matrix}\{1,1,1,1\}&\{1,2,2,2\}\\ \{1,2,2,2\}&\{1,2,3,4\}\end{matrix}$&$SU(3)$&$SU(4)$&$(1,2,1,0)$&$\begin{aligned}{SU(5)}_{8}&\times{SU(2)}_6^2\times {U(1)}^2\\ & +\text{5 free hypers}\end{aligned}$\\
\hline
\end{tabular}

\begin{tabular}{|c|c|c|c|l|}
\hline
4-punctured Sphere&$G_{\text{gauge}}$&$G'_{\text{gauge}}$&$(d_2,d_3,d_4,d_5)$&$G_{\text{global}}$\\
\hline
$\begin{matrix}\{1,1,1,1\}&\{1,2,2,3\}\\ \{1,2,2,2\}&\{1,2,2,3\}\end{matrix}$&$SU(3)$&$SU(3)$&$(1,2,0,0)$&${SU(3)}_6^2\times {U(1)}+\text{4 free hypers}$\\
\hline
$\begin{matrix}\{1,1,1,1\}&\{1,2,3,3\}\\ \{1,2,2,2\}&\{1,2,3,3\}\end{matrix}$&$SU(3)$&$SU(4)$&$(1,2,2,0)$&$\begin{aligned}{SU(4)}_8^2&\times{SU(2)}_6\times {U(1)}^2\\ & +\text{1 free hyper}\end{aligned}$\\
\hline
$\begin{matrix}\{1,1,1,1\}&\{1,2,3,4\}\\ \{1,2,2,2\}&\{1,2,3,4\}\end{matrix}$&$SU(3)$&$SU(5)$&$(1,2,2,2)$&${SU(5)}_{10}^2\times{SU(2)}_6\times {U(1)}^2$\\
\hline
$\begin{matrix}\{1,1,1,1\}&\{1,2,2,3\}\\ \{1,2,2,3\}&\{1,2,3,3\}\end{matrix}$&$SU(3)$&$SU(4)$&$(1,2,1,1)$&${SU(4)}_{10}\times{SU(3)}_8\times {U(1)}^3$\\
\hline
$\begin{matrix}\{1,1,1,1\}&\{1,2,2,3\}\\ \{1,2,2,3\}&\{1,2,3,4\}\end{matrix}$&$SU(3)$&$SU(5)$&$(1,2,1,2)$&${SU(5)}_{10}\times{SU(2)}_{10}^2\times {U(1)}^2$\\
\hline
$\begin{matrix}\{1,1,1,1\}&\{1,2,3,3\}\\ \{1,2,2,3\}&\{1,2,3,3\}\end{matrix}$&$SU(3)$&$SU(4)$&$(1,2,2,1)$&${SU(3)}_{10}\times{SU(3)}_{8}^2\times {U(1)}^3$\\
\hline
$\begin{matrix}\{1,1,1,1\}&\{1,2,3,4\}\\ \{1,2,2,3\}&\{1,2,3,4\}\end{matrix}$&$SU(3)$&$SU(5)$&$(1,2,2,3)$&${SU(5)}_{10}^2\times{SU(2)}_{10}\times {U(1)}^2$\\
\hline
$\begin{matrix}\{1,1,1,1\}&\{1,2,3,3\}\\ \{1,2,3,3\}&\{1,2,3,4\}\end{matrix}$&$SU(4)$&$SU(5)$&$(1,2,3,2)$&${SU(5)}_{10}\times{SU(3)}_{8}^2\times {U(1)}^3$\\
\hline
$\begin{matrix}\{1,1,1,1\}&\{1,2,3,4\}\\ \{1,2,3,3\}&\{1,2,3,4\}\end{matrix}$&$SU(4)$&$SU(5)$&$(1,2,3,3)$&${SU(5)}_{10}^2\times{SU(3)}_{8}\times {U(1)}^2$\\
\hline
$\begin{matrix}\{1,1,2,2\}&\{1,2,2,2\}\\ \{1,1,2,2\}&\{1,2,2,3\}\end{matrix}$&$Sp(2)$&$SU(4)$&$(1,1,1,0)$&$\begin{aligned}{SU(4)}_{8}&\times{Sp(2)}_{6}\times {U(1)}+\\ & \text{2 free hypers}\end{aligned}$\\
\hline
\end{tabular}

\begin{tabular}{|c|c|c|c|l|}
\hline
4-punctured Sphere&$G_{\text{gauge}}$&$G'_{\text{gauge}}$&$(d_2,d_3,d_4,d_5)$&$G_{\text{global}}$\\
\hline
$\begin{matrix}\{1,1,2,2\}&\{1,2,2,2\}\\ \{1,1,2,2\}&\{1,2,3,3\}\end{matrix}$&$Sp(2)$&$SU(4)$&$(1,1,2,0)$&$\begin{aligned}{SU(4)}_{8}&\times{SU(2)}_{8}^2\times {SU(2)}_6\times {U(1)}\\ & + \text{1 free hyper}\end{aligned}$\\
\hline
$\begin{matrix}\{1,1,2,2\}&\{1,2,2,2\}\\ \{1,1,2,2\}&\{1,2,3,4\}\end{matrix}$&$Sp(2)$&$SU(4)$&$(1,1,2,1)$&${SU(6)}_{10}\times {SU(2)}_6\times {U(1)}^2$\\
\hline
$\begin{matrix}\{1,1,2,2\}&\{1,2,2,3\}\\ \{1,1,2,2\}&\{1,2,2,3\}\end{matrix}$&$Sp(2)$&$SU(5)$&$(1,1,1,1)$&${SU(4)}_{10}\times {SU(2)}_{10}\times {U(1)}^2$\\
\hline
$\begin{matrix}\{1,1,2,2\}&\{1,2,2,3\}\\ \{1,1,2,2\}&\{1,2,3,3\}\end{matrix}$&$Sp(2)$&$SU(5)$&$(1,1,2,1)$&${SU(3)}_{10}\times {SU(3)}_{8}\times {U(1)}^3$\\
\hline
$\begin{matrix}\{1,1,2,2\}&\{1,2,2,3\}\\ \{1,1,2,2\}&\{1,2,3,4\}\end{matrix}$&$Sp(2)$&$SU(5)$&$(1,1,2,1)$&${SU(5)}_{10}\times {SU(2)}_{10}\times {U(1)}^3$\\
\hline
$\begin{matrix}\{1,1,2,2\}&\{1,2,3,3\}\\ \{1,1,2,2\}&\{1,2,3,3\}\end{matrix}$&$Sp(2)$&$SU(5)$&$(1,1,3,1)$&${SU(3)}_{8}^2\times {SU(2)}_{10}\times {U(1)}^3$\\
\hline
$\begin{matrix}\{1,1,2,2\}&\{1,2,3,3\}\\ \{1,1,2,2\}&\{1,2,3,4\}\end{matrix}$&$Sp(2)$&$SU(5)$&$(1,1,3,2)$&${SU(5)}_{10}\times {SU(3)}_{8}\times  {U(1)}^3$\\
\hline
$\begin{matrix}\{1,1,2,2\}&\{1,2,3,4\}\\ \{1,1,2,2\}&\{1,2,3,4\}\end{matrix}$&$Sp(2)$&$SU(5)$&$(1,1,3,3)$&${SU(5)}_{10}^2\times  {U(1)}^2$\\
\hline
$\begin{matrix}\{1,1,2,2\}&\{1,2,2,2\}\\ \{1,2,2,2\}&\{1,2,2,3\}\end{matrix}$&$SU(4)$&$SU(4)$&$(1,2,1,0)$&$\begin{aligned}{SU(3)}_{8}&\times{SU(2)}_6^3\times  {U(1)}^2\\ & +\text{2 free hypers}\end{aligned}$\\
\hline
$\begin{matrix}\{1,1,2,2\}&\{1,2,2,2\}\\ \{1,2,2,2\}&\{1,2,3,3\}\end{matrix}$&$SU(4)$&$SU(4)$&$(1,2,2,0)$&$\begin{aligned}{SU(4)}_{8}&\times{SU(2)}_{8}\times {SU(2)}_6^2\times  {U(1)}^2\\ &+\text{1 free hyper}\end{aligned}$\\
\hline
\end{tabular}

\begin{tabular}{|c|c|c|c|l|}
\hline
4-punctured Sphere&$G_{\text{gauge}}$&$G'_{\text{gauge}}$&$(d_2,d_3,d_4,d_5)$&$G_{\text{global}}$\\
\hline
$\begin{matrix}\{1,1,2,2\}&\{1,2,2,2\}\\ \{1,2,2,2\}&\{1,2,3,4\}\end{matrix}$&$SU(4)$&$SU(4)$&$(1,2,2,1)$&${SU(6)}_{10}\times {SU(2)}_6^2\times  {U(1)}^2$\\
\hline
$\begin{matrix}\{1,1,2,2\}&\{1,2,3,3\}\\ \{1,2,2,2\}&\{1,2,3,3\}\end{matrix}$&$SU(4)$&$SU(5)$&$(1,2,3,1)$&${SU(3)}_{8}^2\times {SU(2)}_{10}\times {SU(2)}_6\times  {U(1)}^3$\\
\hline
$\begin{matrix}\{1,1,2,2\}&\{1,2,3,4\}\\ \{1,2,2,2\}&\{1,2,3,4\}\end{matrix}$&$SU(4)$&$SU(5)$&$(1,2,3,3)$&${SU(5)}_{10}^2\times {SU(2)}_6\times  {U(1)}^2$\\
\hline
$\begin{matrix}\{1,1,2,2\}&\{1,2,2,3\}\\ \{1,2,2,3\}&\{1,2,3,3\}\end{matrix}$&$SU(5)$&$SU(5)$&$(1,2,2,2)$&${SU(3)}_{8}\times {SU(2)}_{10}^2\times {U(1)}^4$\\
\hline
$\begin{matrix}\{1,1,2,2\}&\{1,2,2,3\}\\ \{1,2,2,3\}&\{1,2,3,4\}\end{matrix}$&$SU(5)$&$SU(5)$&$(1,2,2,3)$&${SU(5)}_{10}\times {SU(2)}_{10}^2\times {U(1)}^3$\\
\hline
$\begin{matrix}\{1,1,2,2\}&\{1,2,3,3\}\\ \{1,2,2,3\}&\{1,2,3,3\}\end{matrix}$&$SU(5)$&$SU(5)$&$(1,2,3,2)$&${SU(3)}_{8}^2\times {SU(2)}_{10}\times {U(1)}^4$\\
\hline
$\begin{matrix}\{1,1,2,2\}&\{1,2,3,4\}\\ \{1,2,2,3\}&\{1,2,3,4\}\end{matrix}$&$SU(5)$&$SU(5)$&$(1,2,3,4)$&${SU(5)}_{10}^2\times {SU(2)}_{10}\times {U(1)}^2$\\
\hline
$\begin{matrix}\{1,1,2,2\}&\{1,2,3,3\}\\ \{1,2,3,3\}&\{1,2,3,4\}\end{matrix}$&$SU(5)$&$SU(5)$&$(1,2,4,3)$&${SU(5)}_{10}\times {SU(3)}_{8}^2\times {U(1)}^3$\\
\hline
$\begin{matrix}\{1,1,2,2\}&\{1,2,3,4\}\\ \{1,2,3,3\}&\{1,2,3,4\}\end{matrix}$&$SU(5)$&$SU(5)$&$(1,2,4,4)$&${SU(5)}_{10}^2\times {SU(3)}_{8}\times {U(1)}^2$\\
\hline
$\begin{matrix}\{1,2,2,2\}&\{1,2,2,3\}\\ \{1,2,2,2\}&\{1,2,2,3\}\end{matrix}$&$SU(4)$&$SU(5)$&$(1,3,1,1)$&${SU(4)}_{10}\times {SU(2)}_{6}^2\times {U(1)}^3$\\
\hline
\end{tabular}

\begin{tabular}{|c|c|c|c|l|}
\hline
4-punctured Sphere&$G_{\text{gauge}}$&$G'_{\text{gauge}}$&$(d_2,d_3,d_4,d_5)$&$G_{\text{global}}$\\
\hline
$\begin{matrix}\{1,2,2,2\}&\{1,2,2,3\}\\ \{1,2,2,2\}&\{1,2,3,3\}\end{matrix}$&$SU(4)$&$SU(5)$&$(1,3,2,1)$&${SU(3)}_{10}\times {SU(3)}_8\times {SU(2)}_{6}^2\times {U(1)}^3$\\
\hline
$\begin{matrix}\{1,2,2,2\}&\{1,2,2,3\}\\ \{1,2,2,2\}&\{1,2,3,4\}\end{matrix}$&$SU(4)$&$SU(5)$&$(1,3,2,2)$&${SU(5)}_{10}\times {SU(2)}_{10}\times {SU(2)}_{6}^2\times {U(1)}^3$\\
\hline
$\begin{matrix}\{1,2,2,2\}&\{1,2,3,3\}\\ \{1,2,2,2\}&\{1,2,3,3\}\end{matrix}$&$SU(4)$&$SU(5)$&$(1,3,3,1)$&${SU(3)}_{8}^2\times {SU(2)}_{6}^2\times {U(1)}^3$\\
\hline
$\begin{matrix}\{1,2,2,2\}&\{1,2,3,3\}\\ \{1,2,2,2\}&\{1,2,3,4\}\end{matrix}$&$SU(4)$&$SU(5)$&$(1,3,3,2)$&${SU(5)}_{10}\times {SU(3)}_{8}\times {SU(2)}_{6}^2\times {U(1)}^2$\\
\hline
$\begin{matrix}\{1,2,2,2\}&\{1,2,3,4\}\\ \{1,2,2,2\}&\{1,2,3,4\}\end{matrix}$&$SU(4)$&$SU(5)$&$(1,3,3,3)$&${SU(5)}_{10}^2\times {SU(2)}_{10}\times {SU(2)}_{6}^2\times {U(1)}^3$\\
\hline
$\begin{matrix}\{1,2,2,2\}&\{1,2,2,3\}\\ \{1,2,2,3\}&\{1,2,3,3\}\end{matrix}$&$SU(4)$&$SU(5)$&$(1,3,2,2)$&${SU(3)}_{8}\times {SU(2)}_{10}^2\times {SU(2)}_{6}\times {U(1)}^4$\\
\hline
$\begin{matrix}\{1,2,2,2\}&\{1,2,2,3\}\\ \{1,2,2,3\}&\{1,2,3,4\}\end{matrix}$&$SU(5)$&$SU(5)$&$(1,3,2,3)$&${SU(5)}_{10}\times {SU(2)}_{10}^2\times {SU(2)}_{6}\times {U(1)}^3$\\
\hline
$\begin{matrix}\{1,2,2,2\}&\{1,2,3,3\}\\ \{1,2,2,3\}&\{1,2,3,3\}\end{matrix}$&$SU(5)$&$SU(5)$&$(1,3,3,2)$&${SU(3)}_{8}^2\times {SU(2)}_{10}\times {SU(2)}_{6}\times {U(1)}^3$\\
\hline
$\begin{matrix}\{1,2,2,2\}&\{1,2,3,4\}\\ \{1,2,2,3\}&\{1,2,3,4\}\end{matrix}$&$SU(5)$&$SU(5)$&$(1,3,3,5)$&${SU(5)}_{10}^2\times {SU(2)}_{10}\times {SU(2)}_{6}\times {U(1)}^3$\\
\hline
$\begin{matrix}\{1,2,2,2\}&\{1,2,3,3\}\\ \{1,2,3,3\}&\{1,2,3,4\}\end{matrix}$&$SU(5)$&$SU(5)$&$(1,3,4,3)$&${SU(5)}_{10}\times {SU(3)}_{8}^2\times {SU(2)}_{6}\times {U(1)}^3$\\
\hline
\end{tabular}

\begin{tabular}{|c|c|c|c|l|}
\hline
4-punctured Sphere&$G_{\text{gauge}}$&$G'_{\text{gauge}}$&$(d_2,d_3,d_4,d_5)$&$G_{\text{global}}$\\
\hline
$\begin{matrix}\{1,2,2,2\}&\{1,2,3,4\}\\ \{1,2,3,3\}&\{1,2,3,4\}\end{matrix}$&$SU(5)$&$SU(5)$&$(1,3,4,4)$&${SU(5)}_{10}^2\times {SU(3)}_{8}\times {SU(2)}_{6}\times {U(1)}^2$\\
\hline
$\begin{matrix}\{1,2,2,3\}&\{1,2,3,3\}\\ \{1,2,2,3\}&\{1,2,3,4\}\end{matrix}$&$SU(5)$&$SU(5)$&$(1,3,3,4)$&${SU(5)}_{10}\times {SU(3)}_{8}\times {SU(2)}_{10}^2\times {U(1)}^3$\\
\hline
$\begin{matrix}\{1,2,2,3\}&\{1,2,3,4\}\\ \{1,2,2,3\}&\{1,2,3,4\}\end{matrix}$&$SU(5)$&$SU(5)$&$(1,3,3,5)$&${SU(5)}_{10}^2\times {SU(2)}_{10}^2\times {U(1)}^2$\\
\hline
$\begin{matrix}\{1,2,2,3\}&\{1,2,3,3\}\\ \{1,2,3,3\}&\{1,2,3,4\}\end{matrix}$&$SU(5)$&$SU(5)$&$(1,3,4,4)$&${SU(5)}_{10}\times {SU(3)}_8^2\times {SU(2)}_{10}\times {U(1)}^3$\\
\hline
$\begin{matrix}\{1,2,2,3\}&\{1,2,3,4\}\\ \{1,2,3,3\}&\{1,2,3,4\}\end{matrix}$&$SU(5)$&$SU(5)$&$(1,3,4,5)$&${SU(5)}_{10}^2\times {SU(3)}_8\times {SU(2)}_{10}\times {U(1)}^2$\\
\hline
$\begin{matrix}\{1,2,3,3\}&\{1,2,3,4\}\\ \{1,2,3,3\}&\{1,2,3,4\}\end{matrix}$&$SU(5)$&$SU(5)$&$(1,3,5,5)$&${SU(5)}_{10}^2\times {SU(3)}_8^2\times {U(1)}^2$\\
\hline
\end{tabular}
\bigskip

\begin{center}
Theories with three distinct S-duality frames
\end{center}

\begin{tabular}{|c|c|c|c|c|l|}
\hline
4-punctured Sphere&$G_{\text{gauge}}$&$G'_{\text{gauge}}$&$G''_{\text{gauge}}$&$(d_2,d_3,d_4,d_5)$&$G_{\text{global}}$\\
\hline
$\begin{matrix}\{1,1,1,1\}&\{1,2,2,2\}\\ \{1,1,2,2\}&\{1,2,3,4\}\end{matrix}$&$SU(2)$&$SU(3)$&$SU(4)$&$(1,1,1,0)$&$\begin{aligned}{SU(6)}_8&\times {SU(2)}_6\times U(1)+\\ & \text{5 free hypers}\end{aligned}$\\
\hline
$\begin{matrix}\{1,1,1,1\}&\{1,2,2,3\}\\ \{1,1,2,2\}&\{1,2,3,3\}\end{matrix}$&$SU(2)$&$SU(3)$&$SU(4)$&$(1,1,1,0)$&$\begin{aligned}{SU(6)}_8&\times {SU(2)}_6\times U(1)+\\ & \text{2 free hypers}\end{aligned}$\\
\hline
\end{tabular}

\begin{tabular}{|c|c|c|c|c|l|}
\hline
4-punctured Sphere&$G_{\text{gauge}}$&$G'_{\text{gauge}}$&$G''_{\text{gauge}}$&$(d_2,d_3,d_4,d_5)$&$G_{\text{global}}$\\
\hline
$\begin{matrix}\{1,1,1,1\}&\{1,2,2,3\}\\ \{1,1,2,2\}&\{1,2,3,4\}\end{matrix}$&$SU(2)$&$SU(3)$&$SU(5)$&$(1,1,1,1)$&${SU(7)}_{10}\times {U(1)}^2$\\
\hline
$\begin{matrix}\{1,1,1,1\}&\{1,2,3,3\}\\ \{1,1,2,2\}&\{1,2,3,4\}\end{matrix}$&$SU(2)$&$SU(4)$&$SU(5)$&$(1,1,2,1)$&${SU(6)}_{10}\times{SU(3)}_8\times {U(1)}^2$\\
\hline
$\begin{matrix}\{1,1,1,1\}&\{1,2,2,3\}\\ \{1,2,2,2\}&\{1,2,3,4\}\end{matrix}$&$SU(3)$&$SU(3)$&$SU(5)$&$(1,2,1,1)$&${SU(7)}_{10}\times{SU(2)}_6\times {U(1)}^2$\\
\hline
$\begin{matrix}\{1,1,1,1\}&\{1,2,3,3\}\\ \{1,2,2,2\}&\{1,2,3,4\}\end{matrix}$&$SU(3)$&$SU(4)$&$SU(5)$&$(1,2,2,1)$&$\begin{aligned}{SU(6)}_{10}&\times{SU(3)}_8\\ &\times {SU(2)}_6\times {U(1)}^2\end{aligned}$\\
\hline
$\begin{matrix}\{1,1,1,1\}&\{1,2,3,3\}\\ \{1,2,2,3\}&\{1,2,3,4\}\end{matrix}$&$SU(3)$&$SU(4)$&$SU(5)$&$(1,2,2,2)$&$\begin{aligned}{SU(5)}_{10}&\times{SU(3)}_8\\ &\times {SU(2)}_{10}\times {U(1)}^3\end{aligned}$\\
\hline
$\begin{matrix}\{1,1,2,2\}&\{1,2,2,3\}\\ \{1,2,2,2\}&\{1,2,3,4\}\end{matrix}$&$SU(4)$&$SU(5)$&$SU(5)$&$(1,2,2,2)$&$\begin{aligned}{SU(5)}_{10}&\times{SU(2)}_{10}\\ &\times {SU(2)}_{6}\times {U(1)}^3\end{aligned}$\\
\hline
$\begin{matrix}\{1,1,2,2\}&\{1,2,3,3\}\\ \{1,2,2,2\}&\{1,2,3,4\}\end{matrix}$&$SU(4)$&$SU(5)$&$SU(5)$&$(1,2,3,2)$&$\begin{aligned}{SU(5)}_{10}&\times{SU(3)}_{8}\\ &\times {SU(2)}_{6}\times {U(1)}^3\end{aligned}$\\
\hline
$\begin{matrix}\{1,1,2,2\}&\{1,2,3,3\}\\ \{1,2,2,3\}&\{1,2,3,4\}\end{matrix}$&$SU(5)$&$SU(5)$&$SU(5)$&$(1,2,3,3)$&$\begin{aligned}{SU(5)}_{10}&\times{SU(3)}_{8}\\ &\times {SU(2)}_{10}\times {U(1)}^3\end{aligned}$\\
\hline
$\begin{matrix}\{1,2,2,2\}&\{1,2,3,3\}\\ \{1,2,2,3\}&\{1,2,3,4\}\end{matrix}$&$SU(5)$&$SU(5)$&$SU(5)$&$(1,3,3,3)$&$\begin{aligned}{SU(5)}_{10}&\times{SU(3)}_{8}\times {SU(2)}_{10}\\ &\times {SU(2)}_6\times {U(1)}^2\end{aligned}$\\
\hline
$\begin{matrix}\{1,2,2,2\}&\{1,2,3,3\}\\ \{1,2,2,3\}&\{1,2,3,4\}\end{matrix}$&$SU(5)$&$SU(5)$&$SU(5)$&$(1,3,3,4)$&$\begin{aligned}{SU(5)}_{10}&\times{SU(3)}_{8}\times {SU(2)}_{10}\\ &\times {SU(2)}_6\times {U(1)}^3\end{aligned}$\\
\hline
\end{tabular}

\vfill\eject
\bibliographystyle{utphys}
\bibliography{gaiotto}

\end{document}